\documentclass[prl,aps,twocolumn,10pt,floatfix,nofootinbib,superscriptaddress]{revtex4-2}

\usepackage{amsmath}
\usepackage{amssymb}
\usepackage{mathtools}

\usepackage{amsthm}
\usepackage{stackrel}
\usepackage[normalem]{ulem}
\usepackage[flushleft]{threeparttable}

\usepackage{amsfonts}
\usepackage{txfonts}
\usepackage{dsfont}   
\usepackage{pifont}                
\usepackage{bbold}                  

\usepackage{array}
\usepackage{dcolumn}                
\usepackage{makecell}               
\usepackage{multirow}

\usepackage{enumerate}
\usepackage[shortlabels]{enumitem}

\usepackage[usenames,dvipsnames]{xcolor}
\usepackage{graphicx}
\graphicspath{{figs/}} 

\usepackage[version=4]{mhchem}      
\usepackage{physics}                
\usepackage{csquotes}               

\usepackage[colorlinks=true,citecolor=cyan,linkcolor=magenta,filecolor=magenta]{hyperref}
\usepackage[capitalize]{cleveref}
\usepackage{orcidlink}
\usepackage{mathrsfs}
\usepackage{placeins} 

\DeclareMathAlphabet{\mathbbold}{U}{bbold}{m}{n}

\newcommand{\nn}{\nonumber}

\renewcommand{\c}[1]{\mathcal{#1}}

\newcommand{\cnsg}{\tilde{G}} 

\definecolor{AC}{rgb}{1,0.5,0}

\definecolor{IB}{rgb}{0,0,1}

\definecolor{JM}{rgb}{0,0.5,0}

\begin{document}

\title{Anderson localization transition in disordered hyperbolic lattices}

\author{Anffany Chen\,\orcidlink{0000-0002-0926-5801}}
\email{anffany@ualberta.ca}
\affiliation{Theoretical Physics Institute, University of Alberta, Edmonton, Alberta T6G 2E1, Canada}
\affiliation{Department of Physics, University of Alberta, Edmonton, Alberta T6G 2E1, Canada}

\author{Joseph Maciejko\,\orcidlink{0000-0002-6946-1492}}
\affiliation{Theoretical Physics Institute, University of Alberta, Edmonton, Alberta T6G 2E1, Canada}
\affiliation{Department of Physics, University of Alberta, Edmonton, Alberta T6G 2E1, Canada}

\author{Igor Boettcher\,\orcidlink{0000-0002-1634-4022}}
\affiliation{Theoretical Physics Institute, University of Alberta, Edmonton, Alberta T6G 2E1, Canada}
\affiliation{Department of Physics, University of Alberta, Edmonton, Alberta T6G 2E1, Canada}

\date{\today}

\begin{abstract}
We study Anderson localization in disordered
tight-binding models on
hyperbolic lattices. Such lattices are geometries intermediate between ordinary two-dimensional crystalline lattices, which localize at infinitesimal disorder, and Bethe lattices, which localize at strong disorder. Using state-of-the-art computational group theory methods to create large systems, we approximate the thermodynamic limit through appropriate periodic boundary conditions and numerically demonstrate the existence of an Anderson localization transition on the $\{8,3\}$ and $\{8,8\}$ lattices. We find unusually large critical disorder strengths, determine critical exponents, and observe a strong finite-size effect in the level statistics.
\end{abstract}

\maketitle

\emph{Introduction.}---Two-dimensional (2D) hyperbolic lattices with constant negative spatial curvature have recently been realized experimentally in circuit quantum electrodynamics (QED)~\cite{Kollar:2019}, topolectrical circuits ~\cite{Lenggenhager:2021,Zhang:2022,Chen2023,Zhang:2023}, topological photonics \cite{Huang2024}, and scattering wave networks \cite{Chen2024}. A hyperbolic $\{p,q\}$ lattice is a regular tiling of 2D hyperbolic space by $p$-sided polygonal faces and vertices of coordination number $q$, such that $(p-2)(q-2)>4$. The scientific relevance of hyperbolic lattices ranges from testing fundamental principles such as the anti-de Sitter/conformal field theory correspondence in tabletop experiments~\cite{Boyle2020,PhysRevD.102.034511,PhysRevD.103.094507,PhysRevLett.130.091604,10.21468/SciPostPhys.13.5.103,BasteiroArxiv,chen2023ads,dey2024simulating}, to applications in quantum computation \cite{PhysRevLett.99.220405,PhysRevLett.101.110501,PhysRevD.86.065007,PhysRevLett.110.100402,Bao_2017} and quantum error-correction \cite{pastawski2015holographic,Breuckmann_2016,Breuckmann_2017,Lavasani_2019,jahn2021holographic,AliError}. As a new class of synthetic materials, they host a plethora of exotic physical properties beyond those identified in conventional Euclidean lattices, such as nontrivial crystalline symmetries~\cite{Boettcher:2020,Boettcher:2022,Chen2023symmetry}, generalized Bloch states~\cite{Maciejko:2021,Maciejko:2022,Cheng:2022,Lenggenhager:2023,shankar2023,kienzle2022,nagy2023,HinrichsenArxiv}, modified role of interactions \cite{Bienias:2022,Bitan1,Bitan2}, unusual flat bands~\cite{Bzdusek:2022,Mosseri:2022}, and novel topological phenomena~\cite{Yu:2020,Zhang:2022,Zhang:2023,Chen2023symmetry,Urwyler:2022,Liu:2022,Liu:2022b,pei2023,Tao:2022,Tummuru2023}.

Inevitable in real experiments, disorder in lattice systems can be detrimental to the performance of quantum devices, but also lead to novel physical phenomena such as Anderson localization~\cite{Anderson1958,Evers2008,Lagendijk2009}. Prior investigations of disorder-induced localization primarily focused on tight-binding models on Euclidean lattices, where  single-particle states  in 2D become localized upon presence of arbitrarily weak quenched disorder, an effect known as weak localization~\cite{bergmann1984}. Anderson localization has also been studied for tight-binding models on tree-like Bethe lattices ~\cite{abou-chacra1973,abou-chacra1974,mirlin1991} and random regular graphs~\cite{Tikhonov2016,Biroli2018arxiv,Tikhonov2019,Parisi2020,Tikhonov2021,Herre2023,Sierant2023}, the latter being the finite-sized counterparts of the former with periodic boundary conditions. These lattices can be viewed as the $p\to\infty$ limit of $\{p,q\}$ hyperbolic lattices~\cite{kollar2019line} and exhibit a localization transition at finite disorder strength. Other non-Euclidean graphs exhibiting the Anderson localization transition include small-world networks \cite{Mata2017,Mata2022} and Erd\H{o}s-R\'enyi graphs \cite{Sade2005,Mard2017, Alt2021}.

\begin{figure}
\includegraphics[width=\linewidth]{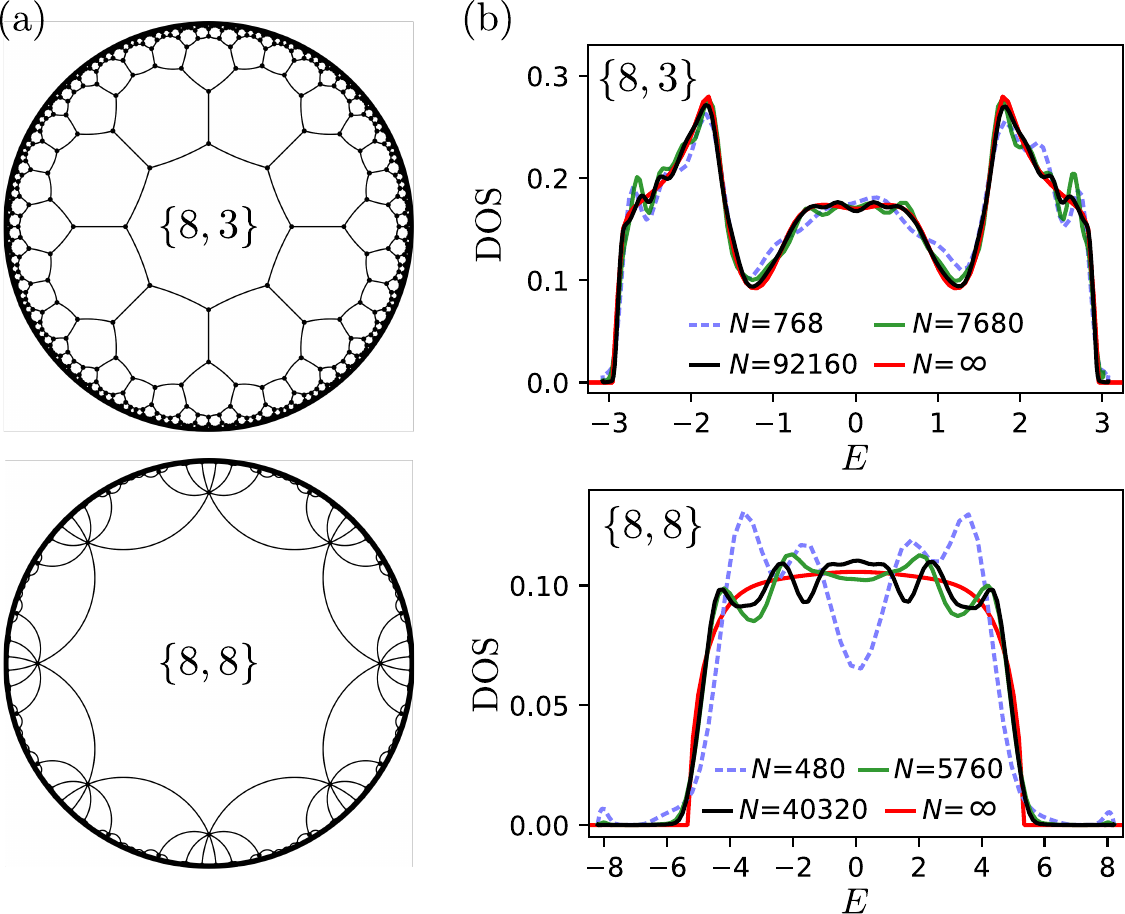}
\caption{\textbf{Approaching the thermodynamic-limit with finite-sized hyperbolic lattices.} (a) To study the localization phenomenon in 
hyperbolic space, we implement the Anderson model 
\eqref{eq:anderson} on $\{8,3\}$ and $\{8,8\}$ hyperbolic lattices, here shown as embedded in the Poincar\'{e} disk. (b) To avoid boundary localization, we construct finite-sized hyperbolic lattices with periodic boundary conditions, dubbed PBC clusters. As system size $N$ increases, the DOS of the disorder-free system $(u_i=0$) on our PBC clusters
converges to the thermodynamic-limit DOS.
The latter (red curve) is captured accurately by the continued-fraction method~\cite{Mosseri2023}.}
\label{fig:cluster_dos} 
\end{figure}

Hyperbolic $\{p,q\}$ lattices with $p$ finite are naturally considered two-dimensional since they correspond to regular tilings of the 2D hyperbolic plane, as shown in Fig.~\ref{fig:cluster_dos}(a). At the same time, much like Bethe lattices, the number of $n$-walks starting from a given site grows exponentially with $n$. Thus, hyperbolic lattices share aspects of both conventional 2D lattices and tree-like lattices.
However, their localization properties under disorder are mostly uncharted. The robustness of certain topological features against disorder has been investigated in Refs. \onlinecite{Yu:2020,Urwyler:2022}.
For continuum models, it has recently been argued that negative curvature prevents weak localization~\cite{curtis2023absence}, hinting at the possibility of an Anderson transition in hyperbolic lattices.

In this Letter, we explicitly demonstrate the existence of an Anderson localization transition on hyperbolic lattices and characterize its properties. We first present a heuristic argument based on classical random walks that disordered hyperbolic $\{p,q\}$ lattices should exhibit a localization transition at a finite critical disorder strength. We then verify this hypothesis by numerical simulations of the Anderson model [Eq.~(\ref{eq:anderson})] on finite $\{8,3\}$ and $\{8,8\}$  lattices with up to $\c{O}(10^4)$ sites and periodic boundary conditions (PBC). These so-called PBC clusters provide a reliable approximation of the infinite lattice and prevent spurious localization on the boundary, which, unlike a Euclidean boundary, would contain a macroscopic number of sites. We present our state-of-the-art technique for constructing large PBC clusters using computational group theory and benchmark it against the known thermodynamic-limit density of states (DOS) in the disorder-free system. For the disordered system, we compute the level statistics and inverse participation ratio (IPR) averaged over many disorder realizations. We conduct a finite-size scaling analysis to determine, for the first time, the critical disorder strengths $W_c$ and scaling length exponents $\nu$ on the $\{8,3\}$ and $\{8,8\}$ lattices. We find that $W_c/t \approx 15$ and $100$, respectively, revealing that hyperbolic lattices are very robust towards disorder, in contrast to 2D Euclidean lattices that exhibit weak localization. Furthermore, we observe a strong finite-size effect in the level statistics, which is a key feature of localization in random regular graphs~\cite{Tikhonov2016,Biroli2018arxiv,Tikhonov2021}.

\emph{Random walks on hyperbolic lattices.}---We first argue that disordered hyperbolic lattices should exhibit a localization transition at a finite critical disorder strength based on the theory of random walks~\cite{Einstein1905}.  Starting at an arbitrary site, a random walker has an equal probability to proceed to each of its neighbors at each time step. By P\'{o}lya's theorem \cite{Polya1921}, the expected number of returns (to the starting point) of a random walk in $d$-dimensional integer lattices $\mathbb{Z}^{d}$ is infinite for $d\le 2$ and finite for $d> 2$, implying that a Euclidean 1D or 2D random walk contains an infinite number of  loops. These loops have crucial implications  for transport properties, because the leading quantum correction to the Drude formula of conductivity is attributed to the quantum interference of clockwise and counterclockwise electronic trajectories along each loop \cite{Datta1995}. This so-called weak localization correction is large in 1D and 2D disordered Euclidean lattices, which greatly suppresses conduction. In contrast, we show below that the expected number of returns on a 2D  hyperbolic $\{p,q\}$ lattice is finite, corresponding to only a small weak-localization correction. This is more akin to the 3D Euclidean case, which is known to display a localization transition~\cite{Anderson1958}.

Consider a random walk on an infinite $\{p,q\}$ lattice, starting at an arbitrary site $i$. The expected number of returns is $\mu=\sum_{n=0}^\infty P_n$, where $P_{n}$ is the probability that an $n$-step walk (or $n$-walk) starts and ends at site $i$  (see \cite{supp} for derivation). $P_n$ is also the fraction of $n$-cycles among all $n$-walks.
The total number of $n$-walks is $q^{n}$. By graph theory, the number of $n$-cycles based at site $i$ is $\left(A^{n}\right)_{ii}$, with $A$ the adjacency matrix of the infinite lattice. Diagonalizing $A$ such that $A=\sum_a\lambda_{a}|\psi_{a}\rangle\langle\psi_{a}|$,
we have $\left(A^{n}\right)_{ii}=\sum_a\lambda_{a}^{n}|\langle i|\psi_{a}\rangle|^{2}$, where $|i\rangle$ is the localized state at site $i$ in the position basis. Denoting by $\lambda_{\rm r}=\max_a|\lambda_a|$ the spectral radius
of $A$, we then find:
\begin{equation}
\begin{aligned}
\left(A^{n}\right)_{ii}\le\sum_a |\lambda_{a}|^{n}\; |\langle i|\psi_{a}\rangle|^{2} 
\leq \lambda_{\rm r}^{n}\sum_a|\langle i|\psi_{a}\rangle|^{2}=\lambda_{\rm r}^{n}.
\end{aligned}
\end{equation}
The last equality follows from completeness of the $|\psi_a\rangle$ basis.
Therefore, the fraction of $n$-cycles is bounded from above according to $P_{n}=\left(A^{n}\right)_{ii}/q^{n}\leq(\lambda_{\rm r}/q)^{n}$. Unlike in Euclidean lattices, the symmetry group of a hyperbolic $\{p,q\}$ lattice exhibits the mathematical property of non-amenability \cite{Kesten1959,Brooks1982} and, as a result, its spectral radius $\lambda_{\rm r}$ is strictly less than $q$~\cite{Woess2000,kollar2019line}. Hence the expected number of returns,
\begin{equation}
\mu=\stackrel[n=0]{\infty}{\sum}P_{n}\le\stackrel[n=0]{\infty}{\sum}\left(\frac{\lambda_{\rm r}}{q}\right)^{n}=\frac{1}{1-\lambda_{\rm r}/q}<\infty,
\end{equation}
is finite. This suggests that localization on disordered hyperbolic lattices occurs at a nonzero critical disorder strength.

\emph{Hyperbolic Anderson model}.---We formulate the Anderson model~\cite{Anderson1958} on a hyperbolic $\{p,q\}$ lattice by the tight-binding Hamiltonian
\begin{align}
\label{eq:anderson}
H=-t\sum_{\langle i,j\rangle}(c_{i}^{\dagger}c_{j}+c_{j}^{\dagger}c_{i})+\sum_iu_{i}c_{i}^{\dagger}c_{i},
\end{align}
where $c_i^\dag$ ($c_i$) creates (annihilates) a particle on site $i$, $t$ is the nearest-neighbor hopping amplitude, and the on-site potentials $u_{i}$ are randomly drawn from a uniform distribution over the interval $[-\frac{W}{2},\frac{W}{2}]$. The Hamiltonian is motivated in part by the circuit QED experiments of Ref. \onlinecite{Kollar:2019}, where $t\sim 100\ \text{MHz}$ and the on-site potentials have a mean value of $8\ \text{GHz}$ with $\sim$$10\ \text{MHz}$ variations. The realized system can thus be modeled by Eq.~\eqref{eq:anderson} with $W/t \sim0.1$. Henceforth we set $t=1$ and measure energies and $W$ in units of $t$.

To demonstrate the localization transition on an infinite hyperbolic lattice and characterize its properties, we study the Hamiltonian (\ref{eq:anderson}) on large but finite lattices with PBC. Hyperbolic lattices with open boundary conditions, as shown in Fig.~\ref{fig:cluster_dos}(a), are discussed, for instance, in Ref.~\onlinecite{Chen2023symmetry}. Properly formulated PBC are essential in the hyperbolic context to systematically approach the thermodynamic limit, yet existing methods are limited either in the system size \cite{Conder:2007} or the number of realizations \cite{Lux2022}. In this work, we propose a novel method which can produce unlimited realizations of hyperbolic PBC clusters with arbitrarily large system sizes. Moreover, the resulting clusters are defect-free systems with translation symmetry. To describe our method, we start by considering a finite patch of the infinite lattice and compactify it by identifying pairs of boundary edges, 
resulting in a tessellation of a high-genus Riemann surface~\cite{Maciejko:2022,Boettcher:2022,Stegmaier:2021}. The compactification is achieved through computing the quotient group $C=\Gamma/G$, where $\Gamma$ is the translation symmetry group of the hyperbolic lattice and $G$ is a normal subgroup of $\Gamma$ (denoted $G\vartriangleleft \Gamma$). In other words, each element in $\Gamma$ corresponds to a site in the infinite lattice, and $G$ contains the equivalence relations which identify each site in the chosen patch with its (infinitely many) equivalent sites outside the patch. The normality constraint ensures that the cluster preserves a notion of translation symmetry and that no dislocations are introduced by the PBC. Each coset $[g]\in\Gamma/G$ represents a site in the PBC cluster, and two sites $[g_{1}],[g_{2}]$ are neighbors if  $[g_{1}]=[g_{2}][\gamma_{i}]$ with $\gamma_{i}$ a generator of $\Gamma$~\cite{Maciejko:2022}. The number of cosets, i.e., the order of the group $C$, corresponds to the number of Bravais unit cells in the cluster, denoted by $N=|C|$.

However, not all choices of normal subgroup $G$ give rise to PBC clusters that can correctly approximate the DOS in the thermodynamic limit~\cite{Lux2022,Lux2023}. For this, we have to construct a so-called \textit{coherent sequence} of finite-index normal subgroups, $\{\cnsg_i\}$, such that $\cnsg_{i}\vartriangleleft\Gamma$ for all $i$ and
\begin{equation} \Gamma\vartriangleright \cnsg_{1}\vartriangleright \cnsg_{2}\vartriangleright \cnsg_{3}\vartriangleright \dots 
\label{eq:nested}
\end{equation} 
In addition, $\bigcap_{i=1}^\infty \cnsg_{i}=\{e\}$,
where $\{e\}$ is the trivial group. Under these conditions, the PBC clusters $C_{i}=\Gamma/\cnsg_i$  approach the thermodynamic limit as $i$ increases. We use the computational algebra software GAP~\cite{LINS,GAP4}  for generating finite coherent sequences through subgroup intersections of low-index normal subgroups \cite{FirthThesis,Conder2005} (see \cite{supp} for methods). We construct four finite coherent sequences of \{8,8\} clusters with up to $N\sim 40\ 000$ sites. By replacing  each vertex of the $\{8,8\}$ cluster with a 16-site unit cell~\cite{Boettcher:2022,supp}, this also yields coherent sequences of $\{8,3\}$ clusters with up to $N\sim 90\ 000$ sites. The adjacency matrices of the generated clusters are available at \cite{Chen2023localization:SDC}. Figure~\ref{fig:cluster_dos}(b) shows the DOS of the disorder-free systems obtained from our first PBC cluster sequence, computed with the kernel polynomial method~\cite{KPM} using the Kwant Python package~\cite{Groth2014}. As the cluster size increases, the cluster DOS gradually approaches the thermodynamic limit~\cite{Mosseri2023} and more exact values of low DOS moments are reproduced (see Tables S2 and S3 of \cite{supp}), indicating convergence.

Having demonstrated that our sequences of PBC clusters accurately capture the thermodynamic limit in the clean limit, we next diagonalize the Anderson model \eqref{eq:anderson} on \{8,8\} clusters with $N=100$ to $40\ 320$
and \{8,3\} clusters with $N=160$ to $92\ 160$. The single-particle Hamiltonian is given by the $N\times N$ adjacency matrix of the cluster (multiplied by $-1$) plus a diagonal matrix with random values in the range $[-\frac{W}{2},\frac{W}{2}]$. We consider 1\ 000 to 100\ 000 disorder realizations, with more realizations for smaller clusters. For each realization, we use the Jacobi--Davidson algorithm through the software code of Ref.~\onlinecite{Bollhofer2007} to obtain the 20 eigenenergies $E_\alpha$ and eigenstates $\psi_\alpha$ closest to the center of the energy spectrum ($E=0$ for the lattices considered here). We focus on such eigenstates because localization generally occurs first at the outer edges of the spectrum and gradually shifts toward the center as $W$ increases, resulting in the so-called ``mobility edge" structure in observables such as the IPR. Therefore, localization at $E \sim 0$ marks the localization phase transition. We verified that hyperbolic Anderson models indeed exhibit a mobility edge in the IPR.

\emph{Level statistics and inverse participation ratio.}---Level statistics, i.e., the distribution of consecutive gaps in an energy spectrum, offers critical insights into wave function localization \cite{Shklovskii1993,Mirlin2000Statisics,Evers2008}. Two delocalized wave functions are coupled due to their spatial overlaps, making degeneracy in their energy eigenvalues unfavorable due to level repulsion. As a result, the level statistics in the delocalized phase follows that of the Wigner--Dyson Gaussian orthogonal ensemble (GOE). In the localized phase, wave functions exhibit minimal overlap. Their energies are independent and resemble random values along a line, with level statistics described by the Poisson distribution.

\begin{figure}
\includegraphics[width=0.85\linewidth]{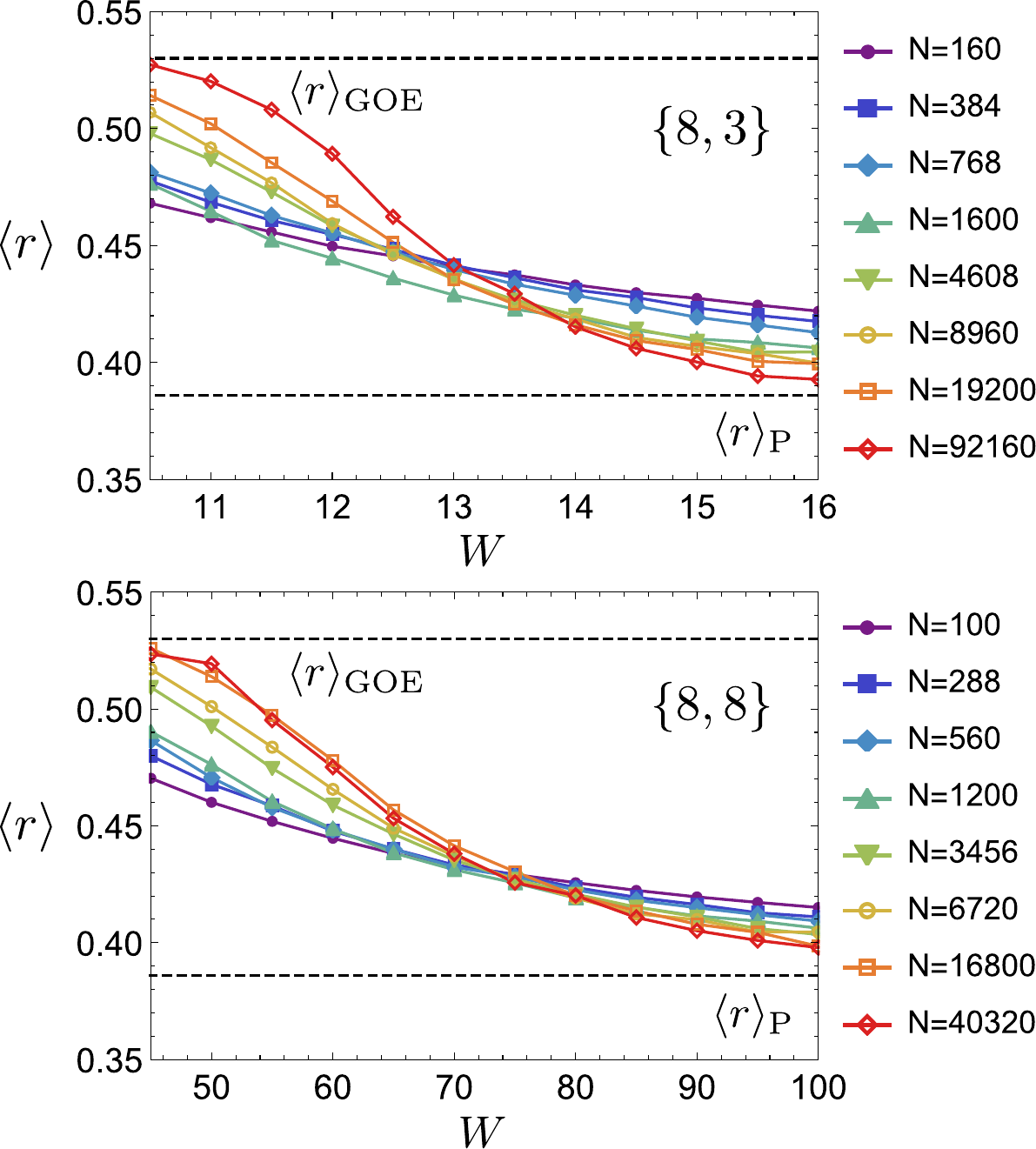}
\caption{\textbf{Finite-size effect in the average gap ratio.} The average gap ratio $\langle r\rangle$, averaged over disorder realizations and 20 eigenstates closest to energy $E=0$, transitions from the GOE ensemble value $\langle r\rangle_{\rm GOE}=0.536$ to the Poissonian value $\langle r\rangle_{\rm P}=0.386$, which signals a localization transition. We observe a strong finite-size effect such that the pairwise crossing of the  $\langle r\rangle$-curves drifts towards stronger $W$ and  $\langle r\rangle_{\rm P}$ as system size $N$ increases.}
\label{fig:r_drift} 
\end{figure}

For each cluster considered, we compute the level statistics of the Anderson model using the near-zero eigenvalues obtained above. Since the gaps between consecutive eigenvalues are strongly affected by the presence of finite-size-induced gaps, we circumvent this issue by considering the ratio of gaps in the sorted spectrum~\cite{Oganesyan2007},
\begin{equation}
0\le r_{\alpha}=\frac{\min\{E_{\alpha+1}-E_{\alpha},E_{\alpha}-E_{\alpha-1}\}}{\max\{E_{\alpha+1}-E_{\alpha},E_{\alpha}-E_{\alpha-1}\}}\le1. \label{eq:r}
\end{equation}
When binned into a histogram, $r_\alpha$ (compiled from different disorder realizations) follows a distribution that transitions from the Wigner surmise of GOE at small $W$ to the Poisson distribution at large $W$ \cite{supp}. This is most easily seen in the disorder-averaged expectation value $\langle r\rangle$, which changes from the GOE value $\langle r\rangle_{\text{GOE}}=4-2\sqrt{3}=0.536$~\cite{Atas2013} to the Poisson value $\langle r\rangle_{\text{P}}=2\ln2-1=0.386$ upon increasing $W$ (see Fig.~\ref{fig:r_drift} and Fig. S2 of \cite{supp}), signaling a localization transition. 

The value $\langle r \rangle$ at the critical disorder $W_c$ is typically independent of $N$ for Euclidean systems, albeit dependent on the boundary condition \cite{Evers2008}, so the intersection of $\langle r \rangle$-curves can be used to locate the critical point. However, here we observe a strong finite-size effect such that the intersection of $\langle r \rangle$-curves drifts toward stronger $W$ and $\langle r\rangle_\text{P}$ as $N$ increases, as shown in Fig.~\ref{fig:r_drift}. Such crossing drift has also been observed in the Anderson model on the $\{\infty,3\}$ lattices~\cite{Tikhonov2016,Biroli2018arxiv,Tikhonov2021}. We extrapolate the disorder strength at the intersection, denoted $W^*$, to the infinite-$N$ limit using the model $W^*=W_c - \beta/\ln(N)$ for some fitting parameter $\beta$ and determine $W_c\sim15$ and $100$ for $\{8,3\}$ and $\{8,8\}$ \cite{supp}.

\begin{figure}[t]
\includegraphics[width=\linewidth]{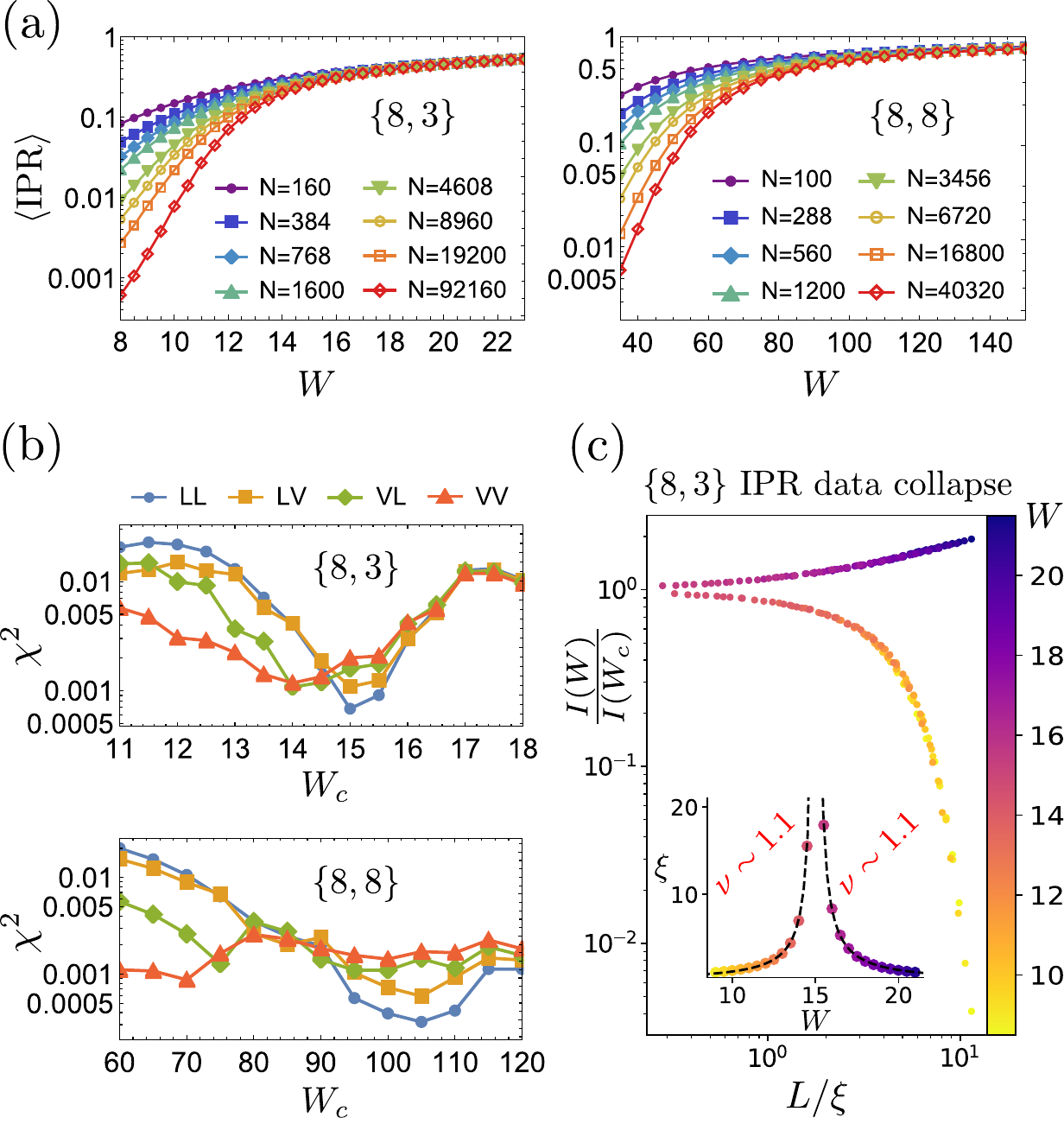}
\caption{\textbf{Finite-size scaling analysis of IPR.} (a) $\langle \rm{IPR}\rangle$ data used for the finite-size scaling analysis. (b) Assuming linear scaling behavior as in Eq.~\eqref{eq:lin_law} yields the optimal data collapse on both the delocalized and localized sides of the transition. This is indicated by the minimal $\chi^2$-values, where we denote LL: linear-linear, LV: linear-volumetric, etc. according to the scaling law used on either side of the transition. The best data collapse occurs at $W_c \approx 15$ for $\{8,3\}$ and $W_c \approx 100$ for $\{8,8\}$. (c) The collapsed IPR data of the $\{8,3\}$ model (see Fig. S5 of \cite{supp} for other collapses). The associated scaling length follows $\xi(W)\propto|W-W_c|^{-\nu}$ with critical exponent $\nu$ as indicated.}
\label{fig:IPR_analysis} 
\end{figure}

The IPR of a normalized wavefunction $\psi(z)$ is given by 
\begin{equation} 
 \text{IPR}(\psi) =\sum_{i=1}^N|\psi(z_{i})|^{4}, 
\end{equation}
where $z_{i}$ denote the site coordinates. If $\psi$ is highly delocalized with finite support on all sites, then $|\psi(z_{i})|^{2}\to 0$ as $N\to \infty$, leading to $\text{IPR}(\psi_{\text{deloc}})\to 0$. At the other extreme, if $\psi$ is localized on a single site $j$ such that $|\psi(z_{i})|^{2}\sim\delta_{ij}$, then $\text{IPR}(\psi_{\text{loc}})\sim1$. We compute the IPR for all $E\approx 0$ eigenstates obtained above, which are then averaged over disorder realizations. We find that the disorder-averaged $\langle \text{IPR}\rangle$ for the Anderson model on various PBC clusters increase from small values to $\sim 1$ at large disorder, suggesting a localization transition for both $\{8,3\}$ and $\{8,8\}$ lattices. Fig.~\ref{fig:IPR_analysis}(a) shows the $\langle \text{IPR}\rangle$ data used  to conduct the following finite-size scaling analysis.

\emph{Finite-size scaling and critical properties.}---Having established the localization transition on hyperbolic lattices, we now use finite-size scaling to extract its critical properties, including the critical disorder strength and critical exponents. We follow Ref.~\onlinecite{Mata2022} to conduct the finite-size scaling analysis of the observables  $\eta(W,N)\equiv(\langle r\rangle-\langle r\rangle_{\text{P}})/(\langle r\rangle_{\text{GOE}}-\langle r\rangle_{\text{P}})$ and $I(W,N)\equiv\langle \text{IPR} \rangle $ (see \cite{supp} for methods). The exponential growth of system size $N$ of hyperbolic and Bethe lattices with graph diameter $L$,  i.e., $N\sim e^{cL}$ for some lattice-dependent constant $c$,  suggests two potential scaling laws for a given observable $O$. Either we have
\begin{equation}
    O(W,N)=O(W_c,N)F_{\text{lin}}(L/\xi(W)), \label{eq:lin_law}
\end{equation}
with an unknown scaling function  $F_{\text{lin}}(L/\xi(W))$  and scaling length $\xi(W)$, or
\begin{equation}
O(W,N)=O(W_c,N)F_{\text{vol}}(N/\Lambda(W)), \label{eq:vol_law}
\end{equation} 
with scaling function $F_{\text{vol}}(N/\Lambda(W))$ and scaling volume $\Lambda(W)$. We refer to the two cases as linear and volumetric scaling, respectively. For $d$-dimensional Euclidean lattices, the two scaling behaviors are equivalent due to $N/\Lambda=(L/\xi)^{d}$. In contrast, in hyperbolic and Bethe lattices,  the ratio $N/\Lambda=e^{c(L-\xi)}$ is a function of $L-\xi$ instead of $L/\xi$. Therefore, we must examine both scaling laws separately and identify which one applies to the data. In the following we omit the lattice-dependent constant $c$ and let $L=\log(N)$.

We prepare curves of $\eta(W,N)$ and $I(W,N)$ as functions of $N$, each curve at a fixed $W$. We then make assumptions on the critical disorder $W_{c}$ and the scaling behaviors (either linear or volumetric) on the delocalized and localized sides of the transition. According to the scaling laws in Eqs.~\eqref{eq:lin_law} and \eqref{eq:vol_law}, all curves rescaled by the critical curve, i.e., $O(W,N)/O(W_{c},N)$, should collapse into a single scaling function. The quality of the collapse is measured by 
the least $\chi^{2}$, indicating an optimal assumption about $W_{c}$ and the scaling behaviors. 

As shown in Fig.~\ref{fig:IPR_analysis}(b), the $\chi^2$ obtained from collapsing $I(W,N)$ reveals a clear local minimum at $W_c\sim 15\ (100)$ for $\{8,3\}$ ($\{8,8\}$), consistent with the crossing-drift analysis.  We find that the $\chi^2$ obtained from collapsing $\eta(W,N)$ is less informative due to noise in the level statistics data \cite{supp}. Linear scaling of $I(W,N)$ in the vicinity of the transition gives the best data collapse on both the delocalized and localized sides. However, since we only consider system sizes up to $\sim$100 000, our result does not exclude the possibility of volumetric scaling for larger systems. Such finite-size crossover has been observed in the Anderson model on random regular graphs~\cite{Tikhonov2016}, such that the delocalized phase is characterized by a correlation volume $N_c(W)$, separating smaller systems $N<N_c(W)$ with linear scaling and larger systems $N>N_c(W)$ with volumetric scaling.

The critical points of $\{\infty,3\}$ and $\{\infty,8\}$ Anderson models have been estimated at $W_c\approx18$~\cite{abou-chacra1973,Tikhonov2019,Parisi2020,Sierant2023} and $W_c\sim110$ \cite{Herre2023} respectively. Comparing the four lattices $\{8,3\}$, $\{\infty,3\}$, $\{8,8\}$, and $\{\infty,8\}$, we find that the critical disorder $W_{c}$ increases with the magnitude of the lattice curvature in units of the lattice constant, which is $0.73$, $1.10$, $3.06$, and $3.23$, respectively \cite{Chen2023symmetry}. This can be attributed to negative curvature acting as an infrared regulator that suppresses the usual logarithmic divergence in the weak-localization correction in 2D~\cite{curtis2023absence,callan1990}. This suppression is more effective for stronger curvature, yielding a higher threshold to observe localization.

Assuming the linear-linear scaling law and $W_c\sim 15\ (100)$ for $\{8,3\}$ ($\{8,8\}$), we plot the collapsed  $\eta$ and $I$ data (Fig.~\ref{fig:IPR_analysis}(c) and Fig. S5 of \cite{supp}) and determine the scaling length exponent $\nu$ by fitting $\xi\propto|W-W_{c}|^{-\nu}$ to find $(\nu^{\{8,3\}}_{\eta,\text{deloc}},\nu^{\{8,3\}}_{\eta,\text{loc}})\approx(0.5,0.4)$,  $(\nu^{\{8,8\}}_{\eta,\text{deloc}},\nu^{\{8,8\}}_{\eta,\text{loc}})\approx(0.9,0.9)$, $(\nu^{\{8,3\}}_{I,\text{deloc}},\nu^{\{8,3\}}_{I,\text{loc}})\approx(1.1,1.1)$,  $(\nu^{\{8,8\}}_{I,\text{deloc}},\nu^{\{8,8\}}_{I,\text{loc}})\approx(1.1,0.9)$ on the delocalized and localized sides, respectively. For comparison, the corresponding critical exponents on random regular graphs and small-world networks with average coordination number of 3 are known to be $(\nu^{\{\infty,3\}}_{\eta,\text{deloc}},\nu^{\{\infty,3\}}_{\eta,\text{loc}})\approx (0.5,0.5)$ and $(\kappa^{\{\infty,3\}}_{I,\text{deloc}},\nu^{\{\infty,3\}}_{I,\text{loc}})\approx(0.5,1)$, where $\kappa^{\{\infty,3\}}_{I,\text{deloc}}$ is the critical exponent of the scaling volume~\cite{Sade2005,Tikhonov2019,Mata2017,Mata2022}.
We also found that the IPR at criticality follows the multifractal scaling $I(W_c) \propto L^{-\tau_2}$ with fractal dimension $\tau_2^{\{8,3\}}\!\sim\!0.3$ and $\tau_2^{\{8,8\}}\!\sim\!0.2$ \cite{supp}.

\emph{Conclusion.---}In this work, we have studied, for the first time, the Anderson localization transition on hyperbolic $\{p,q\}$ lattices. To eliminate 
boundary influence while preserving hyperbolic translation symmetry in the clean limit, we developed an efficient method to create large PBC clusters. We benchmarked the disorder-free system against the known thermodynamic limit and found very good agreement for large systems with $\mathcal{O}(10^4)$ sites and adequate agreement even with $\mathcal{O}(10^2)$ sites.
Through analyzing the level statistics and IPR of the Anderson models, we determined the critical disorder strengths on the $\{8,3\}$ and $\{8,8\}$ lattices to be $W_c\approx 15t$ and $100t$, respectively, implying high resilience against disorder on hyperbolic lattices. This understanding is instrumental in circuit QED applications, where small variations in on-site potentials lead to disorders with $W$ comparable to $t$. We revealed that hyperbolic lattices are genuinely distinct from 2D Euclidean lattices, which exhibit the localization of all eigenstates at infinitesimal disorder strength. Furthermore, they suffer from a strong finite-size effect near the localization transition. In particular, the pairwise intersection of $\langle r\rangle$-curves drifts towards strong disorder and the Poisson distribution, as also seen in the Anderson models on Bethe lattices and random regular graphs. Our results pave the way for future studies of hyperbolic localization.

The localization of wavefunctions can be realized experimentally in (otherwise clean) topolectrical circuits through creating artificial variation in local resistance. Besides localization, our PBC clusters, accessible at Ref.~\onlinecite{Chen2023localization:SDC}, are a powerful tool for various numerical studies of hyperbolic lattices. On the one hand, they are crucial for investigating bulk physics by emulating the thermodynamic limit while eliminating boundary effects. On the other hand, by introducing suitable vacancies, one can design a controlled study of the hyperbolic boundary, or defects in general, to test the bulk-boundary correspondence and emergent boundary phenomena.

\emph{Acknowledgements.---}We thank Jonathan Curtis, Victor Galitski, Alexey Gorshkov, and Canon Sun for valuable discussions. This research was enabled in part by support provided by Compute Ontario (\href{https://www.computeontario.ca/}{computeontario.ca}) and the Digital Research Alliance of Canada (\href{https://alliancecan.ca}{alliancecan.ca}). A.C. was supported by the Avadh Bhatia Fellowship at the University of Alberta. A.C. and I.B. acknowledge support through the University of Alberta startup fund UOFAB Startup Boettcher. J.M. was supported by NSERC Discovery Grants RGPIN-2020-06999 and RGPAS-2020-00064; the Canada Research Chair (CRC) Program; the Government of Alberta's Major Innovation Fund (MIF); and the Pacific Institute for the Mathematical Sciences (PIMS) Collaborative Research Group program. I.B. acknowledges funding from the NSERC Discovery Grants RGPIN-2021-02534 and DGECR2021-00043. 


\let\oldaddcontentsline\addcontentsline
\renewcommand{\addcontentsline}[3]{}
\bibliography{biblio}

\begin{thebibliography}{100}%
\makeatletter
\providecommand \@ifxundefined [1]{%
 \@ifx{#1\undefined}
}%
\providecommand \@ifnum [1]{%
 \ifnum #1\expandafter \@firstoftwo
 \else \expandafter \@secondoftwo
 \fi
}%
\providecommand \@ifx [1]{%
 \ifx #1\expandafter \@firstoftwo
 \else \expandafter \@secondoftwo
 \fi
}%
\providecommand \natexlab [1]{#1}%
\providecommand \enquote  [1]{``#1''}%
\providecommand \bibnamefont  [1]{#1}%
\providecommand \bibfnamefont [1]{#1}%
\providecommand \citenamefont [1]{#1}%
\providecommand \href@noop [0]{\@secondoftwo}%
\providecommand \href [0]{\begingroup \@sanitize@url \@href}%
\providecommand \@href[1]{\@@startlink{#1}\@@href}%
\providecommand \@@href[1]{\endgroup#1\@@endlink}%
\providecommand \@sanitize@url [0]{\catcode `\\12\catcode `\$12\catcode
  `\&12\catcode `\#12\catcode `\^12\catcode `\_12\catcode `\%12\relax}%
\providecommand \@@startlink[1]{}%
\providecommand \@@endlink[0]{}%
\providecommand \url  [0]{\begingroup\@sanitize@url \@url }%
\providecommand \@url [1]{\endgroup\@href {#1}{\urlprefix }}%
\providecommand \urlprefix  [0]{URL }%
\providecommand \Eprint [0]{\href }%
\providecommand \doibase [0]{https://doi.org/}%
\providecommand \selectlanguage [0]{\@gobble}%
\providecommand \bibinfo  [0]{\@secondoftwo}%
\providecommand \bibfield  [0]{\@secondoftwo}%
\providecommand \translation [1]{[#1]}%
\providecommand \BibitemOpen [0]{}%
\providecommand \bibitemStop [0]{}%
\providecommand \bibitemNoStop [0]{.\EOS\space}%
\providecommand \EOS [0]{\spacefactor3000\relax}%
\providecommand \BibitemShut  [1]{\csname bibitem#1\endcsname}%
\let\auto@bib@innerbib\@empty
\bibitem [{\citenamefont {Koll{\'{a}}r}\ \emph {et~al.}(2019)\citenamefont
  {Koll{\'{a}}r}, \citenamefont {Fitzpatrick},\ and\ \citenamefont
  {Houck}}]{Kollar:2019}%
  \BibitemOpen
  \bibfield  {author} {\bibinfo {author} {\bibfnamefont {A.~J.}\ \bibnamefont
  {Koll{\'{a}}r}}, \bibinfo {author} {\bibfnamefont {M.}~\bibnamefont
  {Fitzpatrick}},\ and\ \bibinfo {author} {\bibfnamefont {A.~A.}\ \bibnamefont
  {Houck}},\ }\bibfield  {title} {\bibinfo {title} {{Hyperbolic lattices in
  circuit quantum electrodynamics}},\ }\href
  {https://doi.org/10.1038/s41586-019-1348-3} {\bibfield  {journal} {\bibinfo
  {journal} {Nature}\ }\textbf {\bibinfo {volume} {571}},\ \bibinfo {pages}
  {45} (\bibinfo {year} {2019})}\BibitemShut {NoStop}%
\bibitem [{\citenamefont {Lenggenhager}\ \emph {et~al.}(2022)\citenamefont
  {Lenggenhager}, \citenamefont {Stegmaier}, \citenamefont {Upreti},
  \citenamefont {Hofmann}, \citenamefont {Helbig}, \citenamefont {Vollhardt},
  \citenamefont {Greiter}, \citenamefont {Lee}, \citenamefont {Imhof},
  \citenamefont {Brand}, \citenamefont {Kie{\ss}ling}, \citenamefont
  {Boettcher}, \citenamefont {Neupert}, \citenamefont {Thomale},\ and\
  \citenamefont {Bzdu\v{s}ek}}]{Lenggenhager:2021}%
  \BibitemOpen
  \bibfield  {author} {\bibinfo {author} {\bibfnamefont {P.~M.}\ \bibnamefont
  {Lenggenhager}}, \bibinfo {author} {\bibfnamefont {A.}~\bibnamefont
  {Stegmaier}}, \bibinfo {author} {\bibfnamefont {L.~K.}\ \bibnamefont
  {Upreti}}, \bibinfo {author} {\bibfnamefont {T.}~\bibnamefont {Hofmann}},
  \bibinfo {author} {\bibfnamefont {T.}~\bibnamefont {Helbig}}, \bibinfo
  {author} {\bibfnamefont {A.}~\bibnamefont {Vollhardt}}, \bibinfo {author}
  {\bibfnamefont {M.}~\bibnamefont {Greiter}}, \bibinfo {author} {\bibfnamefont
  {C.~H.}\ \bibnamefont {Lee}}, \bibinfo {author} {\bibfnamefont
  {S.}~\bibnamefont {Imhof}}, \bibinfo {author} {\bibfnamefont
  {H.}~\bibnamefont {Brand}}, \bibinfo {author} {\bibfnamefont
  {T.}~\bibnamefont {Kie{\ss}ling}}, \bibinfo {author} {\bibfnamefont
  {I.}~\bibnamefont {Boettcher}}, \bibinfo {author} {\bibfnamefont
  {T.}~\bibnamefont {Neupert}}, \bibinfo {author} {\bibfnamefont
  {R.}~\bibnamefont {Thomale}},\ and\ \bibinfo {author} {\bibfnamefont
  {T.}~\bibnamefont {Bzdu\v{s}ek}},\ }\bibfield  {title} {\bibinfo {title}
  {Simulating hyperbolic space on a circuit board},\ }\href
  {https://doi.org/10.1038/s41467-022-32042-4} {\bibfield  {journal} {\bibinfo
  {journal} {Nat. Commun.}\ }\textbf {\bibinfo {volume} {13}},\ \bibinfo
  {pages} {4373} (\bibinfo {year} {2022})}\BibitemShut {NoStop}%
\bibitem [{\citenamefont {Zhang}\ \emph {et~al.}(2022)\citenamefont {Zhang},
  \citenamefont {Yuan}, \citenamefont {Sun}, \citenamefont {Sun},\ and\
  \citenamefont {Zhang}}]{Zhang:2022}%
  \BibitemOpen
  \bibfield  {author} {\bibinfo {author} {\bibfnamefont {W.}~\bibnamefont
  {Zhang}}, \bibinfo {author} {\bibfnamefont {H.}~\bibnamefont {Yuan}},
  \bibinfo {author} {\bibfnamefont {N.}~\bibnamefont {Sun}}, \bibinfo {author}
  {\bibfnamefont {H.}~\bibnamefont {Sun}},\ and\ \bibinfo {author}
  {\bibfnamefont {X.}~\bibnamefont {Zhang}},\ }\bibfield  {title} {\bibinfo
  {title} {Observation of novel topological states in hyperbolic lattices},\
  }\href {https://doi.org/10.1038/s41467-022-30631-x} {\bibfield  {journal}
  {\bibinfo  {journal} {Nat. Commun.}\ }\textbf {\bibinfo {volume} {13}},\
  \bibinfo {pages} {2937} (\bibinfo {year} {2022})}\BibitemShut {NoStop}%
\bibitem [{\citenamefont {Chen}\ \emph
  {et~al.}(2023{\natexlab{a}})\citenamefont {Chen}, \citenamefont {Brand},
  \citenamefont {Helbig}, \citenamefont {Hofmann}, \citenamefont {Imhof},
  \citenamefont {Fritzsche}, \citenamefont {Kie{\ss}ling}, \citenamefont
  {Stegmaier}, \citenamefont {Upreti}, \citenamefont {Neupert}, \citenamefont
  {Bzdu\v{s}ek}, \citenamefont {Greiter}, \citenamefont {Thomale},\ and\
  \citenamefont {Boettcher}}]{Chen2023}%
  \BibitemOpen
  \bibfield  {author} {\bibinfo {author} {\bibfnamefont {A.}~\bibnamefont
  {Chen}}, \bibinfo {author} {\bibfnamefont {H.}~\bibnamefont {Brand}},
  \bibinfo {author} {\bibfnamefont {T.}~\bibnamefont {Helbig}}, \bibinfo
  {author} {\bibfnamefont {T.}~\bibnamefont {Hofmann}}, \bibinfo {author}
  {\bibfnamefont {S.}~\bibnamefont {Imhof}}, \bibinfo {author} {\bibfnamefont
  {A.}~\bibnamefont {Fritzsche}}, \bibinfo {author} {\bibfnamefont
  {T.}~\bibnamefont {Kie{\ss}ling}}, \bibinfo {author} {\bibfnamefont
  {A.}~\bibnamefont {Stegmaier}}, \bibinfo {author} {\bibfnamefont {L.~K.}\
  \bibnamefont {Upreti}}, \bibinfo {author} {\bibfnamefont {T.}~\bibnamefont
  {Neupert}}, \bibinfo {author} {\bibfnamefont {T.}~\bibnamefont
  {Bzdu\v{s}ek}}, \bibinfo {author} {\bibfnamefont {M.}~\bibnamefont
  {Greiter}}, \bibinfo {author} {\bibfnamefont {R.}~\bibnamefont {Thomale}},\
  and\ \bibinfo {author} {\bibfnamefont {I.}~\bibnamefont {Boettcher}},\
  }\bibfield  {title} {\bibinfo {title} {{Hyperbolic matter in electrical
  circuits with tunable complex phases}},\ }\href
  {https://doi.org/https://doi.org/10.1038/s41467-023-36359-6} {\bibfield
  {journal} {\bibinfo  {journal} {Nat. Commun.}\ }\textbf {\bibinfo {volume}
  {14}},\ \bibinfo {pages} {622} (\bibinfo {year}
  {2023}{\natexlab{a}})}\BibitemShut {NoStop}%
\bibitem [{\citenamefont {Zhang}\ \emph {et~al.}(2023)\citenamefont {Zhang},
  \citenamefont {Di}, \citenamefont {Zheng}, \citenamefont {Sun},\ and\
  \citenamefont {Zhang}}]{Zhang:2023}%
  \BibitemOpen
  \bibfield  {author} {\bibinfo {author} {\bibfnamefont {W.}~\bibnamefont
  {Zhang}}, \bibinfo {author} {\bibfnamefont {F.}~\bibnamefont {Di}}, \bibinfo
  {author} {\bibfnamefont {X.}~\bibnamefont {Zheng}}, \bibinfo {author}
  {\bibfnamefont {H.}~\bibnamefont {Sun}},\ and\ \bibinfo {author}
  {\bibfnamefont {X.}~\bibnamefont {Zhang}},\ }\bibfield  {title} {\bibinfo
  {title} {{Hyperbolic band topology with non-trivial second Chern numbers}},\
  }\href {https://doi.org/10.1038/s41467-023-36767-8} {\bibfield  {journal}
  {\bibinfo  {journal} {Nat. Commun.}\ }\textbf {\bibinfo {volume} {14}},\
  \bibinfo {pages} {1083} (\bibinfo {year} {2023})}\BibitemShut {NoStop}%
\bibitem [{\citenamefont {Huang}\ \emph {et~al.}(2024)\citenamefont {Huang},
  \citenamefont {He}, \citenamefont {Zhang}, \citenamefont {Zhang},
  \citenamefont {Liu}, \citenamefont {Feng}, \citenamefont {Liu}, \citenamefont
  {Cui}, \citenamefont {Huang}, \citenamefont {Zhang},\ and\ \citenamefont
  {Zhang}}]{Huang2024}%
  \BibitemOpen
  \bibfield  {author} {\bibinfo {author} {\bibfnamefont {L.}~\bibnamefont
  {Huang}}, \bibinfo {author} {\bibfnamefont {L.}~\bibnamefont {He}}, \bibinfo
  {author} {\bibfnamefont {W.}~\bibnamefont {Zhang}}, \bibinfo {author}
  {\bibfnamefont {H.}~\bibnamefont {Zhang}}, \bibinfo {author} {\bibfnamefont
  {D.}~\bibnamefont {Liu}}, \bibinfo {author} {\bibfnamefont {X.}~\bibnamefont
  {Feng}}, \bibinfo {author} {\bibfnamefont {F.}~\bibnamefont {Liu}}, \bibinfo
  {author} {\bibfnamefont {K.}~\bibnamefont {Cui}}, \bibinfo {author}
  {\bibfnamefont {Y.}~\bibnamefont {Huang}}, \bibinfo {author} {\bibfnamefont
  {W.}~\bibnamefont {Zhang}},\ and\ \bibinfo {author} {\bibfnamefont
  {X.}~\bibnamefont {Zhang}},\ }\bibfield  {title} {\bibinfo {title}
  {Hyperbolic photonic topological insulators},\ }\href
  {https://doi.org/10.1038/s41467-024-46035-y} {\bibfield  {journal} {\bibinfo
  {journal} {Nat. Commun.}\ }\textbf {\bibinfo {volume} {15}},\ \bibinfo
  {pages} {1647} (\bibinfo {year} {2024})}\BibitemShut {NoStop}%
\bibitem [{\citenamefont {Chen}\ \emph
  {et~al.}(2024{\natexlab{a}})\citenamefont {Chen}, \citenamefont {Zhang},
  \citenamefont {Qin}, \citenamefont {Bossart}, \citenamefont {Yang},
  \citenamefont {Chen},\ and\ \citenamefont {Fleury}}]{Chen2024}%
  \BibitemOpen
  \bibfield  {author} {\bibinfo {author} {\bibfnamefont {Q.}~\bibnamefont
  {Chen}}, \bibinfo {author} {\bibfnamefont {Z.}~\bibnamefont {Zhang}},
  \bibinfo {author} {\bibfnamefont {H.}~\bibnamefont {Qin}}, \bibinfo {author}
  {\bibfnamefont {A.}~\bibnamefont {Bossart}}, \bibinfo {author} {\bibfnamefont
  {Y.}~\bibnamefont {Yang}}, \bibinfo {author} {\bibfnamefont {H.}~\bibnamefont
  {Chen}},\ and\ \bibinfo {author} {\bibfnamefont {R.}~\bibnamefont {Fleury}},\
  }\bibfield  {title} {\bibinfo {title} {{Anomalous and Chern topological waves
  in hyperbolic networks}},\ }\href
  {https://doi.org/10.1038/s41467-024-46551-x} {\bibfield  {journal} {\bibinfo
  {journal} {Nat. Commun.}\ }\textbf {\bibinfo {volume} {15}},\ \bibinfo
  {pages} {2293} (\bibinfo {year} {2024}{\natexlab{a}})}\BibitemShut {NoStop}%
\bibitem [{\citenamefont {Boyle}\ \emph {et~al.}(2020)\citenamefont {Boyle},
  \citenamefont {Dickens},\ and\ \citenamefont {Flicker}}]{Boyle2020}%
  \BibitemOpen
  \bibfield  {author} {\bibinfo {author} {\bibfnamefont {L.}~\bibnamefont
  {Boyle}}, \bibinfo {author} {\bibfnamefont {M.}~\bibnamefont {Dickens}},\
  and\ \bibinfo {author} {\bibfnamefont {F.}~\bibnamefont {Flicker}},\
  }\bibfield  {title} {\bibinfo {title} {{Conformal Quasicrystals and
  Holography}},\ }\href {https://doi.org/10.1103/PhysRevX.10.011009} {\bibfield
   {journal} {\bibinfo  {journal} {Phys. Rev. X}\ }\textbf {\bibinfo {volume}
  {10}},\ \bibinfo {pages} {011009} (\bibinfo {year} {2020})}\BibitemShut
  {NoStop}%
\bibitem [{\citenamefont {Asaduzzaman}\ \emph {et~al.}(2020)\citenamefont
  {Asaduzzaman}, \citenamefont {Catterall}, \citenamefont {Hubisz},
  \citenamefont {Nelson},\ and\ \citenamefont
  {Unmuth-Yockey}}]{PhysRevD.102.034511}%
  \BibitemOpen
  \bibfield  {author} {\bibinfo {author} {\bibfnamefont {M.}~\bibnamefont
  {Asaduzzaman}}, \bibinfo {author} {\bibfnamefont {S.}~\bibnamefont
  {Catterall}}, \bibinfo {author} {\bibfnamefont {J.}~\bibnamefont {Hubisz}},
  \bibinfo {author} {\bibfnamefont {R.}~\bibnamefont {Nelson}},\ and\ \bibinfo
  {author} {\bibfnamefont {J.}~\bibnamefont {Unmuth-Yockey}},\ }\bibfield
  {title} {\bibinfo {title} {{Holography on tessellations of hyperbolic
  space}},\ }\href {https://doi.org/10.1103/PhysRevD.102.034511} {\bibfield
  {journal} {\bibinfo  {journal} {Phys. Rev. D}\ }\textbf {\bibinfo {volume}
  {102}},\ \bibinfo {pages} {034511} (\bibinfo {year} {2020})}\BibitemShut
  {NoStop}%
\bibitem [{\citenamefont {Brower}\ \emph {et~al.}(2021)\citenamefont {Brower},
  \citenamefont {Cogburn}, \citenamefont {Fitzpatrick}, \citenamefont
  {Howarth},\ and\ \citenamefont {Tan}}]{PhysRevD.103.094507}%
  \BibitemOpen
  \bibfield  {author} {\bibinfo {author} {\bibfnamefont {R.~C.}\ \bibnamefont
  {Brower}}, \bibinfo {author} {\bibfnamefont {C.~V.}\ \bibnamefont {Cogburn}},
  \bibinfo {author} {\bibfnamefont {A.~L.}\ \bibnamefont {Fitzpatrick}},
  \bibinfo {author} {\bibfnamefont {D.}~\bibnamefont {Howarth}},\ and\ \bibinfo
  {author} {\bibfnamefont {C.-I.}\ \bibnamefont {Tan}},\ }\bibfield  {title}
  {\bibinfo {title} {{Lattice setup for quantum field theory in
  ${\mathrm{AdS}}_{2}$}},\ }\href {https://doi.org/10.1103/PhysRevD.103.094507}
  {\bibfield  {journal} {\bibinfo  {journal} {Phys. Rev. D}\ }\textbf {\bibinfo
  {volume} {103}},\ \bibinfo {pages} {094507} (\bibinfo {year}
  {2021})}\BibitemShut {NoStop}%
\bibitem [{\citenamefont {Basteiro}\ \emph {et~al.}(2023)\citenamefont
  {Basteiro}, \citenamefont {Dusel}, \citenamefont {Erdmenger}, \citenamefont
  {Herdt}, \citenamefont {Hinrichsen}, \citenamefont {Meyer},\ and\
  \citenamefont {Schrauth}}]{PhysRevLett.130.091604}%
  \BibitemOpen
  \bibfield  {author} {\bibinfo {author} {\bibfnamefont {P.}~\bibnamefont
  {Basteiro}}, \bibinfo {author} {\bibfnamefont {F.}~\bibnamefont {Dusel}},
  \bibinfo {author} {\bibfnamefont {J.}~\bibnamefont {Erdmenger}}, \bibinfo
  {author} {\bibfnamefont {D.}~\bibnamefont {Herdt}}, \bibinfo {author}
  {\bibfnamefont {H.}~\bibnamefont {Hinrichsen}}, \bibinfo {author}
  {\bibfnamefont {R.}~\bibnamefont {Meyer}},\ and\ \bibinfo {author}
  {\bibfnamefont {M.}~\bibnamefont {Schrauth}},\ }\bibfield  {title} {\bibinfo
  {title} {{Breitenlohner-Freedman Bound on Hyperbolic Tilings}},\ }\href
  {https://doi.org/10.1103/PhysRevLett.130.091604} {\bibfield  {journal}
  {\bibinfo  {journal} {Phys. Rev. Lett.}\ }\textbf {\bibinfo {volume} {130}},\
  \bibinfo {pages} {091604} (\bibinfo {year} {2023})}\BibitemShut {NoStop}%
\bibitem [{\citenamefont {Basteiro}\ \emph
  {et~al.}(2022{\natexlab{a}})\citenamefont {Basteiro}, \citenamefont {Giulio},
  \citenamefont {Erdmenger}, \citenamefont {Karl}, \citenamefont {Meyer},\ and\
  \citenamefont {Xian}}]{10.21468/SciPostPhys.13.5.103}%
  \BibitemOpen
  \bibfield  {author} {\bibinfo {author} {\bibfnamefont {P.}~\bibnamefont
  {Basteiro}}, \bibinfo {author} {\bibfnamefont {G.~D.}\ \bibnamefont
  {Giulio}}, \bibinfo {author} {\bibfnamefont {J.}~\bibnamefont {Erdmenger}},
  \bibinfo {author} {\bibfnamefont {J.}~\bibnamefont {Karl}}, \bibinfo {author}
  {\bibfnamefont {R.}~\bibnamefont {Meyer}},\ and\ \bibinfo {author}
  {\bibfnamefont {Z.-Y.}\ \bibnamefont {Xian}},\ }\bibfield  {title} {\bibinfo
  {title} {{Towards explicit discrete holography: Aperiodic spin chains from
  hyperbolic tilings}},\ }\href {https://doi.org/10.21468/SciPostPhys.13.5.103}
  {\bibfield  {journal} {\bibinfo  {journal} {SciPost Phys.}\ }\textbf
  {\bibinfo {volume} {13}},\ \bibinfo {pages} {103} (\bibinfo {year}
  {2022}{\natexlab{a}})}\BibitemShut {NoStop}%
\bibitem [{\citenamefont {Basteiro}\ \emph
  {et~al.}(2022{\natexlab{b}})\citenamefont {Basteiro}, \citenamefont {Das},
  \citenamefont {Di~Giulio},\ and\ \citenamefont {Erdmenger}}]{BasteiroArxiv}%
  \BibitemOpen
  \bibfield  {author} {\bibinfo {author} {\bibfnamefont {P.}~\bibnamefont
  {Basteiro}}, \bibinfo {author} {\bibfnamefont {R.~N.}\ \bibnamefont {Das}},
  \bibinfo {author} {\bibfnamefont {G.}~\bibnamefont {Di~Giulio}},\ and\
  \bibinfo {author} {\bibfnamefont {J.}~\bibnamefont {Erdmenger}},\ }\bibfield
  {title} {\bibinfo {title} {Aperiodic spin chains at the boundary of
  hyperbolic tilings},\ }\href {https://arxiv.org/abs/2212.11292} {\bibfield
  {journal} {\bibinfo  {journal} {arXiv:2212.11292}\ } (\bibinfo {year}
  {2022}{\natexlab{b}})}\BibitemShut {NoStop}%
\bibitem [{\citenamefont {Chen}\ \emph
  {et~al.}(2023{\natexlab{b}})\citenamefont {Chen}, \citenamefont {Chen},
  \citenamefont {Yang}, \citenamefont {Yang}, \citenamefont {Chen},
  \citenamefont {Meng}, \citenamefont {Yan}, \citenamefont {Xi}, \citenamefont
  {Zhu}, \citenamefont {Liu}, \citenamefont {Shum}, \citenamefont {Chen},
  \citenamefont {Cai}, \citenamefont {Yang}, \citenamefont {Yang},\ and\
  \citenamefont {Gao}}]{chen2023ads}%
  \BibitemOpen
  \bibfield  {author} {\bibinfo {author} {\bibfnamefont {J.}~\bibnamefont
  {Chen}}, \bibinfo {author} {\bibfnamefont {F.}~\bibnamefont {Chen}}, \bibinfo
  {author} {\bibfnamefont {Y.}~\bibnamefont {Yang}}, \bibinfo {author}
  {\bibfnamefont {L.}~\bibnamefont {Yang}}, \bibinfo {author} {\bibfnamefont
  {Z.}~\bibnamefont {Chen}}, \bibinfo {author} {\bibfnamefont {Y.}~\bibnamefont
  {Meng}}, \bibinfo {author} {\bibfnamefont {B.}~\bibnamefont {Yan}}, \bibinfo
  {author} {\bibfnamefont {X.}~\bibnamefont {Xi}}, \bibinfo {author}
  {\bibfnamefont {Z.}~\bibnamefont {Zhu}}, \bibinfo {author} {\bibfnamefont
  {G.-G.}\ \bibnamefont {Liu}}, \bibinfo {author} {\bibfnamefont {P.~P.}\
  \bibnamefont {Shum}}, \bibinfo {author} {\bibfnamefont {H.}~\bibnamefont
  {Chen}}, \bibinfo {author} {\bibfnamefont {R.-G.}\ \bibnamefont {Cai}},
  \bibinfo {author} {\bibfnamefont {R.-Q.}\ \bibnamefont {Yang}}, \bibinfo
  {author} {\bibfnamefont {Y.}~\bibnamefont {Yang}},\ and\ \bibinfo {author}
  {\bibfnamefont {Z.}~\bibnamefont {Gao}},\ }\bibfield  {title} {\bibinfo
  {title} {{AdS/CFT Correspondence in Hyperbolic Lattices}},\ }\href
  {https://arxiv.org/abs/2305.04862} {\bibfield  {journal} {\bibinfo  {journal}
  {arXiv:2305.04862}\ } (\bibinfo {year} {2023}{\natexlab{b}})}\BibitemShut
  {NoStop}%
\bibitem [{\citenamefont {Dey}\ \emph {et~al.}(2024)\citenamefont {Dey},
  \citenamefont {Chen}, \citenamefont {Basteiro}, \citenamefont {Fritzsche},
  \citenamefont {Greiter}, \citenamefont {Kaminski}, \citenamefont
  {Lenggenhager}, \citenamefont {Meyer}, \citenamefont {Sorbello},
  \citenamefont {Stegmaier}, \citenamefont {Thomale}, \citenamefont
  {Erdmenger},\ and\ \citenamefont {Boettcher}}]{dey2024simulating}%
  \BibitemOpen
  \bibfield  {author} {\bibinfo {author} {\bibfnamefont {S.}~\bibnamefont
  {Dey}}, \bibinfo {author} {\bibfnamefont {A.}~\bibnamefont {Chen}}, \bibinfo
  {author} {\bibfnamefont {P.}~\bibnamefont {Basteiro}}, \bibinfo {author}
  {\bibfnamefont {A.}~\bibnamefont {Fritzsche}}, \bibinfo {author}
  {\bibfnamefont {M.}~\bibnamefont {Greiter}}, \bibinfo {author} {\bibfnamefont
  {M.}~\bibnamefont {Kaminski}}, \bibinfo {author} {\bibfnamefont {P.~M.}\
  \bibnamefont {Lenggenhager}}, \bibinfo {author} {\bibfnamefont
  {R.}~\bibnamefont {Meyer}}, \bibinfo {author} {\bibfnamefont
  {R.}~\bibnamefont {Sorbello}}, \bibinfo {author} {\bibfnamefont
  {A.}~\bibnamefont {Stegmaier}}, \bibinfo {author} {\bibfnamefont
  {R.}~\bibnamefont {Thomale}}, \bibinfo {author} {\bibfnamefont
  {J.}~\bibnamefont {Erdmenger}},\ and\ \bibinfo {author} {\bibfnamefont
  {I.}~\bibnamefont {Boettcher}},\ }\bibfield  {title} {\bibinfo {title}
  {Simulating holographic conformal field theories on hyperbolic lattices},\
  }\href {https://arxiv.org/abs/2404.03062} {\bibfield  {journal} {\bibinfo
  {journal} {arXiv:2404.03062}\ } (\bibinfo {year} {2024})}\BibitemShut
  {NoStop}%
\bibitem [{\citenamefont {Vidal}(2007)}]{PhysRevLett.99.220405}%
  \BibitemOpen
  \bibfield  {author} {\bibinfo {author} {\bibfnamefont {G.}~\bibnamefont
  {Vidal}},\ }\bibfield  {title} {\bibinfo {title} {{Entanglement
  Renormalization}},\ }\href {https://doi.org/10.1103/PhysRevLett.99.220405}
  {\bibfield  {journal} {\bibinfo  {journal} {Phys. Rev. Lett.}\ }\textbf
  {\bibinfo {volume} {99}},\ \bibinfo {pages} {220405} (\bibinfo {year}
  {2007})}\BibitemShut {NoStop}%
\bibitem [{\citenamefont {Vidal}(2008)}]{PhysRevLett.101.110501}%
  \BibitemOpen
  \bibfield  {author} {\bibinfo {author} {\bibfnamefont {G.}~\bibnamefont
  {Vidal}},\ }\bibfield  {title} {\bibinfo {title} {{Class of Quantum Many-Body
  States That Can Be Efficiently Simulated}},\ }\href
  {https://doi.org/10.1103/PhysRevLett.101.110501} {\bibfield  {journal}
  {\bibinfo  {journal} {Phys. Rev. Lett.}\ }\textbf {\bibinfo {volume} {101}},\
  \bibinfo {pages} {110501} (\bibinfo {year} {2008})}\BibitemShut {NoStop}%
\bibitem [{\citenamefont {Swingle}(2012)}]{PhysRevD.86.065007}%
  \BibitemOpen
  \bibfield  {author} {\bibinfo {author} {\bibfnamefont {B.}~\bibnamefont
  {Swingle}},\ }\bibfield  {title} {\bibinfo {title} {{Entanglement
  renormalization and holography}},\ }\href
  {https://doi.org/10.1103/PhysRevD.86.065007} {\bibfield  {journal} {\bibinfo
  {journal} {Phys. Rev. D}\ }\textbf {\bibinfo {volume} {86}},\ \bibinfo
  {pages} {065007} (\bibinfo {year} {2012})}\BibitemShut {NoStop}%
\bibitem [{\citenamefont {Haegeman}\ \emph {et~al.}(2013)\citenamefont
  {Haegeman}, \citenamefont {Osborne}, \citenamefont {Verschelde},\ and\
  \citenamefont {Verstraete}}]{PhysRevLett.110.100402}%
  \BibitemOpen
  \bibfield  {author} {\bibinfo {author} {\bibfnamefont {J.}~\bibnamefont
  {Haegeman}}, \bibinfo {author} {\bibfnamefont {T.~J.}\ \bibnamefont
  {Osborne}}, \bibinfo {author} {\bibfnamefont {H.}~\bibnamefont
  {Verschelde}},\ and\ \bibinfo {author} {\bibfnamefont {F.}~\bibnamefont
  {Verstraete}},\ }\bibfield  {title} {\bibinfo {title} {{Entanglement
  Renormalization for Quantum Fields in Real Space}},\ }\href
  {https://doi.org/10.1103/PhysRevLett.110.100402} {\bibfield  {journal}
  {\bibinfo  {journal} {Phys. Rev. Lett.}\ }\textbf {\bibinfo {volume} {110}},\
  \bibinfo {pages} {100402} (\bibinfo {year} {2013})}\BibitemShut {NoStop}%
\bibitem [{\citenamefont {Bao}\ \emph {et~al.}(2017)\citenamefont {Bao},
  \citenamefont {Cao}, \citenamefont {Carroll},\ and\ \citenamefont
  {Chatwin-Davies}}]{Bao_2017}%
  \BibitemOpen
  \bibfield  {author} {\bibinfo {author} {\bibfnamefont {N.}~\bibnamefont
  {Bao}}, \bibinfo {author} {\bibfnamefont {C.}~\bibnamefont {Cao}}, \bibinfo
  {author} {\bibfnamefont {S.~M.}\ \bibnamefont {Carroll}},\ and\ \bibinfo
  {author} {\bibfnamefont {A.}~\bibnamefont {Chatwin-Davies}},\ }\bibfield
  {title} {\bibinfo {title} {{de Sitter space as a tensor network: Cosmic
  no-hair, complementarity, and complexity}},\ }\href
  {https://doi.org/10.1103/PhysRevD.96.123536} {\bibfield  {journal} {\bibinfo
  {journal} {Phys. Rev. D}\ }\textbf {\bibinfo {volume} {96}},\ \bibinfo
  {pages} {123536} (\bibinfo {year} {2017})}\BibitemShut {NoStop}%
\bibitem [{\citenamefont {Pastawski}\ \emph {et~al.}()\citenamefont
  {Pastawski}, \citenamefont {Yoshida}, \citenamefont {Harlow},\ and\
  \citenamefont {Preskill}}]{pastawski2015holographic}%
  \BibitemOpen
  \bibfield  {author} {\bibinfo {author} {\bibfnamefont {F.}~\bibnamefont
  {Pastawski}}, \bibinfo {author} {\bibfnamefont {B.}~\bibnamefont {Yoshida}},
  \bibinfo {author} {\bibfnamefont {D.}~\bibnamefont {Harlow}},\ and\ \bibinfo
  {author} {\bibfnamefont {J.}~\bibnamefont {Preskill}},\ }\bibfield  {title}
  {\bibinfo {title} {{Holographic quantum error-correcting codes: Toy models
  for the bulk/boundary correspondence}},\ }\href
  {https://doi.org/10.1007/JHEP06(2015)149} {\bibfield  {journal} {\bibinfo
  {journal} {JHEP}\ }\textbf {\bibinfo {volume} {06}},\ \bibinfo {pages} {149
  (2015)}}\BibitemShut {NoStop}%
\bibitem [{\citenamefont {Breuckmann}\ and\ \citenamefont
  {Terhal}(2016)}]{Breuckmann_2016}%
  \BibitemOpen
  \bibfield  {author} {\bibinfo {author} {\bibfnamefont {N.~P.}\ \bibnamefont
  {Breuckmann}}\ and\ \bibinfo {author} {\bibfnamefont {B.~M.}\ \bibnamefont
  {Terhal}},\ }\bibfield  {title} {\bibinfo {title} {{Constructions and Noise
  Threshold of Hyperbolic Surface Codes}},\ }\href
  {https://doi.org/10.1109/tit.2016.2555700} {\bibfield  {journal} {\bibinfo
  {journal} {IEEE Trans. Inf. Theory}\ }\textbf {\bibinfo {volume} {62}},\
  \bibinfo {pages} {3731} (\bibinfo {year} {2016})}\BibitemShut {NoStop}%
\bibitem [{\citenamefont {Breuckmann}\ \emph {et~al.}(2017)\citenamefont
  {Breuckmann}, \citenamefont {Vuillot}, \citenamefont {Campbell},
  \citenamefont {Krishna},\ and\ \citenamefont {Terhal}}]{Breuckmann_2017}%
  \BibitemOpen
  \bibfield  {author} {\bibinfo {author} {\bibfnamefont {N.~P.}\ \bibnamefont
  {Breuckmann}}, \bibinfo {author} {\bibfnamefont {C.}~\bibnamefont {Vuillot}},
  \bibinfo {author} {\bibfnamefont {E.}~\bibnamefont {Campbell}}, \bibinfo
  {author} {\bibfnamefont {A.}~\bibnamefont {Krishna}},\ and\ \bibinfo {author}
  {\bibfnamefont {B.~M.}\ \bibnamefont {Terhal}},\ }\bibfield  {title}
  {\bibinfo {title} {{Hyperbolic and semi-hyperbolic surface codes for quantum
  storage}},\ }\href {https://doi.org/10.1088/2058-9565/aa7d3b} {\bibfield
  {journal} {\bibinfo  {journal} {Quantum Sci. Technol.}\ }\textbf {\bibinfo
  {volume} {2}},\ \bibinfo {pages} {035007} (\bibinfo {year}
  {2017})}\BibitemShut {NoStop}%
\bibitem [{\citenamefont {Lavasani}\ \emph {et~al.}(2019)\citenamefont
  {Lavasani}, \citenamefont {Zhu},\ and\ \citenamefont
  {Barkeshli}}]{Lavasani_2019}%
  \BibitemOpen
  \bibfield  {author} {\bibinfo {author} {\bibfnamefont {A.}~\bibnamefont
  {Lavasani}}, \bibinfo {author} {\bibfnamefont {G.}~\bibnamefont {Zhu}},\ and\
  \bibinfo {author} {\bibfnamefont {M.}~\bibnamefont {Barkeshli}},\ }\bibfield
  {title} {\bibinfo {title} {{Universal logical gates with constant overhead:
  instantaneous Dehn twists for hyperbolic quantum codes}},\ }\href
  {https://doi.org/10.22331/q-2019-08-26-180} {\bibfield  {journal} {\bibinfo
  {journal} {Quantum}\ }\textbf {\bibinfo {volume} {3}},\ \bibinfo {pages}
  {180} (\bibinfo {year} {2019})}\BibitemShut {NoStop}%
\bibitem [{\citenamefont {Jahn}\ and\ \citenamefont
  {Eisert}(2021)}]{jahn2021holographic}%
  \BibitemOpen
  \bibfield  {author} {\bibinfo {author} {\bibfnamefont {A.}~\bibnamefont
  {Jahn}}\ and\ \bibinfo {author} {\bibfnamefont {J.}~\bibnamefont {Eisert}},\
  }\bibfield  {title} {\bibinfo {title} {{Holographic tensor network models and
  quantum error correction: a topical review}},\ }\href
  {https://doi.org/10.1088/2058-9565/ac0293} {\bibfield  {journal} {\bibinfo
  {journal} {Quantum Sci. Technol.}\ }\textbf {\bibinfo {volume} {6}},\
  \bibinfo {pages} {033002} (\bibinfo {year} {2021})}\BibitemShut {NoStop}%
\bibitem [{\citenamefont {Fahimniya}\ \emph {et~al.}(2023)\citenamefont
  {Fahimniya}, \citenamefont {Dehghani}, \citenamefont {Bharti}, \citenamefont
  {Mathew}, \citenamefont {Koll{\'a}r}, \citenamefont {Gorshkov},\ and\
  \citenamefont {Gullans}}]{AliError}%
  \BibitemOpen
  \bibfield  {author} {\bibinfo {author} {\bibfnamefont {A.}~\bibnamefont
  {Fahimniya}}, \bibinfo {author} {\bibfnamefont {H.}~\bibnamefont {Dehghani}},
  \bibinfo {author} {\bibfnamefont {K.}~\bibnamefont {Bharti}}, \bibinfo
  {author} {\bibfnamefont {S.}~\bibnamefont {Mathew}}, \bibinfo {author}
  {\bibfnamefont {A.~J.}\ \bibnamefont {Koll{\'a}r}}, \bibinfo {author}
  {\bibfnamefont {A.~V.}\ \bibnamefont {Gorshkov}},\ and\ \bibinfo {author}
  {\bibfnamefont {M.~J.}\ \bibnamefont {Gullans}},\ }\bibfield  {title}
  {\bibinfo {title} {{Fault-tolerant hyperbolic {Floquet} quantum error
  correcting codes}},\ }\href {https://arxiv.org/abs/2309.10033} {\bibfield
  {journal} {\bibinfo  {journal} {arXiv:2309.10033}\ } (\bibinfo {year}
  {2023})}\BibitemShut {NoStop}%
\bibitem [{\citenamefont {Boettcher}\ \emph {et~al.}(2020)\citenamefont
  {Boettcher}, \citenamefont {Bienias}, \citenamefont {Belyansky},
  \citenamefont {Koll\'ar},\ and\ \citenamefont {Gorshkov}}]{Boettcher:2020}%
  \BibitemOpen
  \bibfield  {author} {\bibinfo {author} {\bibfnamefont {I.}~\bibnamefont
  {Boettcher}}, \bibinfo {author} {\bibfnamefont {P.}~\bibnamefont {Bienias}},
  \bibinfo {author} {\bibfnamefont {R.}~\bibnamefont {Belyansky}}, \bibinfo
  {author} {\bibfnamefont {A.~J.}\ \bibnamefont {Koll\'ar}},\ and\ \bibinfo
  {author} {\bibfnamefont {A.~V.}\ \bibnamefont {Gorshkov}},\ }\bibfield
  {title} {\bibinfo {title} {Quantum simulation of hyperbolic space with
  circuit quantum electrodynamics: From graphs to geometry},\ }\href
  {https://doi.org/10.1103/PhysRevA.102.032208} {\bibfield  {journal} {\bibinfo
   {journal} {Phys. Rev. A}\ }\textbf {\bibinfo {volume} {102}},\ \bibinfo
  {pages} {032208} (\bibinfo {year} {2020})}\BibitemShut {NoStop}%
\bibitem [{\citenamefont {Boettcher}\ \emph {et~al.}(2022)\citenamefont
  {Boettcher}, \citenamefont {Gorshkov}, \citenamefont {Koll{\'a}r},
  \citenamefont {Maciejko}, \citenamefont {Rayan},\ and\ \citenamefont
  {Thomale}}]{Boettcher:2022}%
  \BibitemOpen
  \bibfield  {author} {\bibinfo {author} {\bibfnamefont {I.}~\bibnamefont
  {Boettcher}}, \bibinfo {author} {\bibfnamefont {A.~V.}\ \bibnamefont
  {Gorshkov}}, \bibinfo {author} {\bibfnamefont {A.~J.}\ \bibnamefont
  {Koll{\'a}r}}, \bibinfo {author} {\bibfnamefont {J.}~\bibnamefont
  {Maciejko}}, \bibinfo {author} {\bibfnamefont {S.}~\bibnamefont {Rayan}},\
  and\ \bibinfo {author} {\bibfnamefont {R.}~\bibnamefont {Thomale}},\
  }\bibfield  {title} {\bibinfo {title} {Crystallography of hyperbolic
  lattices},\ }\href {https://doi.org/10.1103/PhysRevB.105.125118} {\bibfield
  {journal} {\bibinfo  {journal} {Phys. Rev. B}\ }\textbf {\bibinfo {volume}
  {105}},\ \bibinfo {pages} {125118} (\bibinfo {year} {2022})}\BibitemShut
  {NoStop}%
\bibitem [{\citenamefont {Chen}\ \emph
  {et~al.}(2023{\natexlab{c}})\citenamefont {Chen}, \citenamefont {Guan},
  \citenamefont {Lenggenhager}, \citenamefont {Maciejko}, \citenamefont
  {Boettcher},\ and\ \citenamefont {Bzdu\v{s}ek}}]{Chen2023symmetry}%
  \BibitemOpen
  \bibfield  {author} {\bibinfo {author} {\bibfnamefont {A.}~\bibnamefont
  {Chen}}, \bibinfo {author} {\bibfnamefont {Y.}~\bibnamefont {Guan}}, \bibinfo
  {author} {\bibfnamefont {P.~M.}\ \bibnamefont {Lenggenhager}}, \bibinfo
  {author} {\bibfnamefont {J.}~\bibnamefont {Maciejko}}, \bibinfo {author}
  {\bibfnamefont {I.}~\bibnamefont {Boettcher}},\ and\ \bibinfo {author}
  {\bibfnamefont {T.}~\bibnamefont {Bzdu\v{s}ek}},\ }\bibfield  {title}
  {\bibinfo {title} {{Symmetry and topology of hyperbolic Haldane models}},\
  }\href {https://doi.org/10.1103/PhysRevB.108.085114} {\bibfield  {journal}
  {\bibinfo  {journal} {Phys. Rev. B}\ }\textbf {\bibinfo {volume} {108}},\
  \bibinfo {pages} {085114} (\bibinfo {year} {2023}{\natexlab{c}})}\BibitemShut
  {NoStop}%
\bibitem [{\citenamefont {Maciejko}\ and\ \citenamefont
  {Rayan}(2021)}]{Maciejko:2021}%
  \BibitemOpen
  \bibfield  {author} {\bibinfo {author} {\bibfnamefont {J.}~\bibnamefont
  {Maciejko}}\ and\ \bibinfo {author} {\bibfnamefont {S.}~\bibnamefont
  {Rayan}},\ }\bibfield  {title} {\bibinfo {title} {{Hyperbolic band theory}},\
  }\href {https://doi.org/10.1126/sciadv.abe9170} {\bibfield  {journal}
  {\bibinfo  {journal} {Sci. Adv.}\ }\textbf {\bibinfo {volume} {7}},\ \bibinfo
  {pages} {eabe9170} (\bibinfo {year} {2021})}\BibitemShut {NoStop}%
\bibitem [{\citenamefont {Maciejko}\ and\ \citenamefont
  {Rayan}(2022)}]{Maciejko:2022}%
  \BibitemOpen
  \bibfield  {author} {\bibinfo {author} {\bibfnamefont {J.}~\bibnamefont
  {Maciejko}}\ and\ \bibinfo {author} {\bibfnamefont {S.}~\bibnamefont
  {Rayan}},\ }\bibfield  {title} {\bibinfo {title} {Automorphic {Bloch}
  theorems for hyperbolic lattices},\ }\href
  {http://www.pnas.org/content/119/9/e2116869119} {\bibfield  {journal}
  {\bibinfo  {journal} {Proc. Natl. Acad. Sci. U.S.A.}\ }\textbf {\bibinfo
  {volume} {119}},\ \bibinfo {pages} {e2116869119} (\bibinfo {year}
  {2022})}\BibitemShut {NoStop}%
\bibitem [{\citenamefont {Cheng}\ \emph {et~al.}(2022)\citenamefont {Cheng},
  \citenamefont {Serafin}, \citenamefont {{McInerney}}, \citenamefont
  {Rocklin}, \citenamefont {Sun},\ and\ \citenamefont {Mao}}]{Cheng:2022}%
  \BibitemOpen
  \bibfield  {author} {\bibinfo {author} {\bibfnamefont {N.}~\bibnamefont
  {Cheng}}, \bibinfo {author} {\bibfnamefont {F.}~\bibnamefont {Serafin}},
  \bibinfo {author} {\bibfnamefont {J.}~\bibnamefont {{McInerney}}}, \bibinfo
  {author} {\bibfnamefont {Z.}~\bibnamefont {Rocklin}}, \bibinfo {author}
  {\bibfnamefont {K.}~\bibnamefont {Sun}},\ and\ \bibinfo {author}
  {\bibfnamefont {X.}~\bibnamefont {Mao}},\ }\bibfield  {title} {\bibinfo
  {title} {Band {Theory} and {Boundary} {Modes} of {High}-{Dimensional}
  {Representations} of {Infinite} {Hyperbolic} {Lattices}},\ }\href
  {https://doi.org/10.1103/PhysRevLett.129.088002} {\bibfield  {journal}
  {\bibinfo  {journal} {Phys. Rev. Lett.}\ }\textbf {\bibinfo {volume} {129}},\
  \bibinfo {pages} {088002} (\bibinfo {year} {2022})}\BibitemShut {NoStop}%
\bibitem [{\citenamefont {Lenggenhager}\ \emph {et~al.}(2023)\citenamefont
  {Lenggenhager}, \citenamefont {Maciejko},\ and\ \citenamefont
  {Bzdušek}}]{Lenggenhager:2023}%
  \BibitemOpen
  \bibfield  {author} {\bibinfo {author} {\bibfnamefont {P.~M.}\ \bibnamefont
  {Lenggenhager}}, \bibinfo {author} {\bibfnamefont {J.}~\bibnamefont
  {Maciejko}},\ and\ \bibinfo {author} {\bibfnamefont {T.}~\bibnamefont
  {Bzdušek}},\ }\bibfield  {title} {\bibinfo {title} {Non-{Abelian}
  {Hyperbolic} {Band} {Theory} from {Supercells}},\ }\href
  {https://doi.org/10.1103/PhysRevLett.131.226401} {\bibfield  {journal}
  {\bibinfo  {journal} {Phys. Rev. Lett.}\ }\textbf {\bibinfo {volume} {131}},\
  \bibinfo {pages} {226401} (\bibinfo {year} {2023})}\BibitemShut {NoStop}%
\bibitem [{\citenamefont {Shankar}\ and\ \citenamefont
  {Maciejko}(2023)}]{shankar2023}%
  \BibitemOpen
  \bibfield  {author} {\bibinfo {author} {\bibfnamefont {G.}~\bibnamefont
  {Shankar}}\ and\ \bibinfo {author} {\bibfnamefont {J.}~\bibnamefont
  {Maciejko}},\ }\bibfield  {title} {\bibinfo {title} {Hyperbolic lattices and
  two-dimensional {Yang}-{Mills} theory},\ }\href
  {http://arxiv.org/abs/2309.03857} {\bibfield  {journal} {\bibinfo  {journal}
  {arXiv:2309.03857}\ } (\bibinfo {year} {2023})}\BibitemShut {NoStop}%
\bibitem [{\citenamefont {Kienzle}\ and\ \citenamefont
  {Rayan}(2022)}]{kienzle2022}%
  \BibitemOpen
  \bibfield  {author} {\bibinfo {author} {\bibfnamefont {E.}~\bibnamefont
  {Kienzle}}\ and\ \bibinfo {author} {\bibfnamefont {S.}~\bibnamefont
  {Rayan}},\ }\bibfield  {title} {\bibinfo {title} {Hyperbolic band theory
  through {Higgs} bundles},\ }\href {https://doi.org/10.1016/j.aim.2022.108664}
  {\bibfield  {journal} {\bibinfo  {journal} {Adv. Math.}\ }\textbf {\bibinfo
  {volume} {409}},\ \bibinfo {pages} {108664} (\bibinfo {year}
  {2022})}\BibitemShut {NoStop}%
\bibitem [{\citenamefont {Nagy}\ and\ \citenamefont {Rayan}(2024)}]{nagy2023}%
  \BibitemOpen
  \bibfield  {author} {\bibinfo {author} {\bibfnamefont {{\'A}.}~\bibnamefont
  {Nagy}}\ and\ \bibinfo {author} {\bibfnamefont {S.}~\bibnamefont {Rayan}},\
  }\bibfield  {title} {\bibinfo {title} {On the {Hyperbolic} {Bloch}
  {Transform}},\ }\href {https://doi.org/10.1007/s00023-023-01336-8} {\bibfield
   {journal} {\bibinfo  {journal} {Ann. Henri Poincaré}\ }\textbf {\bibinfo
  {volume} {25}},\ \bibinfo {pages} {1713} (\bibinfo {year}
  {2024})}\BibitemShut {NoStop}%
\bibitem [{\citenamefont {Petermann}\ and\ \citenamefont
  {Hinrichsen}(2023)}]{HinrichsenArxiv}%
  \BibitemOpen
  \bibfield  {author} {\bibinfo {author} {\bibfnamefont {E.}~\bibnamefont
  {Petermann}}\ and\ \bibinfo {author} {\bibfnamefont {H.}~\bibnamefont
  {Hinrichsen}},\ }\bibfield  {title} {\bibinfo {title} {{Eigenmodes of the
  Laplacian on Hyperbolic Lattices}},\ }\href
  {https://arxiv.org/abs/2306.08248} {\bibfield  {journal} {\bibinfo  {journal}
  {arXiv:2306.08248}\ } (\bibinfo {year} {2023})}\BibitemShut {NoStop}%
\bibitem [{\citenamefont {Bienias}\ \emph {et~al.}(2022)\citenamefont
  {Bienias}, \citenamefont {Boettcher}, \citenamefont {Belyansky},
  \citenamefont {Koll{\'a}r},\ and\ \citenamefont {Gorshkov}}]{Bienias:2022}%
  \BibitemOpen
  \bibfield  {author} {\bibinfo {author} {\bibfnamefont {P.}~\bibnamefont
  {Bienias}}, \bibinfo {author} {\bibfnamefont {I.}~\bibnamefont {Boettcher}},
  \bibinfo {author} {\bibfnamefont {R.}~\bibnamefont {Belyansky}}, \bibinfo
  {author} {\bibfnamefont {A.~J.}\ \bibnamefont {Koll{\'a}r}},\ and\ \bibinfo
  {author} {\bibfnamefont {A.~V.}\ \bibnamefont {Gorshkov}},\ }\bibfield
  {title} {\bibinfo {title} {Circuit {Quantum} {Electrodynamics} in
  {Hyperbolic} {Space}: {From} {Photon} {Bound} {States} to {Frustrated} {Spin}
  {Models}},\ }\href {https://doi.org/10.1103/PhysRevLett.128.013601}
  {\bibfield  {journal} {\bibinfo  {journal} {Phys. Rev. Lett.}\ }\textbf
  {\bibinfo {volume} {128}},\ \bibinfo {pages} {013601} (\bibinfo {year}
  {2022})}\BibitemShut {NoStop}%
\bibitem [{\citenamefont {Gluscevich}\ \emph {et~al.}(2023)\citenamefont
  {Gluscevich}, \citenamefont {Samanta}, \citenamefont {Manna},\ and\
  \citenamefont {Roy}}]{Bitan1}%
  \BibitemOpen
  \bibfield  {author} {\bibinfo {author} {\bibfnamefont {N.}~\bibnamefont
  {Gluscevich}}, \bibinfo {author} {\bibfnamefont {A.}~\bibnamefont {Samanta}},
  \bibinfo {author} {\bibfnamefont {S.}~\bibnamefont {Manna}},\ and\ \bibinfo
  {author} {\bibfnamefont {B.}~\bibnamefont {Roy}},\ }\bibfield  {title}
  {\bibinfo {title} {Dynamic mass generation on two-dimensional electronic
  hyperbolic lattices},\ }\href {https://arxiv.org/abs/2302.04864} {\bibfield
  {journal} {\bibinfo  {journal} {arXiv:2302.04864}\ } (\bibinfo {year}
  {2023})}\BibitemShut {NoStop}%
\bibitem [{\citenamefont {Gluscevich}\ and\ \citenamefont
  {Roy}(2023)}]{Bitan2}%
  \BibitemOpen
  \bibfield  {author} {\bibinfo {author} {\bibfnamefont {N.}~\bibnamefont
  {Gluscevich}}\ and\ \bibinfo {author} {\bibfnamefont {B.}~\bibnamefont
  {Roy}},\ }\bibfield  {title} {\bibinfo {title} {{Magnetic catalysis in weakly
  interacting hyperbolic Dirac materials}},\ }\href
  {https://arxiv.org/abs/2305.11174} {\bibfield  {journal} {\bibinfo  {journal}
  {arXiv:2305.11174}\ } (\bibinfo {year} {2023})}\BibitemShut {NoStop}%
\bibitem [{\citenamefont {Bzdu\v{s}ek}\ and\ \citenamefont
  {Maciejko}(2022)}]{Bzdusek:2022}%
  \BibitemOpen
  \bibfield  {author} {\bibinfo {author} {\bibfnamefont {T.}~\bibnamefont
  {Bzdu\v{s}ek}}\ and\ \bibinfo {author} {\bibfnamefont {J.}~\bibnamefont
  {Maciejko}},\ }\bibfield  {title} {\bibinfo {title} {Flat bands and
  band-touching from real-space topology in hyperbolic lattices},\ }\href
  {https://doi.org/10.1103/PhysRevB.106.155146} {\bibfield  {journal} {\bibinfo
   {journal} {Phys. Rev. B}\ }\textbf {\bibinfo {volume} {106}},\ \bibinfo
  {pages} {155146} (\bibinfo {year} {2022})}\BibitemShut {NoStop}%
\bibitem [{\citenamefont {Mosseri}\ \emph {et~al.}(2022)\citenamefont
  {Mosseri}, \citenamefont {Vogeler},\ and\ \citenamefont
  {Vidal}}]{Mosseri:2022}%
  \BibitemOpen
  \bibfield  {author} {\bibinfo {author} {\bibfnamefont {R.}~\bibnamefont
  {Mosseri}}, \bibinfo {author} {\bibfnamefont {R.}~\bibnamefont {Vogeler}},\
  and\ \bibinfo {author} {\bibfnamefont {J.}~\bibnamefont {Vidal}},\ }\bibfield
   {title} {\bibinfo {title} {Aharonov-{Bohm} cages, flat bands, and gap
  labeling in hyperbolic tilings},\ }\href
  {https://doi.org/10.1103/PhysRevB.106.155120} {\bibfield  {journal} {\bibinfo
   {journal} {Phys. Rev. B}\ }\textbf {\bibinfo {volume} {106}},\ \bibinfo
  {pages} {155120} (\bibinfo {year} {2022})}\BibitemShut {NoStop}%
\bibitem [{\citenamefont {Yu}\ \emph {et~al.}(2020)\citenamefont {Yu},
  \citenamefont {Piao},\ and\ \citenamefont {Park}}]{Yu:2020}%
  \BibitemOpen
  \bibfield  {author} {\bibinfo {author} {\bibfnamefont {S.}~\bibnamefont
  {Yu}}, \bibinfo {author} {\bibfnamefont {X.}~\bibnamefont {Piao}},\ and\
  \bibinfo {author} {\bibfnamefont {N.}~\bibnamefont {Park}},\ }\bibfield
  {title} {\bibinfo {title} {Topological {Hyperbolic} {Lattices}},\ }\href
  {https://doi.org/10.1103/PhysRevLett.125.053901} {\bibfield  {journal}
  {\bibinfo  {journal} {Phys. Rev. Lett.}\ }\textbf {\bibinfo {volume} {125}},\
  \bibinfo {pages} {053901} (\bibinfo {year} {2020})}\BibitemShut {NoStop}%
\bibitem [{\citenamefont {Urwyler}\ \emph {et~al.}(2022)\citenamefont
  {Urwyler}, \citenamefont {Lenggenhager}, \citenamefont {Boettcher},
  \citenamefont {Thomale}, \citenamefont {Neupert},\ and\ \citenamefont
  {Bzdu\v{s}ek}}]{Urwyler:2022}%
  \BibitemOpen
  \bibfield  {author} {\bibinfo {author} {\bibfnamefont {D.~M.}\ \bibnamefont
  {Urwyler}}, \bibinfo {author} {\bibfnamefont {P.~M.}\ \bibnamefont
  {Lenggenhager}}, \bibinfo {author} {\bibfnamefont {I.}~\bibnamefont
  {Boettcher}}, \bibinfo {author} {\bibfnamefont {R.}~\bibnamefont {Thomale}},
  \bibinfo {author} {\bibfnamefont {T.}~\bibnamefont {Neupert}},\ and\ \bibinfo
  {author} {\bibfnamefont {T.}~\bibnamefont {Bzdu\v{s}ek}},\ }\bibfield
  {title} {\bibinfo {title} {{Hyperbolic Topological Band Insulators}},\ }\href
  {https://doi.org/10.1103/PhysRevLett.129.246402} {\bibfield  {journal}
  {\bibinfo  {journal} {Phys. Rev. Lett.}\ }\textbf {\bibinfo {volume} {129}},\
  \bibinfo {pages} {246402} (\bibinfo {year} {2022})}\BibitemShut {NoStop}%
\bibitem [{\citenamefont {Liu}\ \emph {et~al.}(2022)\citenamefont {Liu},
  \citenamefont {Hua}, \citenamefont {Peng},\ and\ \citenamefont
  {Zhou}}]{Liu:2022}%
  \BibitemOpen
  \bibfield  {author} {\bibinfo {author} {\bibfnamefont {Z.-R.}\ \bibnamefont
  {Liu}}, \bibinfo {author} {\bibfnamefont {C.-B.}\ \bibnamefont {Hua}},
  \bibinfo {author} {\bibfnamefont {T.}~\bibnamefont {Peng}},\ and\ \bibinfo
  {author} {\bibfnamefont {B.}~\bibnamefont {Zhou}},\ }\bibfield  {title}
  {\bibinfo {title} {Chern insulator in a hyperbolic lattice},\ }\href
  {https://doi.org/10.1103/PhysRevB.105.245301} {\bibfield  {journal} {\bibinfo
   {journal} {Phys. Rev. B}\ }\textbf {\bibinfo {volume} {105}},\ \bibinfo
  {pages} {245301} (\bibinfo {year} {2022})}\BibitemShut {NoStop}%
\bibitem [{\citenamefont {Liu}\ \emph {et~al.}(2023)\citenamefont {Liu},
  \citenamefont {Hua}, \citenamefont {Peng}, \citenamefont {Chen},\ and\
  \citenamefont {Zhou}}]{Liu:2022b}%
  \BibitemOpen
  \bibfield  {author} {\bibinfo {author} {\bibfnamefont {Z.-R.}\ \bibnamefont
  {Liu}}, \bibinfo {author} {\bibfnamefont {C.-B.}\ \bibnamefont {Hua}},
  \bibinfo {author} {\bibfnamefont {T.}~\bibnamefont {Peng}}, \bibinfo {author}
  {\bibfnamefont {R.}~\bibnamefont {Chen}},\ and\ \bibinfo {author}
  {\bibfnamefont {B.}~\bibnamefont {Zhou}},\ }\bibfield  {title} {\bibinfo
  {title} {Higher-order topological insulators in hyperbolic lattices},\ }\href
  {https://doi.org/10.1103/PhysRevB.107.125302} {\bibfield  {journal} {\bibinfo
   {journal} {Phys. Rev. B}\ }\textbf {\bibinfo {volume} {107}},\ \bibinfo
  {pages} {125302} (\bibinfo {year} {2023})}\BibitemShut {NoStop}%
\bibitem [{\citenamefont {Pei}\ \emph {et~al.}(2023)\citenamefont {Pei},
  \citenamefont {Yuan}, \citenamefont {Zhang},\ and\ \citenamefont
  {Zhang}}]{pei2023}%
  \BibitemOpen
  \bibfield  {author} {\bibinfo {author} {\bibfnamefont {Q.}~\bibnamefont
  {Pei}}, \bibinfo {author} {\bibfnamefont {H.}~\bibnamefont {Yuan}}, \bibinfo
  {author} {\bibfnamefont {W.}~\bibnamefont {Zhang}},\ and\ \bibinfo {author}
  {\bibfnamefont {X.}~\bibnamefont {Zhang}},\ }\bibfield  {title} {\bibinfo
  {title} {Engineering boundary-dominated topological states in defective
  hyperbolic lattices},\ }\href {https://doi.org/10.1103/PhysRevB.107.165145}
  {\bibfield  {journal} {\bibinfo  {journal} {Phys. Rev. B}\ }\textbf {\bibinfo
  {volume} {107}},\ \bibinfo {pages} {165145} (\bibinfo {year}
  {2023})}\BibitemShut {NoStop}%
\bibitem [{\citenamefont {Tao}\ and\ \citenamefont {Xu}(2023)}]{Tao:2022}%
  \BibitemOpen
  \bibfield  {author} {\bibinfo {author} {\bibfnamefont {Y.-L.}\ \bibnamefont
  {Tao}}\ and\ \bibinfo {author} {\bibfnamefont {Y.}~\bibnamefont {Xu}},\
  }\bibfield  {title} {\bibinfo {title} {Higher-order topological hyperbolic
  lattices},\ }\href {https://doi.org/10.1103/PhysRevB.107.184201} {\bibfield
  {journal} {\bibinfo  {journal} {Phys. Rev. B}\ }\textbf {\bibinfo {volume}
  {107}},\ \bibinfo {pages} {184201} (\bibinfo {year} {2023})}\BibitemShut
  {NoStop}%
\bibitem [{\citenamefont {Tummuru}\ \emph {et~al.}(2023)\citenamefont
  {Tummuru}, \citenamefont {Chen}, \citenamefont {Lenggenhager}, \citenamefont
  {Neupert}, \citenamefont {Maciejko},\ and\ \citenamefont
  {Bzdušek}}]{Tummuru2023}%
  \BibitemOpen
  \bibfield  {author} {\bibinfo {author} {\bibfnamefont {T.}~\bibnamefont
  {Tummuru}}, \bibinfo {author} {\bibfnamefont {A.}~\bibnamefont {Chen}},
  \bibinfo {author} {\bibfnamefont {P.~M.}\ \bibnamefont {Lenggenhager}},
  \bibinfo {author} {\bibfnamefont {T.}~\bibnamefont {Neupert}}, \bibinfo
  {author} {\bibfnamefont {J.}~\bibnamefont {Maciejko}},\ and\ \bibinfo
  {author} {\bibfnamefont {T.}~\bibnamefont {Bzdušek}},\ }\bibfield  {title}
  {\bibinfo {title} {Hyperbolic non-{Abelian} semimetal},\ }\href
  {http://arxiv.org/abs/2307.09876} {\bibfield  {journal} {\bibinfo  {journal}
  {arXiv:2307.09876}\ } (\bibinfo {year} {2023})}\BibitemShut {NoStop}%
\bibitem [{\citenamefont {Anderson}(1958)}]{Anderson1958}%
  \BibitemOpen
  \bibfield  {author} {\bibinfo {author} {\bibfnamefont {P.~W.}\ \bibnamefont
  {Anderson}},\ }\bibfield  {title} {\bibinfo {title} {{Absence of Diffusion in
  Certain Random Lattices}},\ }\href {https://doi.org/10.1103/PhysRev.109.1492}
  {\bibfield  {journal} {\bibinfo  {journal} {Phys. Rev.}\ }\textbf {\bibinfo
  {volume} {109}},\ \bibinfo {pages} {1492} (\bibinfo {year}
  {1958})}\BibitemShut {NoStop}%
\bibitem [{\citenamefont {Evers}\ and\ \citenamefont
  {Mirlin}(2008)}]{Evers2008}%
  \BibitemOpen
  \bibfield  {author} {\bibinfo {author} {\bibfnamefont {F.}~\bibnamefont
  {Evers}}\ and\ \bibinfo {author} {\bibfnamefont {A.~D.}\ \bibnamefont
  {Mirlin}},\ }\bibfield  {title} {\bibinfo {title} {Anderson transitions},\
  }\href {https://doi.org/10.1103/RevModPhys.80.1355} {\bibfield  {journal}
  {\bibinfo  {journal} {Rev. Mod. Phys.}\ }\textbf {\bibinfo {volume} {80}},\
  \bibinfo {pages} {1355} (\bibinfo {year} {2008})}\BibitemShut {NoStop}%
\bibitem [{\citenamefont {Lagendijk}\ \emph {et~al.}(2009)\citenamefont
  {Lagendijk}, \citenamefont {Tiggelen},\ and\ \citenamefont
  {Wiersma}}]{Lagendijk2009}%
  \BibitemOpen
  \bibfield  {author} {\bibinfo {author} {\bibfnamefont {A.}~\bibnamefont
  {Lagendijk}}, \bibinfo {author} {\bibfnamefont {B.}~\bibnamefont
  {Tiggelen}},\ and\ \bibinfo {author} {\bibfnamefont {D.}~\bibnamefont
  {Wiersma}},\ }\bibfield  {title} {\bibinfo {title} {{Fifty years of Anderson
  Localization}},\ }\href {https://doi.org/10.1063/1.3206091} {\bibfield
  {journal} {\bibinfo  {journal} {Phys. Today}\ }\textbf {\bibinfo {volume}
  {62}},\ \bibinfo {pages} {24} (\bibinfo {year} {2009})}\BibitemShut {NoStop}%
\bibitem [{\citenamefont {Bergmann}(1984)}]{bergmann1984}%
  \BibitemOpen
  \bibfield  {author} {\bibinfo {author} {\bibfnamefont {G.}~\bibnamefont
  {Bergmann}},\ }\bibfield  {title} {\bibinfo {title} {Weak localization in
  thin films: a time-of-flight experiment with conduction electrons},\ }\href
  {https://doi.org/10.1016/0370-1573(84)90103-0} {\bibfield  {journal}
  {\bibinfo  {journal} {Phys. Rep.}\ }\textbf {\bibinfo {volume} {107}},\
  \bibinfo {pages} {1} (\bibinfo {year} {1984})}\BibitemShut {NoStop}%
\bibitem [{\citenamefont {Abou-Chacra}\ \emph {et~al.}(1973)\citenamefont
  {Abou-Chacra}, \citenamefont {Thouless},\ and\ \citenamefont
  {Anderson}}]{abou-chacra1973}%
  \BibitemOpen
  \bibfield  {author} {\bibinfo {author} {\bibfnamefont {R.}~\bibnamefont
  {Abou-Chacra}}, \bibinfo {author} {\bibfnamefont {D.~J.}\ \bibnamefont
  {Thouless}},\ and\ \bibinfo {author} {\bibfnamefont {P.~W.}\ \bibnamefont
  {Anderson}},\ }\bibfield  {title} {\bibinfo {title} {A selfconsistent theory
  of localization},\ }\href {https://doi.org/10.1088/0022-3719/6/10/009}
  {\bibfield  {journal} {\bibinfo  {journal} {J. Phys. C}\ }\textbf {\bibinfo
  {volume} {6}},\ \bibinfo {pages} {1734} (\bibinfo {year} {1973})}\BibitemShut
  {NoStop}%
\bibitem [{\citenamefont {Abou-Chacra}\ and\ \citenamefont
  {Thouless}(1974)}]{abou-chacra1974}%
  \BibitemOpen
  \bibfield  {author} {\bibinfo {author} {\bibfnamefont {R.}~\bibnamefont
  {Abou-Chacra}}\ and\ \bibinfo {author} {\bibfnamefont {D.~J.}\ \bibnamefont
  {Thouless}},\ }\bibfield  {title} {\bibinfo {title} {Self-consistent theory
  of localization. {II}. {Localization} near the band edges},\ }\href
  {https://doi.org/10.1088/0022-3719/7/1/015} {\bibfield  {journal} {\bibinfo
  {journal} {J. Phys. C}\ }\textbf {\bibinfo {volume} {7}},\ \bibinfo {pages}
  {65} (\bibinfo {year} {1974})}\BibitemShut {NoStop}%
\bibitem [{\citenamefont {Mirlin}\ and\ \citenamefont
  {Fyodorov}(1991)}]{mirlin1991}%
  \BibitemOpen
  \bibfield  {author} {\bibinfo {author} {\bibfnamefont {A.~D.}\ \bibnamefont
  {Mirlin}}\ and\ \bibinfo {author} {\bibfnamefont {Y.~V.}\ \bibnamefont
  {Fyodorov}},\ }\bibfield  {title} {\bibinfo {title} {Localization transition
  in the {Anderson} model on the {Bethe} lattice: {Spontaneous} symmetry
  breaking and correlation functions},\ }\href
  {https://doi.org/10.1016/0550-3213(91)90028-V} {\bibfield  {journal}
  {\bibinfo  {journal} {Nucl. Phys. B}\ }\textbf {\bibinfo {volume} {366}},\
  \bibinfo {pages} {507} (\bibinfo {year} {1991})}\BibitemShut {NoStop}%
\bibitem [{\citenamefont {Tikhonov}\ \emph {et~al.}(2016)\citenamefont
  {Tikhonov}, \citenamefont {Mirlin},\ and\ \citenamefont
  {Skvortsov}}]{Tikhonov2016}%
  \BibitemOpen
  \bibfield  {author} {\bibinfo {author} {\bibfnamefont {K.~S.}\ \bibnamefont
  {Tikhonov}}, \bibinfo {author} {\bibfnamefont {A.~D.}\ \bibnamefont
  {Mirlin}},\ and\ \bibinfo {author} {\bibfnamefont {M.~A.}\ \bibnamefont
  {Skvortsov}},\ }\bibfield  {title} {\bibinfo {title} {Anderson localization
  and ergodicity on random regular graphs},\ }\href
  {https://doi.org/10.1103/PhysRevB.94.220203} {\bibfield  {journal} {\bibinfo
  {journal} {Phys. Rev. B}\ }\textbf {\bibinfo {volume} {94}},\ \bibinfo
  {pages} {220203} (\bibinfo {year} {2016})}\BibitemShut {NoStop}%
\bibitem [{\citenamefont {Biroli}\ and\ \citenamefont
  {Tarzia}(2018)}]{Biroli2018arxiv}%
  \BibitemOpen
  \bibfield  {author} {\bibinfo {author} {\bibfnamefont {G.}~\bibnamefont
  {Biroli}}\ and\ \bibinfo {author} {\bibfnamefont {M.}~\bibnamefont
  {Tarzia}},\ }\bibfield  {title} {\bibinfo {title} {Delocalization and
  ergodicity of the {Anderson} model on {Bethe} lattices},\ }\href
  {https://arxiv.org/abs/1810.07545} {\bibfield  {journal} {\bibinfo  {journal}
  {arXiv:1810.07545}\ } (\bibinfo {year} {2018})}\BibitemShut {NoStop}%
\bibitem [{\citenamefont {Tikhonov}\ and\ \citenamefont
  {Mirlin}(2019)}]{Tikhonov2019}%
  \BibitemOpen
  \bibfield  {author} {\bibinfo {author} {\bibfnamefont {K.~S.}\ \bibnamefont
  {Tikhonov}}\ and\ \bibinfo {author} {\bibfnamefont {A.~D.}\ \bibnamefont
  {Mirlin}},\ }\bibfield  {title} {\bibinfo {title} {Critical behavior at the
  localization transition on random regular graphs},\ }\href
  {https://doi.org/10.1103/PhysRevB.99.214202} {\bibfield  {journal} {\bibinfo
  {journal} {Phys. Rev. B}\ }\textbf {\bibinfo {volume} {99}},\ \bibinfo
  {pages} {214202} (\bibinfo {year} {2019})}\BibitemShut {NoStop}%
\bibitem [{\citenamefont {Parisi}\ \emph {et~al.}(2019)\citenamefont {Parisi},
  \citenamefont {Pascazio}, \citenamefont {Pietracaprina}, \citenamefont
  {Ros},\ and\ \citenamefont {Scardicchio}}]{Parisi2020}%
  \BibitemOpen
  \bibfield  {author} {\bibinfo {author} {\bibfnamefont {G.}~\bibnamefont
  {Parisi}}, \bibinfo {author} {\bibfnamefont {S.}~\bibnamefont {Pascazio}},
  \bibinfo {author} {\bibfnamefont {F.}~\bibnamefont {Pietracaprina}}, \bibinfo
  {author} {\bibfnamefont {V.}~\bibnamefont {Ros}},\ and\ \bibinfo {author}
  {\bibfnamefont {A.}~\bibnamefont {Scardicchio}},\ }\bibfield  {title}
  {\bibinfo {title} {Anderson transition on the bethe lattice: an approach with
  real energies},\ }\href {https://doi.org/10.1088/1751-8121/ab56e8} {\bibfield
   {journal} {\bibinfo  {journal} {J. Phys. A: Math. Theor.}\ }\textbf
  {\bibinfo {volume} {53}},\ \bibinfo {pages} {014003} (\bibinfo {year}
  {2019})}\BibitemShut {NoStop}%
\bibitem [{\citenamefont {Tikhonov}\ and\ \citenamefont
  {Mirlin}(2021)}]{Tikhonov2021}%
  \BibitemOpen
  \bibfield  {author} {\bibinfo {author} {\bibfnamefont {K.}~\bibnamefont
  {Tikhonov}}\ and\ \bibinfo {author} {\bibfnamefont {A.}~\bibnamefont
  {Mirlin}},\ }\bibfield  {title} {\bibinfo {title} {From {Anderson}
  localization on random regular graphs to many-body localization},\ }\href
  {https://doi.org/https://doi.org/10.1016/j.aop.2021.168525} {\bibfield
  {journal} {\bibinfo  {journal} {Ann. Phys.}\ }\textbf {\bibinfo {volume}
  {435}},\ \bibinfo {pages} {168525} (\bibinfo {year} {2021})}\BibitemShut
  {NoStop}%
\bibitem [{\citenamefont {Herre}\ \emph {et~al.}(2023)\citenamefont {Herre},
  \citenamefont {Karcher}, \citenamefont {Tikhonov},\ and\ \citenamefont
  {Mirlin}}]{Herre2023}%
  \BibitemOpen
  \bibfield  {author} {\bibinfo {author} {\bibfnamefont {J.-N.}\ \bibnamefont
  {Herre}}, \bibinfo {author} {\bibfnamefont {J.~F.}\ \bibnamefont {Karcher}},
  \bibinfo {author} {\bibfnamefont {K.~S.}\ \bibnamefont {Tikhonov}},\ and\
  \bibinfo {author} {\bibfnamefont {A.~D.}\ \bibnamefont {Mirlin}},\ }\bibfield
   {title} {\bibinfo {title} {Ergodicity-to-localization transition on random
  regular graphs with large connectivity and in many-body quantum dots},\
  }\href {https://doi.org/10.1103/PhysRevB.108.014203} {\bibfield  {journal}
  {\bibinfo  {journal} {Phys. Rev. B}\ }\textbf {\bibinfo {volume} {108}},\
  \bibinfo {pages} {014203} (\bibinfo {year} {2023})}\BibitemShut {NoStop}%
\bibitem [{\citenamefont {Sierant}\ \emph {et~al.}(2023)\citenamefont
  {Sierant}, \citenamefont {Lewenstein},\ and\ \citenamefont
  {Scardicchio}}]{Sierant2023}%
  \BibitemOpen
  \bibfield  {author} {\bibinfo {author} {\bibfnamefont {P.}~\bibnamefont
  {Sierant}}, \bibinfo {author} {\bibfnamefont {M.}~\bibnamefont
  {Lewenstein}},\ and\ \bibinfo {author} {\bibfnamefont {A.}~\bibnamefont
  {Scardicchio}},\ }\bibfield  {title} {\bibinfo {title} {{Universality in
  Anderson localization on random graphs with varying connectivity}},\ }\href
  {https://doi.org/10.21468/SciPostPhys.15.2.045} {\bibfield  {journal}
  {\bibinfo  {journal} {SciPost Phys.}\ }\textbf {\bibinfo {volume} {15}},\
  \bibinfo {pages} {045} (\bibinfo {year} {2023})}\BibitemShut {NoStop}%
\bibitem [{\citenamefont {Koll{\'a}r}\ \emph {et~al.}(2020)\citenamefont
  {Koll{\'a}r}, \citenamefont {Fitzpatrick}, \citenamefont {Sarnak},\ and\
  \citenamefont {Houck}}]{kollar2019line}%
  \BibitemOpen
  \bibfield  {author} {\bibinfo {author} {\bibfnamefont {A.~J.}\ \bibnamefont
  {Koll{\'a}r}}, \bibinfo {author} {\bibfnamefont {M.}~\bibnamefont
  {Fitzpatrick}}, \bibinfo {author} {\bibfnamefont {P.}~\bibnamefont
  {Sarnak}},\ and\ \bibinfo {author} {\bibfnamefont {A.~A.}\ \bibnamefont
  {Houck}},\ }\bibfield  {title} {\bibinfo {title} {{Line-graph lattices:
  Euclidean and non-Euclidean flat bands, and implementations in circuit
  quantum electrodynamics}},\ }\href
  {https://doi.org/10.1007/s00220-019-03645-8} {\bibfield  {journal} {\bibinfo
  {journal} {Commun. Math. Phys.}\ }\textbf {\bibinfo {volume} {376}},\
  \bibinfo {pages} {1909} (\bibinfo {year} {2020})}\BibitemShut {NoStop}%
\bibitem [{\citenamefont {Garcia-Mata}\ \emph {et~al.}(2017)\citenamefont
  {Garcia-Mata}, \citenamefont {Giraud}, \citenamefont {Georgeot},
  \citenamefont {Martin}, \citenamefont {Dubertrand},\ and\ \citenamefont
  {{Lemari\'e}}}]{Mata2017}%
  \BibitemOpen
  \bibfield  {author} {\bibinfo {author} {\bibfnamefont {I.}~\bibnamefont
  {Garcia-Mata}}, \bibinfo {author} {\bibfnamefont {O.}~\bibnamefont {Giraud}},
  \bibinfo {author} {\bibfnamefont {B.}~\bibnamefont {Georgeot}}, \bibinfo
  {author} {\bibfnamefont {J.}~\bibnamefont {Martin}}, \bibinfo {author}
  {\bibfnamefont {R.}~\bibnamefont {Dubertrand}},\ and\ \bibinfo {author}
  {\bibfnamefont {G.}~\bibnamefont {{Lemari\'e}}},\ }\bibfield  {title}
  {\bibinfo {title} {Scaling theory of the {Anderson} transition in random
  graphs: Ergodicity and universality},\ }\href
  {https://doi.org/10.1103/PhysRevLett.118.166801} {\bibfield  {journal}
  {\bibinfo  {journal} {Phys. Rev. Lett.}\ }\textbf {\bibinfo {volume} {118}},\
  \bibinfo {pages} {166801} (\bibinfo {year} {2017})}\BibitemShut {NoStop}%
\bibitem [{\citenamefont {Garcia-Mata}\ \emph {et~al.}(2022)\citenamefont
  {Garcia-Mata}, \citenamefont {Martin}, \citenamefont {Giraud}, \citenamefont
  {Georgeot}, \citenamefont {Dubertrand},\ and\ \citenamefont
  {{Lemari\'e}}}]{Mata2022}%
  \BibitemOpen
  \bibfield  {author} {\bibinfo {author} {\bibfnamefont {I.}~\bibnamefont
  {Garcia-Mata}}, \bibinfo {author} {\bibfnamefont {J.}~\bibnamefont {Martin}},
  \bibinfo {author} {\bibfnamefont {O.}~\bibnamefont {Giraud}}, \bibinfo
  {author} {\bibfnamefont {B.}~\bibnamefont {Georgeot}}, \bibinfo {author}
  {\bibfnamefont {R.}~\bibnamefont {Dubertrand}},\ and\ \bibinfo {author}
  {\bibfnamefont {G.}~\bibnamefont {{Lemari\'e}}},\ }\bibfield  {title}
  {\bibinfo {title} {{Critical properties of the Anderson transition on random
  graphs: Two-parameter scaling theory, Kosterlitz-Thouless type flow, and
  many-body localization}},\ }\href
  {https://doi.org/10.1103/PhysRevB.106.214202} {\bibfield  {journal} {\bibinfo
   {journal} {Phys. Rev. B}\ }\textbf {\bibinfo {volume} {106}},\ \bibinfo
  {pages} {214202} (\bibinfo {year} {2022})}\BibitemShut {NoStop}%
\bibitem [{\citenamefont {Sade}\ \emph {et~al.}(2005)\citenamefont {Sade},
  \citenamefont {Kalisky}, \citenamefont {Havlin},\ and\ \citenamefont
  {Berkovits}}]{Sade2005}%
  \BibitemOpen
  \bibfield  {author} {\bibinfo {author} {\bibfnamefont {M.}~\bibnamefont
  {Sade}}, \bibinfo {author} {\bibfnamefont {T.}~\bibnamefont {Kalisky}},
  \bibinfo {author} {\bibfnamefont {S.}~\bibnamefont {Havlin}},\ and\ \bibinfo
  {author} {\bibfnamefont {R.}~\bibnamefont {Berkovits}},\ }\bibfield  {title}
  {\bibinfo {title} {Localization transition on complex networks via spectral
  statistics},\ }\href {https://doi.org/10.1103/PhysRevE.72.066123} {\bibfield
  {journal} {\bibinfo  {journal} {Phys. Rev. E}\ }\textbf {\bibinfo {volume}
  {72}},\ \bibinfo {pages} {066123} (\bibinfo {year} {2005})}\BibitemShut
  {NoStop}%
\bibitem [{\citenamefont {Mard}\ \emph {et~al.}(2017)\citenamefont {Mard},
  \citenamefont {Hoyos}, \citenamefont {Miranda},\ and\ \citenamefont
  {Dobrosavljevi\ifmmode~\acute{c}\else \'{c}\fi{}}}]{Mard2017}%
  \BibitemOpen
  \bibfield  {author} {\bibinfo {author} {\bibfnamefont {H.~J.}\ \bibnamefont
  {Mard}}, \bibinfo {author} {\bibfnamefont {J.~A.}\ \bibnamefont {Hoyos}},
  \bibinfo {author} {\bibfnamefont {E.}~\bibnamefont {Miranda}},\ and\ \bibinfo
  {author} {\bibfnamefont {V.}~\bibnamefont
  {Dobrosavljevi\ifmmode~\acute{c}\else \'{c}\fi{}}},\ }\bibfield  {title}
  {\bibinfo {title} {{Strong-disorder approach for the Anderson localization
  transition}},\ }\href {https://doi.org/10.1103/PhysRevB.96.045143} {\bibfield
   {journal} {\bibinfo  {journal} {Phys. Rev. B}\ }\textbf {\bibinfo {volume}
  {96}},\ \bibinfo {pages} {045143} (\bibinfo {year} {2017})}\BibitemShut
  {NoStop}%
\bibitem [{\citenamefont {Alt}\ \emph {et~al.}(2021)\citenamefont {Alt},
  \citenamefont {Ducatez},\ and\ \citenamefont {Knowles}}]{Alt2021}%
  \BibitemOpen
  \bibfield  {author} {\bibinfo {author} {\bibfnamefont {J.}~\bibnamefont
  {Alt}}, \bibinfo {author} {\bibfnamefont {R.}~\bibnamefont {Ducatez}},\ and\
  \bibinfo {author} {\bibfnamefont {A.}~\bibnamefont {Knowles}},\ }\bibfield
  {title} {\bibinfo {title} {{Delocalization Transition for Critical
  Erdős–Rényi Graphs}},\ }\href
  {https://doi.org/10.1007/s00220-021-04167-y} {\bibfield  {journal} {\bibinfo
  {journal} {Commun. Math. Phys.}\ }\textbf {\bibinfo {volume} {388}},\
  \bibinfo {pages} {507} (\bibinfo {year} {2021})}\BibitemShut {NoStop}%
\bibitem [{\citenamefont {Curtis}\ \emph {et~al.}(2023)\citenamefont {Curtis},
  \citenamefont {Narang},\ and\ \citenamefont {Galitski}}]{curtis2023absence}%
  \BibitemOpen
  \bibfield  {author} {\bibinfo {author} {\bibfnamefont {J.~B.}\ \bibnamefont
  {Curtis}}, \bibinfo {author} {\bibfnamefont {P.}~\bibnamefont {Narang}},\
  and\ \bibinfo {author} {\bibfnamefont {V.}~\bibnamefont {Galitski}},\
  }\bibfield  {title} {\bibinfo {title} {{Absence of Weak Localization on
  Negative Curvature Surfaces}},\ }\href {https://arxiv.org/abs/2308.01351}
  {\bibfield  {journal} {\bibinfo  {journal} {arXiv:2308.01351}\ } (\bibinfo
  {year} {2023})}\BibitemShut {NoStop}%
\bibitem [{\citenamefont {Einstein}(1905)}]{Einstein1905}%
  \BibitemOpen
  \bibfield  {author} {\bibinfo {author} {\bibfnamefont {A.}~\bibnamefont
  {Einstein}},\ }\bibfield  {title} {\bibinfo {title} {{Über die von der
  molekularkinetischen Theorie der Wärme geforderte Bewegung von in ruhenden
  Flüssigkeiten suspendierten Teilchen}},\ }\href
  {https://doi.org/https://doi.org/10.1002/andp.19053220806} {\bibfield
  {journal} {\bibinfo  {journal} {Ann. Phys.}\ }\textbf {\bibinfo {volume}
  {322}},\ \bibinfo {pages} {549} (\bibinfo {year} {1905})}\BibitemShut
  {NoStop}%
\bibitem [{\citenamefont {Pólya}(1921)}]{Polya1921}%
  \BibitemOpen
  \bibfield  {author} {\bibinfo {author} {\bibfnamefont {G.}~\bibnamefont
  {Pólya}},\ }\bibfield  {title} {\bibinfo {title} {{Über eine Aufgabe der
  Wahrscheinlichkeitsrechnung betreffend die Irrfahrt im Straßennetz}},\
  }\href {http://eudml.org/doc/158886} {\bibfield  {journal} {\bibinfo
  {journal} {Math. Ann.}\ }\textbf {\bibinfo {volume} {84}},\ \bibinfo {pages}
  {149} (\bibinfo {year} {1921})}\BibitemShut {NoStop}%
\bibitem [{\citenamefont {Datta}(1995)}]{Datta1995}%
  \BibitemOpen
  \bibfield  {author} {\bibinfo {author} {\bibfnamefont {S.}~\bibnamefont
  {Datta}},\ }\href {https://doi.org/10.1017/CBO9780511805776} {\emph {\bibinfo
  {title} {{Electronic Transport in Mesoscopic Systems}}}}\ (\bibinfo
  {publisher} {Cambridge University Press},\ \bibinfo {address} {Cambridge},\
  \bibinfo {year} {1995})\BibitemShut {NoStop}%
\bibitem [{sup()}]{supp}%
  \BibitemOpen
  \href@noop {} {}\bibinfo {note} {See Supplemental Material, which cites
  additional
  Ref.~\cite{AURICH199191,Bolte1993,BOGOMOLNY1997219,PhysRevE.106.034114,Katok}.}\BibitemShut
  {Stop}%
\bibitem [{\citenamefont {Kesten}(1959)}]{Kesten1959}%
  \BibitemOpen
  \bibfield  {author} {\bibinfo {author} {\bibfnamefont {H.}~\bibnamefont
  {Kesten}},\ }\bibfield  {title} {\bibinfo {title} {{Symmetric random walks on
  groups}},\ }\href@noop {} {\bibfield  {journal} {\bibinfo  {journal} {Trans.
  Am. Math. Soc.}\ }\textbf {\bibinfo {volume} {92}},\ \bibinfo {pages}
  {336–354} (\bibinfo {year} {1959})}\BibitemShut {NoStop}%
\bibitem [{\citenamefont {Brooks}(1982)}]{Brooks1982}%
  \BibitemOpen
  \bibfield  {author} {\bibinfo {author} {\bibfnamefont {R.}~\bibnamefont
  {Brooks}},\ }\bibfield  {title} {\bibinfo {title} {{Amenability and the
  spectrum of the Laplacian}},\ }\href@noop {} {\bibfield  {journal} {\bibinfo
  {journal} {Bull. Am. Math. Soc.}\ }\textbf {\bibinfo {volume} {6}},\ \bibinfo
  {pages} {87–89} (\bibinfo {year} {1982})}\BibitemShut {NoStop}%
\bibitem [{\citenamefont {Woess}(2000)}]{Woess2000}%
  \BibitemOpen
  \bibfield  {author} {\bibinfo {author} {\bibfnamefont {W.}~\bibnamefont
  {Woess}},\ }\href@noop {} {\emph {\bibinfo {title} {Random Walks on Infinite
  Graphs and Groups}}}\ (\bibinfo  {publisher} {Cambridge University Press},\
  \bibinfo {address} {Cambridge},\ \bibinfo {year} {2000})\BibitemShut
  {NoStop}%
\bibitem [{\citenamefont {Mosseri}\ and\ \citenamefont
  {Vidal}(2023)}]{Mosseri2023}%
  \BibitemOpen
  \bibfield  {author} {\bibinfo {author} {\bibfnamefont {R.}~\bibnamefont
  {Mosseri}}\ and\ \bibinfo {author} {\bibfnamefont {J.}~\bibnamefont
  {Vidal}},\ }\bibfield  {title} {\bibinfo {title} {Density of states of
  tight-binding models in the hyperbolic plane},\ }\href
  {https://doi.org/10.1103/PhysRevB.108.035154} {\bibfield  {journal} {\bibinfo
   {journal} {Phys. Rev. B}\ }\textbf {\bibinfo {volume} {108}},\ \bibinfo
  {pages} {035154} (\bibinfo {year} {2023})}\BibitemShut {NoStop}%
\bibitem [{\citenamefont {Conder}(2007)}]{Conder:2007}%
  \BibitemOpen
  \bibfield  {author} {\bibinfo {author} {\bibfnamefont {M.}~\bibnamefont
  {Conder}},\ }\href
  {https://www.math.auckland.ac.nz/~conder/TriangleGroupQuotients101.txt}
  {\bibinfo {title} {Quotients of triangle groups acting on surfaces of genus 2
  to 101}} (\bibinfo {year} {2007})\BibitemShut {NoStop}%
\bibitem [{\citenamefont {Lux}\ and\ \citenamefont
  {Prodan}(2023{\natexlab{a}})}]{Lux2022}%
  \BibitemOpen
  \bibfield  {author} {\bibinfo {author} {\bibfnamefont {F.~R.}\ \bibnamefont
  {Lux}}\ and\ \bibinfo {author} {\bibfnamefont {E.}~\bibnamefont {Prodan}},\
  }\bibfield  {title} {\bibinfo {title} {Spectral and {Combinatorial} {Aspects}
  of {Cayley}-{Crystals}},\ }\href {https://doi.org/10.1007/s00023-023-01373-3}
  {\bibfield  {journal} {\bibinfo  {journal} {Ann. Henri Poincaré}\ }
  (\bibinfo {year} {2023}{\natexlab{a}})}\BibitemShut {NoStop}%
\bibitem [{\citenamefont {Stegmaier}\ \emph {et~al.}(2022)\citenamefont
  {Stegmaier}, \citenamefont {Upreti}, \citenamefont {Thomale},\ and\
  \citenamefont {Boettcher}}]{Stegmaier:2021}%
  \BibitemOpen
  \bibfield  {author} {\bibinfo {author} {\bibfnamefont {A.}~\bibnamefont
  {Stegmaier}}, \bibinfo {author} {\bibfnamefont {L.~K.}\ \bibnamefont
  {Upreti}}, \bibinfo {author} {\bibfnamefont {R.}~\bibnamefont {Thomale}},\
  and\ \bibinfo {author} {\bibfnamefont {I.}~\bibnamefont {Boettcher}},\
  }\bibfield  {title} {\bibinfo {title} {Universality of {Hofstadter}
  {Butterflies} on {Hyperbolic} {Lattices}},\ }\href
  {https://doi.org/10.1103/PhysRevLett.128.166402} {\bibfield  {journal}
  {\bibinfo  {journal} {Phys. Rev. Lett.}\ }\textbf {\bibinfo {volume} {128}},\
  \bibinfo {pages} {166402} (\bibinfo {year} {2022})}\BibitemShut {NoStop}%
\bibitem [{\citenamefont {Lux}\ and\ \citenamefont
  {Prodan}(2023{\natexlab{b}})}]{Lux2023}%
  \BibitemOpen
  \bibfield  {author} {\bibinfo {author} {\bibfnamefont {F.~R.}\ \bibnamefont
  {Lux}}\ and\ \bibinfo {author} {\bibfnamefont {E.}~\bibnamefont {Prodan}},\
  }\bibfield  {title} {\bibinfo {title} {Converging {Periodic} {Boundary}
  {Conditions} and {Detection} of {Topological} {Gaps} on {Regular}
  {Hyperbolic} {Tessellations}},\ }\href
  {https://doi.org/10.1103/PhysRevLett.131.176603} {\bibfield  {journal}
  {\bibinfo  {journal} {Phys. Rev. Lett.}\ }\textbf {\bibinfo {volume} {131}},\
  \bibinfo {pages} {176603} (\bibinfo {year} {2023}{\natexlab{b}})}\BibitemShut
  {NoStop}%
\bibitem [{\citenamefont {Rober}(2020)}]{LINS}%
  \BibitemOpen
  \bibfield  {author} {\bibinfo {author} {\bibfnamefont {F.}~\bibnamefont
  {Rober}},\ }\href@noop {} {\bibinfo {title} {The {GAP} package {LINS}}},\
  \bibinfo {howpublished} {\url{https://github.com/FriedrichRober/LINS}}
  (\bibinfo {year} {2020})\BibitemShut {NoStop}%
\bibitem [{GAP()}]{GAP4}%
  \BibitemOpen
  GAP,\ \href {https://www.gap-system.org} {\emph {\bibinfo {title} {{GAP --
  Groups, Algorithms, and Programming, Version 4.11.1}}}},\ \bibinfo
  {organization} {The GAP~Group} (\bibinfo {year} {2021})\BibitemShut {NoStop}%
\bibitem [{\citenamefont {Firth}(2004)}]{FirthThesis}%
  \BibitemOpen
  \bibfield  {author} {\bibinfo {author} {\bibfnamefont {D.}~\bibnamefont
  {Firth}},\ }\emph {\bibinfo {title} {An Algorithm to Find Normal Subgroups of
  a Finitely Presented Group, up to a Given Finite Index}},\ \href@noop {}
  {Ph.D. thesis},\ \bibinfo  {school} {University of Warwick} (\bibinfo {year}
  {2004})\BibitemShut {NoStop}%
\bibitem [{\citenamefont {Conder}\ and\ \citenamefont
  {{Dobcs\'anyi}}(2005)}]{Conder2005}%
  \BibitemOpen
  \bibfield  {author} {\bibinfo {author} {\bibfnamefont {M.}~\bibnamefont
  {Conder}}\ and\ \bibinfo {author} {\bibfnamefont {P.}~\bibnamefont
  {{Dobcs\'anyi}}},\ }\bibfield  {title} {\bibinfo {title} {Applications and
  adaptations of the low index subgroups procedure},\ }\href
  {https://doi.org/10.1090/S0025-5718-04-01647-3} {\bibfield  {journal}
  {\bibinfo  {journal} {Math. Comp.}\ }\textbf {\bibinfo {volume} {74}},\
  \bibinfo {pages} {485} (\bibinfo {year} {2005})}\BibitemShut {NoStop}%
\bibitem [{\citenamefont {Chen}\ \emph
  {et~al.}(2024{\natexlab{b}})\citenamefont {Chen}, \citenamefont {Maciejko},\
  and\ \citenamefont {Boettcher}}]{Chen2023localization:SDC}%
  \BibitemOpen
  \bibfield  {author} {\bibinfo {author} {\bibfnamefont {A.}~\bibnamefont
  {Chen}}, \bibinfo {author} {\bibfnamefont {J.}~\bibnamefont {Maciejko}},\
  and\ \bibinfo {author} {\bibfnamefont {I.}~\bibnamefont {Boettcher}},\ }\href
  {https://doi.org/10.5683/SP3/3LWXHR} {\bibinfo {title} {{Supplemental Data
  for: Anderson localization transition in disordered hyperbolic lattices}}}
  (\bibinfo {year} {2024}{\natexlab{b}})\BibitemShut {NoStop}%
\bibitem [{\citenamefont {Wei\ss{}e}\ \emph {et~al.}(2006)\citenamefont
  {Wei\ss{}e}, \citenamefont {Wellein}, \citenamefont {Alvermann},\ and\
  \citenamefont {Fehske}}]{KPM}%
  \BibitemOpen
  \bibfield  {author} {\bibinfo {author} {\bibfnamefont {A.}~\bibnamefont
  {Wei\ss{}e}}, \bibinfo {author} {\bibfnamefont {G.}~\bibnamefont {Wellein}},
  \bibinfo {author} {\bibfnamefont {A.}~\bibnamefont {Alvermann}},\ and\
  \bibinfo {author} {\bibfnamefont {H.}~\bibnamefont {Fehske}},\ }\bibfield
  {title} {\bibinfo {title} {The kernel polynomial method},\ }\href
  {https://doi.org/10.1103/RevModPhys.78.275} {\bibfield  {journal} {\bibinfo
  {journal} {Rev. Mod. Phys.}\ }\textbf {\bibinfo {volume} {78}},\ \bibinfo
  {pages} {275} (\bibinfo {year} {2006})}\BibitemShut {NoStop}%
\bibitem [{\citenamefont {Groth}\ \emph {et~al.}(2014)\citenamefont {Groth},
  \citenamefont {Wimmer}, \citenamefont {Akhmerov},\ and\ \citenamefont
  {Waintal}}]{Groth2014}%
  \BibitemOpen
  \bibfield  {author} {\bibinfo {author} {\bibfnamefont {C.~W.}\ \bibnamefont
  {Groth}}, \bibinfo {author} {\bibfnamefont {M.}~\bibnamefont {Wimmer}},
  \bibinfo {author} {\bibfnamefont {A.~R.}\ \bibnamefont {Akhmerov}},\ and\
  \bibinfo {author} {\bibfnamefont {X.}~\bibnamefont {Waintal}},\ }\bibfield
  {title} {\bibinfo {title} {Kwant: a software package for quantum transport},\
  }\href {https://doi.org/10.1088/1367-2630/16/6/063065} {\bibfield  {journal}
  {\bibinfo  {journal} {New J. Phys.}\ }\textbf {\bibinfo {volume} {16}},\
  \bibinfo {pages} {063065} (\bibinfo {year} {2014})}\BibitemShut {NoStop}%
\bibitem [{\citenamefont {Bollhöfer}\ and\ \citenamefont
  {Notay}(2007)}]{Bollhofer2007}%
  \BibitemOpen
  \bibfield  {author} {\bibinfo {author} {\bibfnamefont {M.}~\bibnamefont
  {Bollhöfer}}\ and\ \bibinfo {author} {\bibfnamefont {Y.}~\bibnamefont
  {Notay}},\ }\bibfield  {title} {\bibinfo {title} {{JADAMILU: a software code
  for computing selected eigenvalues of large sparse symmetric matrices}},\
  }\href {https://doi.org/https://doi.org/10.1016/j.cpc.2007.08.004} {\bibfield
   {journal} {\bibinfo  {journal} {Comput. Phys. Commun.}\ }\textbf {\bibinfo
  {volume} {177}},\ \bibinfo {pages} {951} (\bibinfo {year}
  {2007})}\BibitemShut {NoStop}%
\bibitem [{\citenamefont {Shklovskii}\ \emph {et~al.}(1993)\citenamefont
  {Shklovskii}, \citenamefont {Shapiro}, \citenamefont {Sears}, \citenamefont
  {Lambrianides},\ and\ \citenamefont {Shore}}]{Shklovskii1993}%
  \BibitemOpen
  \bibfield  {author} {\bibinfo {author} {\bibfnamefont {B.~I.}\ \bibnamefont
  {Shklovskii}}, \bibinfo {author} {\bibfnamefont {B.}~\bibnamefont {Shapiro}},
  \bibinfo {author} {\bibfnamefont {B.~R.}\ \bibnamefont {Sears}}, \bibinfo
  {author} {\bibfnamefont {P.}~\bibnamefont {Lambrianides}},\ and\ \bibinfo
  {author} {\bibfnamefont {H.~B.}\ \bibnamefont {Shore}},\ }\bibfield  {title}
  {\bibinfo {title} {Statistics of spectra of disordered systems near the
  metal-insulator transition},\ }\href
  {https://doi.org/10.1103/PhysRevB.47.11487} {\bibfield  {journal} {\bibinfo
  {journal} {Phys. Rev. B}\ }\textbf {\bibinfo {volume} {47}},\ \bibinfo
  {pages} {11487} (\bibinfo {year} {1993})}\BibitemShut {NoStop}%
\bibitem [{\citenamefont {Mirlin}(2000)}]{Mirlin2000Statisics}%
  \BibitemOpen
  \bibfield  {author} {\bibinfo {author} {\bibfnamefont {A.~D.}\ \bibnamefont
  {Mirlin}},\ }\bibfield  {title} {\bibinfo {title} {Statistics of energy
  levels and eigenfunctions in disordered systems},\ }\href
  {https://doi.org/https://doi.org/10.1016/S0370-1573(99)00091-5} {\bibfield
  {journal} {\bibinfo  {journal} {Phys. Rep.}\ }\textbf {\bibinfo {volume}
  {326}},\ \bibinfo {pages} {259} (\bibinfo {year} {2000})}\BibitemShut
  {NoStop}%
\bibitem [{\citenamefont {Oganesyan}\ and\ \citenamefont
  {Huse}(2007)}]{Oganesyan2007}%
  \BibitemOpen
  \bibfield  {author} {\bibinfo {author} {\bibfnamefont {V.}~\bibnamefont
  {Oganesyan}}\ and\ \bibinfo {author} {\bibfnamefont {D.~A.}\ \bibnamefont
  {Huse}},\ }\bibfield  {title} {\bibinfo {title} {Localization of interacting
  fermions at high temperature},\ }\href
  {https://doi.org/10.1103/PhysRevB.75.155111} {\bibfield  {journal} {\bibinfo
  {journal} {Phys. Rev. B}\ }\textbf {\bibinfo {volume} {75}},\ \bibinfo
  {pages} {155111} (\bibinfo {year} {2007})}\BibitemShut {NoStop}%
\bibitem [{\citenamefont {Atas}\ \emph {et~al.}(2013)\citenamefont {Atas},
  \citenamefont {Bogomolny}, \citenamefont {Giraud},\ and\ \citenamefont
  {Roux}}]{Atas2013}%
  \BibitemOpen
  \bibfield  {author} {\bibinfo {author} {\bibfnamefont {Y.~Y.}\ \bibnamefont
  {Atas}}, \bibinfo {author} {\bibfnamefont {E.}~\bibnamefont {Bogomolny}},
  \bibinfo {author} {\bibfnamefont {O.}~\bibnamefont {Giraud}},\ and\ \bibinfo
  {author} {\bibfnamefont {G.}~\bibnamefont {Roux}},\ }\bibfield  {title}
  {\bibinfo {title} {{Distribution of the Ratio of Consecutive Level Spacings
  in Random Matrix Ensembles}},\ }\href
  {https://doi.org/10.1103/PhysRevLett.110.084101} {\bibfield  {journal}
  {\bibinfo  {journal} {Phys. Rev. Lett.}\ }\textbf {\bibinfo {volume} {110}},\
  \bibinfo {pages} {084101} (\bibinfo {year} {2013})}\BibitemShut {NoStop}%
\bibitem [{\citenamefont {Callan}\ and\ \citenamefont
  {Wilczek}(1990)}]{callan1990}%
  \BibitemOpen
  \bibfield  {author} {\bibinfo {author} {\bibfnamefont {C.~G.}\ \bibnamefont
  {Callan}}\ and\ \bibinfo {author} {\bibfnamefont {F.}~\bibnamefont
  {Wilczek}},\ }\bibfield  {title} {\bibinfo {title} {Infrared behavior at
  negative curvature},\ }\href {https://doi.org/10.1016/0550-3213(90)90451-I}
  {\bibfield  {journal} {\bibinfo  {journal} {Nucl. Phys. B}\ }\textbf
  {\bibinfo {volume} {340}},\ \bibinfo {pages} {366} (\bibinfo {year}
  {1990})}\BibitemShut {NoStop}%
\bibitem [{\citenamefont {Aurich}\ \emph {et~al.}(1991)\citenamefont {Aurich},
  \citenamefont {Bogomolny},\ and\ \citenamefont {Steiner}}]{AURICH199191}%
  \BibitemOpen
  \bibfield  {author} {\bibinfo {author} {\bibfnamefont {R.}~\bibnamefont
  {Aurich}}, \bibinfo {author} {\bibfnamefont {E.}~\bibnamefont {Bogomolny}},\
  and\ \bibinfo {author} {\bibfnamefont {F.}~\bibnamefont {Steiner}},\
  }\bibfield  {title} {\bibinfo {title} {{Periodic orbits on the regular
  hyperbolic octagon}},\ }\href
  {https://doi.org/https://doi.org/10.1016/0167-2789(91)90053-C} {\bibfield
  {journal} {\bibinfo  {journal} {Physica D}\ }\textbf {\bibinfo {volume}
  {48}},\ \bibinfo {pages} {91} (\bibinfo {year} {1991})}\BibitemShut {NoStop}%
\bibitem [{\citenamefont {Bolte}(1993)}]{Bolte1993}%
  \BibitemOpen
  \bibfield  {author} {\bibinfo {author} {\bibfnamefont {J.}~\bibnamefont
  {Bolte}},\ }\bibfield  {title} {\bibinfo {title} {Some studies on
  arithmetical chaos in classical and quantum mechanics},\ }\href
  {https://doi.org/10.1142/S0217979293003759} {\bibfield  {journal} {\bibinfo
  {journal} {Int. J. Mod. Phys. B}\ }\textbf {\bibinfo {volume} {07}},\
  \bibinfo {pages} {4451} (\bibinfo {year} {1993})}\BibitemShut {NoStop}%
\bibitem [{\citenamefont {Bogomolny}\ \emph {et~al.}(1997)\citenamefont
  {Bogomolny}, \citenamefont {Georgeot}, \citenamefont {Giannoni},\ and\
  \citenamefont {Schmit}}]{BOGOMOLNY1997219}%
  \BibitemOpen
  \bibfield  {author} {\bibinfo {author} {\bibfnamefont {E.}~\bibnamefont
  {Bogomolny}}, \bibinfo {author} {\bibfnamefont {B.}~\bibnamefont {Georgeot}},
  \bibinfo {author} {\bibfnamefont {M.-J.}\ \bibnamefont {Giannoni}},\ and\
  \bibinfo {author} {\bibfnamefont {C.}~\bibnamefont {Schmit}},\ }\bibfield
  {title} {\bibinfo {title} {Arithmetical chaos},\ }\href
  {https://www.sciencedirect.com/science/article/abs/pii/S0370157397000161}
  {\bibfield  {journal} {\bibinfo  {journal} {Phys. Rep.}\ }\textbf {\bibinfo
  {volume} {291}},\ \bibinfo {pages} {219} (\bibinfo {year}
  {1997})}\BibitemShut {NoStop}%
\bibitem [{\citenamefont {Attar}\ and\ \citenamefont
  {Boettcher}(2022)}]{PhysRevE.106.034114}%
  \BibitemOpen
  \bibfield  {author} {\bibinfo {author} {\bibfnamefont {A.}~\bibnamefont
  {Attar}}\ and\ \bibinfo {author} {\bibfnamefont {I.}~\bibnamefont
  {Boettcher}},\ }\bibfield  {title} {\bibinfo {title} {{Selberg trace formula
  in hyperbolic band theory}},\ }\href
  {https://doi.org/10.1103/PhysRevE.106.034114} {\bibfield  {journal} {\bibinfo
   {journal} {Phys. Rev. E}\ }\textbf {\bibinfo {volume} {106}},\ \bibinfo
  {pages} {034114} (\bibinfo {year} {2022})}\BibitemShut {NoStop}%
\bibitem [{\citenamefont {Katok}(1992)}]{Katok}%
  \BibitemOpen
  \bibfield  {author} {\bibinfo {author} {\bibfnamefont {S.}~\bibnamefont
  {Katok}},\ }\href@noop {} {\emph {\bibinfo {title} {Fuchsian Groups}}}\
  (\bibinfo  {publisher} {The University of Chicago Press},\ \bibinfo {address}
  {Chicago},\ \bibinfo {year} {1992})\BibitemShut {NoStop}%
\end{thebibliography}%

\let\addcontentsline\oldaddcontentsline 

\cleardoublepage

\setcounter{equation}{0}
\setcounter{figure}{0}
\setcounter{table}{0}
\renewcommand{\theequation}{S\arabic{equation}}
\renewcommand{\thefigure}{S\arabic{figure}}
\renewcommand{\thetable}{S\arabic{table}}

\renewcommand\theHtable{Appendix.\thetable}
\renewcommand\theHfigure{Appendix.\thefigure}
\renewcommand\theHequation{Appendix.\theequation}

\begin{center}
\textbf{\Large Supplemental Material}
\end{center}


\tableofcontents

\section{S1. Derivation of expected number of returns} \label{app:random_walk}
Consider a classical random walk on an infinite lattice, starting at an arbitrary site $i$. We show below that the expected number of returns is given by 
\begin{equation}
 \mu=\sum_{n=0}^\infty P_n, \label{eq:expected_returns}
\end{equation}
where $P_{n}$ is the probability that an $n$-step walk (or $n$-walk) starts and ends at site $i$. 

Define a random variable $b_{n}$ such that $b_{n}=1$ if the walker returns to site $i$ after $n$ steps, and $b_{n}=0$ otherwise. The expected number of returns is then
\begin{equation} 
\mu=\overline{\sum_{n=0}^\infty b_{n}}=\sum_{n=0}^\infty \overline{ b_{n}}, \end{equation} 
where the overbar denotes the average over all possible infinite random walks. Since $\overline{b_n}=1\cdot P_{n}+0\cdot(1-P_{n})=P_{n}$, we have $\mu=\sum_{n=0}^\infty P_{n}$ as stated in Eq.~\eqref{eq:expected_returns}.

\section{S2. Coherent sequences of \{8,8\} PBC clusters} \label{app:coherent_seq}

In this section, we detail the numerical procedure for constructing a fast-converging coherent sequence (\ref{eq:nested}) of PBC clusters to best approximate the thermodynamic-limit density of states (DOS). The first step is to identify a large pool of low-index normal subgroups of the translation symmetry group $\Gamma$ of the $\{8,8\}$ hyperbolic lattice \cite{Maciejko:2021,Maciejko:2022}: 
\begin{equation}
\Gamma=\langle\gamma_{1},\gamma_{2},\gamma_{3},\gamma_{4}:\gamma_{1}\gamma_{2}^{-1}\gamma_{3}\gamma_{4}^{-1}\gamma_{1}^{-1}\gamma_{2}\gamma_{3}^{-1}\gamma_{4}=e\rangle, \label{eq:Gamma88}
\end{equation} 
where $e$ denotes the identity element. Note that our method applies to other known hyperbolic Bravais lattices discussed in Refs.~\onlinecite{Boettcher:2022,Chen2023symmetry}. We obtain all normal subgroups of $\Gamma$ with indices up to 25 using the low-index normal subgroups procedure~\cite{FirthThesis,Conder2005} implemented within the computational algebra software GAP~\cite{LINS,GAP4}. The numbers of normal subgroups with indices 1 to 25 are (in order) \{1, 15, 40, 155, 156, 660, 400, 1635, 1210, 2430, 1464, 7300, 2380, 6120, 6240, 16851, 5220, 20745, 7240, 26970, 16640, 22140, 12720, 83400, 20306\}. In particular, it is essential to consider normal subgroups $G$ with non-Abelian quotient groups $\Gamma/G$. The PBC clusters built from normal subgroups with non-Abelian quotient capture the non-commutativity of the translation generators, which is characteristic of hyperbolic lattices. The numbers of normal subgroups with non-Abelian quotient with indices 1 to 25 are (in order) \{0, 0, 0, 0, 0, 60, 0, 240, 0, 90, 0, 1100, 0, 120, 0, 5040, 0, 2595, 0, 2790, 640, 180, 0, 27600, 0\}. 

Given any parent sequence of normal subgroups $\{G_{i}\vartriangleleft\Gamma\}_{i=1}^{i_{\max}}$, selected from the pool of low-index normal subgroups, we can construct a coherent sequence $\{\cnsg_{i}\}_{i=1}^{i_{\max}}$ through
\begin{equation} \cnsg_{i}:= \cnsg_{i-1}\cap G_{i}, \label{eq:coherent_seq2}
\end{equation}
and $\cnsg_{0}:=\Gamma$. For every $i=1,2,...$, we have $\cnsg_{i}\vartriangleleft\Gamma$ by the inductive use of Lemma S.1:\\[1ex]
\noindent \textbf{Lemma S.1} 
Let $H,K\vartriangleleft \Gamma$. Then $H\cap K\vartriangleleft \Gamma$.\\[1ex]
\emph{Proof.}
See the proof of Lemma A.1 of Ref.~\onlinecite{Tummuru2023}.\\[1ex]
It follows that every $\cnsg_{i}$ also satisfies $\cnsg_{i}= \cnsg_{i-1}\cap G_{i} \vartriangleleft \cnsg_{i-1}$ by Lemma S.2:\\[1ex]
\textbf{Lemma S.2} Let $H,K\vartriangleleft \Gamma$. Then $H\cap K\vartriangleleft H$ and $H\cap K\vartriangleleft K$.\\[1ex]
\emph{Proof.} The proof of Lemma A.1 of Ref.~\onlinecite{Tummuru2023}  shows that $H\cap K$ is a group, and since it is contained in both $H$ and $K$, it is a subgroup of both $H$ and $K$. Since by Lemma S.1, $H\cap K$ is normal in $\Gamma$, it is \emph{a fortiori} also normal in both $H$ and $K$, which are subgroups of $\Gamma$. Indeed, since any $h\in H$ is also in $\Gamma$, we have $h(H\cap K)=(H\cap K)h$, and since any $k\in K$ is also in $\Gamma$, we have $k(H\cap K)=(H\cap K)k$.\\[1ex]

Therefore, the first condition for a coherent sequence in Eq.~\eqref{eq:nested} in the main text is satisfied. Note that we only consider finite-length coherent sequences, so the second condition $\bigcap_{i=1}^\infty \cnsg_{i}=\{e\}$ is irrelevant. For each $\cnsg_{i}$, we take the quotient group $\Gamma/\cnsg_{i}$. Each coset $[g]\in\Gamma/\cnsg_{i}$ then represents a site in the corresponding PBC cluster, and two sites $[g_{1}],[g_{2}]$ are neighbors if they are related by  $[g_{1}]=[g_{2}][\gamma_{i}]$ with $\gamma_{i}$ a generator of $\Gamma$~\cite{Maciejko:2022}. The entire computational procedure of taking the intersection $\bigcap_{i} G_{i}$ of any set of normal subgroups $\{G_i\}$, computing the quotient group $\Gamma/\bigcap_{i} G_{i}$, and constructing the adjacency matrix of the PBC cluster given by $\Gamma/\bigcap_{i} G_{i}$ is conducted in GAP and identical to the procedure documented in the Supplementary Material of Ref.~\onlinecite{Tummuru2023}. 

Although any coherent sequence generated using the above procedure should ultimately approach the thermodynamic limit---specifically, the DOS of the clusters eventually converges to the thermodynamic-limit DOS which is well approximated by methods such as continued-fraction expansions~\cite{Mosseri2023} or the supercell method~\cite{Lenggenhager:2023}---our objective is to produce rapidly converging coherent sequences, from which we can obtain relatively small PBC clusters that closely approximate the thermodynamic limit. To this end, we devise a highly selective algorithm to construct the parent sequence. Our algorithm begins with a base step, followed by any desired number of iterative steps:

\begin{itemize}
\item Base step -- We arbitrarily select two small indices, e.g. 6 and 8, and employ GAP to compute the intersections $G_{1}\cap G_{2}$ of all pairs of normal subgroups ($G_{1}$, $G_{2}$) with indices 6 and 8 in $\Gamma$ respectively. To appreciate the scale of the search, there are $\sim$10$^6$ pairs in this case as there are 660 index-6 and 1635 index-8 normal subgroups. Note that by Theorem A.4 of Ref.~\onlinecite{Tummuru2023}, each pair must contain at least one normal subgroup with non-Abelian quotient in order for the intersection to have a non-Abelian quotient, so an additional step to discard intersections with Abelian quotients is required. Alternatively, we can simply demand that both $G_{1}$ and $G_{2}$ have non-Abelian quotient, in which case there are 14400 pairs. For each intersection $G_{1}\cap G_{2}$, we construct an adjacency matrix $A$ based on the quotient group $\Gamma/(G_{1}\cap G_{2})$, and compute the first few DOS moments $\langle A^{n}\rangle$, defined as 
\begin{equation}
\langle A^n \rangle = \frac{1}{N}\text{Tr}(A^{n}), \label{eq:dos_moment}
\end{equation}
where $N$ is the cluster size (total number of sites). The intersection $G_{1}\cap G_{2}$ that gives rise to the best DOS moments (i.e., closest to the exact values~\cite{Mosseri2023}) is kept for the next step and denoted $\cnsg_{2}$. Note that $\cnsg_{0}=\Gamma$ and $\cnsg_{1}=G_1$.

\item Iterative $i^{\text{th}}$ step -- We arbitrarily select a small index, e.g., 10, which can repeat previously used indices. For each index-10 normal subgroup $G_{i}$, we employ GAP to compute the intersection between $G_{i}$ and $\cnsg_{i-1}$. For each intersection, an adjacency matrix $A$ is constructed based on the quotient group $\Gamma/(\cnsg_{i-1}\cap G_{i})$, and then the first few DOS moments $\langle A^{n}\rangle$ are computed. The $G_{i}$ with intersection $\cnsg_{i}:=\cnsg_{i-1}\cap G_{i}$ that gives rise to the best DOS moments is kept for the next step.
\end{itemize}

We constructed a total of four finite coherent sequences for this work. The selected parent normal subgroups $G_i^{(s)}$ in sequence $s=1,2,3,4$ are specified in Table~\ref{table:parent_nsgs} by listing their generators $a,b,c,\ldots\in\Gamma$ such that $G_i^{(s)}$ is the normal closure of $\langle a,b,c,\ldots\rangle$. As detailed in Supplemental Material S3, the \{8,3\} clusters are further constructed by adding sublattice degrees of freedom to the \{8,8\} clusters, since \{8,8\} is the hyperbolic Bravais lattice of the \{8,3\} lattice. The DOS moments of the PBC clusters are listed in Tables~\ref{table:88_cluster_dos} and \ref{table:83_cluster_dos}, in comparison with the exact values~\cite{Mosseri2023}. The clusters used for the finite-size scaling analysis are chosen from these sequences, with the adjacency matrices available in the Supplementary Data  \cite{Chen2023localization:SDC}. Note that the DOS moments of the PBC clusters are greater than or equal to the exact values, which can be understood as follows. Given an adjacency matrix $A$, the diagonal element $(A^{n})_{ii}$ is the local DOS (LDOS) moment at site~$i$, which measures the number of $n$-cycles based at site~$i$. The DOS moment as defined in Eq.~\eqref{eq:dos_moment} is then the total number of $n$-cycles divided by the system size. The periodic boundary conditions of the PBC clusters imply the occasional presence of unexpected short cycles that are absent from the infinite lattice, increasing the DOS moments.

Lastly, we note that our method for constructing PBC clusters is based on the translation group $\Gamma$ rather than the full space group, which is the triangle group $\Delta(2,8,8)$~\cite{Lenggenhager:2023}, so our PBC clusters can sometimes break the point-group or sublattice symmetry. This is evidenced by the fact that the DOS moments of the $\{8,3\}$ clusters can sometimes be non-integers, indicating site-dependent LDOS moments. This occurs when the sublattice symmetry is broken and the sites within the unit cell are no longer related by symmetry.

\makeatletter\onecolumngrid@push\makeatother
\begin{table*}[p]
\begin{tabular}{|c>{\centering}p{13cm}c|}
\hline 
Parent NSGs & Generators & Index\tabularnewline
\hline 
\hline 
$G_{1}^{(1)}$ & $\gamma_{2}\gamma_{1},\gamma_{2}^{-1}\gamma_{1}^{-1},\gamma_{3}^{-2},\gamma_{4}\gamma_{1}^{-1},\gamma_{4}^{-1}\gamma_{1},\gamma_{1}^{3},\gamma_{1}\gamma_{2}^{-1}\gamma_{1},\gamma_{1}\gamma_{3}^{-2}\gamma_{1}^{-1},\gamma_{1}\gamma_{4}\gamma_{1},\gamma_{1}^{-1}\gamma_{3}^{-2}\gamma_{1},\gamma_{3}\gamma_{1}\gamma_{3}^{-1}\gamma_{1},$

$\gamma_{3}\gamma_{1}^{-1}\gamma_{3}^{-1}\gamma_{1}^{-1},\gamma_{3}\gamma_{2}\gamma_{3}^{-1}\gamma_{1}^{-1},\gamma_{3}\gamma_{4}\gamma_{3}^{-1}\gamma_{1}$ & 6\tabularnewline
\hline 
$G_{2}^{(1)}$ & $\gamma_{2},\gamma_{3}\gamma_{1}^{-1},\gamma_{3}^{-1}\gamma_{1},\gamma_{1}\gamma_{2}\gamma_{1}^{-1},\gamma_{1}\gamma_{3}\gamma_{1}^{-2},\gamma_{1}^{-4},\gamma_{1}^{-1}\gamma_{2}\gamma_{1},\gamma_{1}^{-1}\gamma_{3}^{-1}\gamma_{1}^{-2},\gamma_{1}^{-1}\gamma_{4}^{2}\gamma_{1}^{-1},\gamma_{1}^{-1}\gamma_{4}^{-2}\gamma_{1}^{-1},$

$\gamma_{4}\gamma_{1}\gamma_{4}\gamma_{1}^{-1},\gamma_{4}\gamma_{1}^{-1}\gamma_{4}^{-1}\gamma_{1}^{-1},\gamma_{4}\gamma_{2}\gamma_{4}^{-1},\gamma_{4}\gamma_{3}\gamma_{4}\gamma_{1}^{-1},\gamma_{4}^{2}\gamma_{1}^{-2},\gamma_{4}^{-1}\gamma_{1}\gamma_{4}^{-1}\gamma_{1}^{-1},\gamma_{4}^{-1}\gamma_{1}^{-1}\gamma_{4}\gamma_{1}^{-1},$

$\gamma_{4}^{-1}\gamma_{2}\gamma_{4}$ & 8\tabularnewline
\hline 
$G_{3}^{(1)}$ & $\gamma_{2}^{-2},\gamma_{4}\gamma_{3},\gamma_{4}^{-1}\gamma_{3}^{-1},\gamma_{1}^{2}\gamma_{3},\gamma_{1}\gamma_{2}^{-2}\gamma_{1}^{-1},\gamma_{1}\gamma_{3}\gamma_{1},\gamma_{1}\gamma_{3}^{-2},\gamma_{1}\gamma_{4}\gamma_{3}^{-1},\gamma_{1}\gamma_{4}^{-1}\gamma_{1},\gamma_{1}^{-2}\gamma_{3}^{-1},$

$\gamma_{1}^{-1}\gamma_{2}^{-2}\gamma_{1},\gamma_{1}^{-1}\gamma_{3}^{2},\gamma_{1}^{-1}\gamma_{4}^{-1}\gamma_{3},\gamma_{2}\gamma_{1}\gamma_{2}^{-1}\gamma_{1},\gamma_{2}\gamma_{1}^{-1}\gamma_{2}^{-1}\gamma_{1}^{-1},\gamma_{2}\gamma_{4}\gamma_{3}\gamma_{2}^{-1},\gamma_{2}\gamma_{4}^{-1}\gamma_{3}^{-1}\gamma_{2}^{-1},$

$\gamma_{3}\gamma_{1}^{-1}\gamma_{3},\gamma_{3}\gamma_{2}\gamma_{3}\gamma_{2}^{-1},(\gamma_{3}\gamma_{2}^{-1})^{2},\gamma_{3}^{-1}\gamma_{2}\gamma_{3}^{-1}\gamma_{2}^{-1},(\gamma_{3}^{-1}\gamma_{2}^{-1})^{2},\gamma_{1}\gamma_{2}\gamma_{1}^{-1}\gamma_{3}^{-1}\gamma_{2}^{-1}$ & 10\tabularnewline
\hline 
$G_{4}^{(1)}$ & $\gamma_{1}^{3},\gamma_{1}\gamma_{2}\gamma_{4}^{-1},\gamma_{1}\gamma_{3}^{-1}\gamma_{2},\gamma_{1}\gamma_{4}^{-1}\gamma_{3}^{-1},\gamma_{1}^{-1}\gamma_{2}\gamma_{4}^{-1}\gamma_{1}^{-1},\gamma_{1}^{-1}\gamma_{2}^{-1}\gamma_{3},\gamma_{1}^{-1}\gamma_{3}\gamma_{4},\gamma_{1}^{-1}\gamma_{3}^{-1}\gamma_{2}\gamma_{1}^{-1},$

$\gamma_{1}^{-1}\gamma_{4}\gamma_{2}^{-1},\gamma_{1}^{-1}\gamma_{4}^{-1}\gamma_{3}^{-1}\gamma_{1}^{-1},\gamma_{2}\gamma_{1}\gamma_{3}^{-1},(\gamma_{2}\gamma_{1}^{-1})^{2},\gamma_{2}^{3},\gamma_{2}\gamma_{3}\gamma_{4}^{-1}\gamma_{1}^{-1},\gamma_{2}\gamma_{3}^{-1}\gamma_{4}^{-1},\gamma_{2}\gamma_{4}\gamma_{3}^{-1}\gamma_{1}^{-1},$

$\gamma_{2}^{-1}\gamma_{1}\gamma_{4}^{-1}\gamma_{1}^{-1},\gamma_{2}^{-1}\gamma_{1}^{-1}\gamma_{4},\gamma_{2}^{-1}\gamma_{4}\gamma_{3},\gamma_{3}\gamma_{1}\gamma_{2}\gamma_{1}^{-1},\gamma_{3}\gamma_{2}\gamma_{3}^{-1}\gamma_{1}^{-1},\gamma_{3}\gamma_{2}^{-1}\gamma_{4},\gamma_{3}^{3},\gamma_{3}\gamma_{4}^{-2}\gamma_{1}^{-1},$

$(\gamma_{3}^{-1}\gamma_{1}^{-1})^{2},\gamma_{3}^{-1}\gamma_{2}^{-1}\gamma_{4}^{-1}\gamma_{1}^{-1}$ & 12\tabularnewline
\hline 
$G_{5}^{(1)}$ & $\gamma_{2}\gamma_{1}^{-1},\gamma_{2}^{-1}\gamma_{1},\gamma_{3},\gamma_{4}^{-2},\gamma_{1}\gamma_{2}\gamma_{1}^{-2},\gamma_{1}\gamma_{3}\gamma_{1}^{-1},\gamma_{1}\gamma_{4}^{-2}\gamma_{1}^{-1},\gamma_{1}^{-1}\gamma_{2}^{-1}\gamma_{1}^{2},\gamma_{1}^{-1}\gamma_{3}\gamma_{1},\gamma_{1}^{-1}\gamma_{4}^{-2}\gamma_{1},$

$\gamma_{4}\gamma_{1}\gamma_{4}^{-1}\gamma_{1},\gamma_{4}\gamma_{1}^{-1}\gamma_{4}^{-1}\gamma_{1}^{-1},\gamma_{4}\gamma_{2}\gamma_{4}^{-1}\gamma_{1},\gamma_{1}^{2}\gamma_{2}\gamma_{1}^{-3},\gamma_{1}^{2}\gamma_{3}\gamma_{1}^{-2},\gamma_{1}^{2}\gamma_{4}^{-2}\gamma_{1}^{-2},\gamma_{1}\gamma_{4}\gamma_{1}^{-1}\gamma_{4}^{-1}\gamma_{1}^{-2},$

$\gamma_{1}^{-2}\gamma_{2}^{-1}\gamma_{1}^{3},\gamma_{1}^{-2}\gamma_{3}\gamma_{1}^{2},\gamma_{1}^{-2}\gamma_{4}^{-2}\gamma_{1}^{2},\gamma_{1}^{-1}\gamma_{4}\gamma_{1}\gamma_{4}^{-1}\gamma_{1}^{2},\gamma_{1}^{7},\gamma_{1}^{3}\gamma_{2}\gamma_{1}^{3},\gamma_{1}^{3}\gamma_{3}\gamma_{1}^{-3},\gamma_{1}^{3}\gamma_{4}^{-2}\gamma_{1}^{-3},$

$\gamma_{1}^{2}\gamma_{4}\gamma_{1}^{-1}\gamma_{4}^{-1}\gamma_{1}^{-3},\gamma_{1}^{-3}\gamma_{3}\gamma_{1}^{3},\gamma_{1}^{-3}\gamma_{4}^{-2}\gamma_{1}^{3},\gamma_{1}^{-2}\gamma_{4}\gamma_{1}\gamma_{4}^{-1}\gamma_{1}^{3},\gamma_{1}^{3}\gamma_{4}\gamma_{1}^{-1}\gamma_{4}^{-1}\gamma_{1}^{3}$ & 14\tabularnewline
\hline 
\hline 
$G_{1}^{(2)}$ & $\gamma_{3}\gamma_{1}^{-1},\gamma_{3}^{-1}\gamma_{1},\gamma_{4}^{-2},\gamma_{1}^{2}\gamma_{2}^{-1},\gamma_{1}\gamma_{2}^{2},\gamma_{1}\gamma_{2}^{-1}\gamma_{1},\gamma_{1}\gamma_{3}\gamma_{2}^{-1},\gamma_{1}\gamma_{4}^{-2}\gamma_{1}^{-1},\gamma_{1}^{-2}\gamma_{2},\gamma_{1}^{-1}\gamma_{2}^{-2},$

$\gamma_{1}^{-1}\gamma_{3}^{-1}\gamma_{2},\gamma_{1}^{-1}\gamma_{4}^{-2}\gamma_{1},\gamma_{2}\gamma_{1}\gamma_{2},\gamma_{2}\gamma_{3}\gamma_{2},\gamma_{2}\gamma_{4}^{-2}\gamma_{2}^{-1},\gamma_{2}^{-1}\gamma_{4}^{-2}\gamma_{2},\gamma_{4}\gamma_{1}\gamma_{4}^{-1}\gamma_{1},\gamma_{4}\gamma_{1}^{-1}\gamma_{4}^{-1}\gamma_{1}^{-1},$

$\gamma_{4}\gamma_{2}\gamma_{4}^{-1}\gamma_{2},\gamma_{4}\gamma_{2}^{-1}\gamma_{4}^{-1}\gamma_{2}^{-1},\gamma_{4}\gamma_{3}\gamma_{4}^{-1}\gamma_{1},\gamma_{1}\gamma_{4}\gamma_{2}^{-1}\gamma_{4}^{-1}\gamma_{2}$ & 10\tabularnewline
\hline 
$G_{2}^{(2)}$ & $\gamma_{3}^{-2},\gamma_{4}\gamma_{2},\gamma_{4}^{-1}\gamma_{2}^{-1},\gamma_{1}^{2}\gamma_{2},\gamma_{1}\gamma_{2}\gamma_{1},\gamma_{1}\gamma_{2}^{-2},\gamma_{1}\gamma_{3}^{-2}\gamma_{1}^{-1},\gamma_{1}\gamma_{4}\gamma_{2}^{-1},\gamma_{1}\gamma_{4}^{-1}\gamma_{1},\gamma_{1}^{-2}\gamma_{2}^{-1},$

$\gamma_{1}^{-1}\gamma_{2}^{2},\gamma_{1}^{-1}\gamma_{3}^{-2}\gamma_{1},\gamma_{1}^{-1}\gamma_{4}^{-1}\gamma_{2},\gamma_{2}\gamma_{1}^{-1}\gamma_{2},\gamma_{2}\gamma_{3}^{-2}\gamma_{2}^{-1},\gamma_{2}^{-1}\gamma_{3}^{-2}\gamma_{2},\gamma_{3}\gamma_{1}\gamma_{3}^{-1}\gamma_{1},\gamma_{3}\gamma_{1}^{-1}\gamma_{3}^{-1}\gamma_{1}^{-1},$

$\gamma_{3}\gamma_{2}\gamma_{3}^{-1}\gamma_{2},\gamma_{3}\gamma_{2}^{-1}\gamma_{3}^{-1}\gamma_{2}^{-1},\gamma_{3}\gamma_{4}\gamma_{3}^{-1}\gamma_{2}^{-1},\gamma_{1}\gamma_{3}\gamma_{1}^{-1}\gamma_{3}^{-1}\gamma_{2}$ & 10\tabularnewline
\hline 
$G_{3}^{(2)}$ & $\gamma_{1}^{-2},\gamma_{4}^{-2},\gamma_{1}\gamma_{2}\gamma_{3},\gamma_{1}\gamma_{3}^{-1}\gamma_{2}^{-1},\gamma_{1}\gamma_{4}^{-2}\gamma_{1}^{-1},\gamma_{2}\gamma_{1}^{-2}\gamma_{2}^{-1},\gamma_{2}^{3},\gamma_{2}\gamma_{3}^{-2}\gamma_{1}^{-1},\gamma_{2}\gamma_{4}\gamma_{3},\gamma_{2}\gamma_{4}^{-1}\gamma_{3},$

$\gamma_{2}^{-1}\gamma_{1}\gamma_{3}^{-1},\gamma_{2}^{-1}\gamma_{1}^{-1}\gamma_{3}^{-1},\gamma_{2}^{-1}\gamma_{3}\gamma_{1}^{-1}\gamma_{2}^{-1},\gamma_{2}^{-1}\gamma_{3}^{-1}\gamma_{4}^{-1},\gamma_{2}^{-1}\gamma_{4}\gamma_{3}^{-1}\gamma_{1}^{-1},\gamma_{2}^{-1}\gamma_{4}^{-1}\gamma_{3}^{-1}\gamma_{1}^{-1},$

$\gamma_{3}\gamma_{2}\gamma_{4}^{-1},\gamma_{3}^{3},\gamma_{3}\gamma_{4}\gamma_{2}\gamma_{1}^{-1},\gamma_{3}\gamma_{4}^{-1}\gamma_{2}\gamma_{1}^{-1},\gamma_{3}^{-1}\gamma_{1}\gamma_{2}\gamma_{3}^{-1},\gamma_{3}^{-1}\gamma_{1}^{-1}\gamma_{2}\gamma_{3}^{-1},\gamma_{3}^{-1}\gamma_{2}^{2}\gamma_{1}^{-1},$

$\gamma_{4}\gamma_{1}\gamma_{4}^{-1}\gamma_{1}^{-1},\gamma_{4}\gamma_{1}^{-1}\gamma_{4}^{-1}\gamma_{1}^{-1},\gamma_{4}\gamma_{2}^{2}\gamma_{3}^{-1},\gamma_{1}\gamma_{2}^{-1}\gamma_{1}^{-1}\gamma_{3}^{-1}\gamma_{1}^{-1}$ & 12\tabularnewline
\hline 
$G_{4}^{(2)}$ & $\gamma_{2}^{-2},\gamma_{3}\gamma_{2}^{-1},\gamma_{3}^{-1}\gamma_{2}^{-1},\gamma_{4}\gamma_{1}^{-1},\gamma_{4}^{-1}\gamma_{1},\gamma_{1}\gamma_{2}^{-2}\gamma_{1}^{-1},\gamma_{1}\gamma_{3}\gamma_{2}^{-1}\gamma_{1}^{-1},\gamma_{1}\gamma_{3}^{-1}\gamma_{2}^{-1}\gamma_{1}^{-1},\gamma_{1}\gamma_{4}\gamma_{1}^{-2},$

$\gamma_{1}^{-1}\gamma_{2}^{-2}\gamma_{1},\gamma_{1}^{-1}\gamma_{3}^{-1}\gamma_{2}^{-1}\gamma_{1},\gamma_{1}^{-1}\gamma_{4}^{-1}\gamma_{1}^{2},\gamma_{2}\gamma_{1}\gamma_{2}^{-1}\gamma_{1},\gamma_{2}\gamma_{1}^{-1}\gamma_{2}^{-1}\gamma_{1}^{-1},\gamma_{2}\gamma_{4}\gamma_{2}^{-1}\gamma_{1},\gamma_{1}^{2}\gamma_{2}^{-2}\gamma_{1}^{-2},$

$\gamma_{1}^{2}\gamma_{3}\gamma_{2}^{-1}\gamma_{1}^{-2},\gamma_{1}^{2}\gamma_{4}\gamma_{1}^{-3},\gamma_{1}\gamma_{2}\gamma_{1}^{-1}\gamma_{2}^{-1}\gamma_{1}^{-2},\gamma_{1}^{-2}\gamma_{2}^{-2}\gamma_{1}^{2},\gamma_{1}^{-2}\gamma_{3}^{-1}\gamma_{2}^{-1}\gamma_{1}^{2},\gamma_{1}^{-2}\gamma_{4}^{-1}\gamma_{1}^{3},$

$\gamma_{1}^{-1}\gamma_{2}\gamma_{1}\gamma_{2}^{-1}\gamma_{1}^{2},\gamma_{1}^{7},\gamma_{1}^{3}\gamma_{2}^{-2}\gamma_{1}^{-3},\gamma_{1}^{3}\gamma_{3}\gamma_{2}^{-1}\gamma_{1}^{-3},\gamma_{1}^{2}\gamma_{2}\gamma_{1}^{-1}\gamma_{2}^{-1}\gamma_{1}^{-3},\gamma_{1}^{-3}\gamma_{2}^{-2}\gamma_{1}^{3},\gamma_{1}^{-2}\gamma_{2}\gamma_{1}\gamma_{2}^{-1}\gamma_{1}^{3},$

$\gamma_{1}^{3}\gamma_{2}\gamma_{1}^{-1}\gamma_{2}^{-1}\gamma_{1}^{3}$ & 14\tabularnewline
\hline 
\hline 
$G_{1}^{(3)}$ & $\gamma_{4}\gamma_{1}^{-1},\gamma_{4}^{-1}\gamma_{1},\gamma_{1}^{2}\gamma_{3},\gamma_{1}\gamma_{2}^{2},\gamma_{1}\gamma_{2}^{-1}\gamma_{3}^{-1},\gamma_{1}\gamma_{3}\gamma_{1},\gamma_{1}\gamma_{3}^{-1}\gamma_{2}^{-1},\gamma_{1}\gamma_{4}\gamma_{3},\gamma_{1}^{-2}\gamma_{3}^{-1},\gamma_{1}^{-1}\gamma_{2}\gamma_{3},$

$\gamma_{1}^{-1}\gamma_{2}^{-2},\gamma_{1}^{-1}\gamma_{3}\gamma_{2},\gamma_{1}^{-1}\gamma_{4}^{-1}\gamma_{3}^{-1},\gamma_{2}\gamma_{1}\gamma_{2},\gamma_{2}\gamma_{1}^{-1}\gamma_{3},\gamma_{2}\gamma_{4}\gamma_{2}$ & 7\tabularnewline
\hline 
$G_{2}^{(3)}$ & $\gamma_{2},\gamma_{3}\gamma_{1}^{-1},\gamma_{3}^{-1}\gamma_{1},\gamma_{4}^{-2},\gamma_{1}\gamma_{2}\gamma_{1}^{-1},\gamma_{1}\gamma_{3}\gamma_{1}^{-2},\gamma_{1}\gamma_{4}^{-2}\gamma_{1}^{-1},\gamma_{1}^{-4},\gamma_{1}^{-1}\gamma_{2}\gamma_{1},\gamma_{1}^{-1}\gamma_{3}^{-1}\gamma_{1}^{-2},$

$\gamma_{1}^{-1}\gamma_{4}^{-2}\gamma_{1},\gamma_{4}\gamma_{1}\gamma_{4}^{-1}\gamma_{1},\gamma_{4}\gamma_{1}^{-1}\gamma_{4}^{-1}\gamma_{1}^{-1},\gamma_{4}\gamma_{2}\gamma_{4}^{-1},\gamma_{4}\gamma_{3}\gamma_{4}^{-1}\gamma_{1},\gamma_{1}^{2}\gamma_{4}^{-2}\gamma_{1}^{-2},\gamma_{1}\gamma_{4}\gamma_{1}^{-1}\gamma_{4}^{-1}\gamma_{1}^{-2},$

$\gamma_{1}^{-1}\gamma_{4}\gamma_{1}\gamma_{4}^{-1}\gamma_{1}^{-2}$ & 8\tabularnewline
\hline 
$G_{3}^{(3)}$ & $\gamma_{1}^{-2},\gamma_{2}^{-2},\gamma_{3}^{-2},\gamma_{4}^{-2},\gamma_{1}\gamma_{2}^{-2}\gamma_{1}^{-1},\gamma_{1}\gamma_{3}^{-2}\gamma_{1}^{-1},\gamma_{1}\gamma_{4}^{-2}\gamma_{1}^{-1},\gamma_{2}\gamma_{1}\gamma_{3}^{-1}\gamma_{1}^{-1},\gamma_{2}\gamma_{1}^{-1}\gamma_{3}^{-1}\gamma_{1}^{-1},$

$\gamma_{2}\gamma_{3}\gamma_{4}^{-1}\gamma_{1}^{-1},\gamma_{2}\gamma_{4}^{-2}\gamma_{2}^{-1},\gamma_{3}\gamma_{1}\gamma_{2}^{-1}\gamma_{1}^{-1},\gamma_{3}\gamma_{1}^{-1}\gamma_{2}^{-1}\gamma_{1}^{-1},\gamma_{3}\gamma_{2}\gamma_{4}^{-1}\gamma_{2}^{-1},\gamma_{3}\gamma_{2}^{-1}\gamma_{4}^{-1}\gamma_{2}^{-1},$

$\gamma_{3}\gamma_{4}\gamma_{3}^{-1}\gamma_{1}^{-1},\gamma_{4}\gamma_{1}\gamma_{4}^{-1}\gamma_{2}^{-1},\gamma_{4}\gamma_{2}\gamma_{4}^{-1}\gamma_{1}^{-1},\gamma_{4}\gamma_{3}\gamma_{2}^{-1}\gamma_{1}^{-1},\gamma_{1}\gamma_{2}\gamma_{4}^{-2}\gamma_{2}^{-1}\gamma_{1}^{-1},$

$\gamma_{1}\gamma_{3}\gamma_{2}\gamma_{4}^{-1}\gamma_{2}^{-1}\gamma_{1}^{-1},\gamma_{1}\gamma_{3}\gamma_{2}^{-1}\gamma_{4}^{-1}\gamma_{2}^{-1}\gamma_{1}^{-1},\gamma_{1}\gamma_{4}\gamma_{1}\gamma_{4}^{-1}\gamma_{2}^{-1}\gamma_{1}^{-1},\gamma_{2}\gamma_{4}\gamma_{3}\gamma_{4}^{-1}\gamma_{2}^{-1}\gamma_{1}^{-1}$ & 10\tabularnewline
\hline 
$G_{4}^{(3)}$=$G_{4}^{(1)}$ & see above & 12\tabularnewline
\hline 
\hline 
$G_{1}^{(4)}$ & $\gamma_{1}^{-2},\gamma_{2},\gamma_{3}^{-2},\gamma_{4}^{-2},\gamma_{1}\gamma_{2}\gamma_{1}^{-1},\gamma_{1}\gamma_{3}^{-2}\gamma_{1}^{-1},\gamma_{1}\gamma_{4}^{-2}\gamma_{1}^{-1},\gamma_{3}\gamma_{1}\gamma_{4}^{-1}\gamma_{1}^{-1},\gamma_{3}\gamma_{1}^{-1}\gamma_{4}^{-1}\gamma_{1}^{-1},\gamma_{3}\gamma_{2}\gamma_{3}^{-1},$

$\gamma_{3}\gamma_{4}\gamma_{3}^{-1}\gamma_{1}^{-1},\gamma_{3}\gamma_{4}^{-1}\gamma_{3}^{-1}\gamma_{1}^{-1},\gamma_{4}\gamma_{1}\gamma_{3}^{-1}\gamma_{1}^{-1},\gamma_{4}\gamma_{2}\gamma_{4}^{-1}$ & 6\tabularnewline
\hline 
$G_{2}^{(4)}$ & $\gamma_{4}\gamma_{3}^{-1},\gamma_{4}^{-1}\gamma_{3},\gamma_{1}^{2}\gamma_{2}^{-1},\gamma_{1}\gamma_{2}\gamma_{3},\gamma_{1}\gamma_{2}^{-1}\gamma_{1},\gamma_{1}\gamma_{3}\gamma_{2},\gamma_{1}\gamma_{4}\gamma_{2},\gamma_{1}^{-2}\gamma_{2},\gamma_{1}^{-1}\gamma_{2}^{-1}\gamma_{3}^{-1},\gamma_{1}^{-1}\gamma_{3}^{2}\gamma_{1}^{-1},$

$\gamma_{1}^{-1}\gamma_{3}^{-1}\gamma_{2}^{-1},\gamma_{1}^{-1}\gamma_{4}^{-1}\gamma_{2}^{-1},\gamma_{2}\gamma_{1}\gamma_{3},\gamma_{2}^{2}\gamma_{3}\gamma_{1}^{-1},\gamma_{2}\gamma_{3}^{-2},\gamma_{2}^{-1}\gamma_{1}^{-1}\gamma_{3}^{-1},\gamma_{2}^{-2}\gamma_{3}\gamma_{1}^{-1},\gamma_{2}^{-1}\gamma_{3}^{2}$ & 8\tabularnewline
\hline 
$G_{3}^{(4)}$ & $\gamma_{1}^{-2},\gamma_{1}\gamma_{2}\gamma_{4},\gamma_{1}\gamma_{3}^{-1}\gamma_{4}^{-1},\gamma_{1}\gamma_{4}\gamma_{3},\gamma_{1}\gamma_{4}^{-1}\gamma_{2}^{-1},\gamma_{2}\gamma_{1}\gamma_{3}^{-1}\gamma_{1}^{-1},\gamma_{2}\gamma_{1}^{-1}\gamma_{3}^{-1}\gamma_{1}^{-1},\gamma_{2}^{3},\gamma_{2}\gamma_{3}\gamma_{2}\gamma_{1}^{-1},$

$\gamma_{2}\gamma_{4}^{-1}\gamma_{3},\gamma_{2}^{-1}\gamma_{1}\gamma_{4}^{-1},\gamma_{2}^{-1}\gamma_{1}^{-1}\gamma_{4}^{-1},\gamma_{2}^{-1}\gamma_{3}^{-1}\gamma_{4},\gamma_{2}^{-1}\gamma_{4}\gamma_{3}^{-1}\gamma_{1}^{-1},\gamma_{2}^{-1}\gamma_{4}^{-1}\gamma_{3}\gamma_{2}^{-1},$

$\gamma_{3}\gamma_{1}\gamma_{4},\gamma_{3}\gamma_{1}^{-1}\gamma_{4},\gamma_{3}\gamma_{2}\gamma_{4}^{-1},(\gamma_{3}\gamma_{2}^{-1})^{2},\gamma_{3}\gamma_{4}\gamma_{3}^{-1}\gamma_{2},\gamma_{3}\gamma_{4}^{-1}\gamma_{2}\gamma_{1}^{-1},\gamma_{3}^{-1}\gamma_{1}\gamma_{2}\gamma_{1}^{-1},$

$\gamma_{3}^{-1}\gamma_{1}^{-1}\gamma_{2}\gamma_{1}^{-1},\gamma_{3}^{-1}\gamma_{2}^{-1}\gamma_{3}^{-1}\gamma_{1}^{-1},\gamma_{4}\gamma_{2}\gamma_{3}\gamma_{2}^{-1},\gamma_{4}\gamma_{3}^{-2}\gamma_{1}^{-1}$ & 12\tabularnewline
\hline 
$G_{4}^{(4)}$ & $\gamma_{2}^{-2},\gamma_{4}^{-2},\gamma_{1}^{3},\gamma_{1}\gamma_{2}^{-2}\gamma_{1}^{-1},\gamma_{1}\gamma_{3}\gamma_{4}^{-1},\gamma_{1}\gamma_{4}^{-2}\gamma_{1}^{-1},\gamma_{1}^{-1}\gamma_{2}^{-2}\gamma_{1},\gamma_{1}^{-1}\gamma_{3}\gamma_{4}^{-1}\gamma_{1}^{-1},\gamma_{1}^{-1}\gamma_{4}\gamma_{3}^{-1},$

$\gamma_{1}^{-1}\gamma_{4}^{-1}\gamma_{3}^{-1},\gamma_{2}\gamma_{1}\gamma_{4}^{-1}\gamma_{1}^{-1},\gamma_{2}\gamma_{1}^{-1}\gamma_{3}\gamma_{1}^{-1},\gamma_{2}\gamma_{3}\gamma_{2}^{-1}\gamma_{1},\gamma_{2}\gamma_{3}^{-1}\gamma_{2}^{-1}\gamma_{1}^{-1},\gamma_{2}\gamma_{4}\gamma_{3}\gamma_{1},\gamma_{2}\gamma_{4}^{-1}\gamma_{3}\gamma_{1},$

$(\gamma_{3}\gamma_{1})^{2},\gamma_{3}\gamma_{1}^{-1}\gamma_{2}^{-1}\gamma_{1}^{-1},\gamma_{3}\gamma_{2}\gamma_{3}\gamma_{1}^{-1},\gamma_{3}\gamma_{2}^{-1}\gamma_{3}\gamma_{1}^{-1},\gamma_{3}^{3},\gamma_{3}^{-1}\gamma_{1}\gamma_{2}^{-1}\gamma_{1},\gamma_{3}^{-1}\gamma_{1}^{-1}\gamma_{4}^{-1},$

$\gamma_{3}^{-1}\gamma_{2}\gamma_{4}^{-1}\gamma_{1}^{-1},\gamma_{3}^{-1}\gamma_{2}^{-1}\gamma_{4}^{-1}\gamma_{1}^{-1},\gamma_{4}\gamma_{2}\gamma_{3}\gamma_{1},\gamma_{4}\gamma_{2}^{-1}\gamma_{3}\gamma_{1}$ & 12\tabularnewline
\hline 
\end{tabular}
\caption{We select the above normal subgroups of $\Gamma$, the translation symmetry group of lattice $\{8,8\}$    as defined in Eq.~\eqref{eq:Gamma88}, to form the parent sequences $\{G^{(s)}_i\}_i$, from which we construct the coherent sequences $\{\cnsg^{(s)}_i\}_i$ according to Eq.~\eqref{eq:coherent_seq2}. The superscript $s=1,...,4$ is the sequence label. Here we define the normal subgroups by their generators, which are products of the generators of $\Gamma$.} \label{table:parent_nsgs}
\end{table*}
\clearpage
\makeatletter\onecolumngrid@pop\makeatother

\begin{table}
\begin{tabular}{|cccccc|}
\hline 
Cluster & $N$ & $\langle A^{2}\rangle$ & $\langle A^{4}\rangle$ & $\langle A^{6}\rangle$ & $\langle A^{8}\rangle$\tabularnewline
\hline 
\multicolumn{6}{|c|}{\ coherent sequence \#1 \ }\\
\hline 
$C_{1}^{(1)}$ & 6 & 22 & 934 & 49582 & \multicolumn{1}{c|}{2937334}\tabularnewline
\hline 
$C_{2}^{(1)}$ & 48 & \textbf{8} & 170 & 6838 & 375470\tabularnewline
\hline 
$C_{3}^{(1)}$ & 480 & \textbf{8} & \textbf{120} & 2818 & 104380\tabularnewline
\hline 
$C_{4}^{(1)}$ & 5760 & \textbf{8} & \textbf{120} & 2224 & 47960\tabularnewline
\hline 
$C_{5}^{(1)}$ & 40320 & \textbf{8} & \textbf{120} & \textbf{2192} & 44972\tabularnewline
\hline 
\multicolumn{6}{|c|}{\ coherent sequence \#2\ }\\
\hline 
$C_{1}^{(2)}$ & {10} & {14} & {534} & {28604} & {1725862}\tabularnewline
\hline 
$C_{2}^{(2)}$ & 100 & \textbf{8} & 144 & 4364 & 200672\tabularnewline
\hline 
$C_{3}^{(2)}$ & 1200 & \textbf{8} & \textbf{120} & 2294 & 55120\tabularnewline
\hline 
$C_{4}^{(2)}$ & 16800 & \textbf{8} & \textbf{120} & 2216 & 46664\tabularnewline
\hline 
\multicolumn{6}{|c|}{\ coherent sequence \#3 \ }\\
\hline 
$C_{1}^{(3)}$ & 7 & 12  & 604 & 37590 & 2397836\tabularnewline
\hline 
$C_{2}^{(3)}$ & 56 & \textbf{8} & 156 & 5828 & 316476\tabularnewline
\hline 
$C_{3}^{(3)}$ & 560 & \textbf{8} & \textbf{120} & 2720 & 94232\tabularnewline
\hline 
$C_{4}^{(3)}$ & 6720 & \textbf{8} & \textbf{120} & 2234 & 48552\tabularnewline
\hline 
\multicolumn{6}{|c|}{\ coherent sequence \#4 \ }\\
\hline 
$C_{1}^{(4)}$ & 6 & 16  & 736 & 44416 & 2807296\tabularnewline
\hline 
$C_{2}^{(4)}$ & 24 & \textbf{8} & 210 & 11292 & 703256\tabularnewline
\hline 
$C_{3}^{(4)}$ & 288 & \textbf{8} & \textbf{120} & 2592 & 87114\tabularnewline
\hline 
$C_{4}^{(4)}$ & 3456 & \textbf{8} & \textbf{120} & 2202 & 47176\tabularnewline
\hline 
\multicolumn{6}{|c|}{\ exact \ }\\
\hline 
 & $\infty$ & \textbf{8} & \textbf{120} & \textbf{2192} & \textbf{44264}\tabularnewline
\hline 
\end{tabular}
\caption{DOS moments of the \{8,8\} PBC clusters $C^{(s)}_i=\Gamma/\cnsg^{(s)}_{i}$  where $s$ is the sequence label and the normal subgroups $\cnsg^{(s)}_i$ in the coherent sequences are constructed from the parent normal subgroups $G^{(s)}_i$ (listed in Table~\ref{table:parent_nsgs}) according to Eq.~\eqref{eq:coherent_seq2}. The exact DOS moments (printed in boldface) are obtained from Ref.~\onlinecite{Mosseri2023}. Odd moments vanish because \{8,8\} is bipartite.} \label{table:88_cluster_dos}
\end{table}

\section{S3. Building \{8, 3\}  clusters from \{8, 8\}  clusters } \label{app:83clusters}
Given the adjacency matrix $A_{\{8,8\}}$ of a $\{8,8\}$ PBC cluster with $N$ sites, we can construct the adjacency matrix $A_{\{8,3\}}$ of a $\{8,3\}$ cluster by introducing sublattice degrees of freedom to $A_{\{8,8\}}$, since $\{8,8\}$ is the hyperbolic Bravais lattice of $\{8,3\}$. Each site in $A_{\{8,8\}}$ is mapped to a corresponding unit cell of 16 sites. The first part of $A_{\{8,3\}}$ is the tensor product $\mathbb{1}_{N\times N}\otimes V$, where $\mathbb{1}_{N\times N}$ is the identity matrix and $V$ is the adjacency matrix of the unit cell (with open boundary), dubbed the \textit{intracell matrix}:
\begin{equation}
V=\left(\begin{array}{cccccccccccccccc}
0 & 1 & 0 & 0 & 0 & 0 & 0 & 1 & 1 & 0 & 0 & 0 & 0 & 0 & 0 & 0\\
1 & 0 & 1 & 0 & 0 & 0 & 0 & 0 & 0 & 1 & 0 & 0 & 0 & 0 & 0 & 0\\
0 & 1 & 0 & 1 & 0 & 0 & 0 & 0 & 0 & 0 & 1 & 0 & 0 & 0 & 0 & 0\\
0 & 0 & 1 & 0 & 1 & 0 & 0 & 0 & 0 & 0 & 0 & 1 & 0 & 0 & 0 & 0\\
0 & 0 & 0 & 1 & 0 & 1 & 0 & 0 & 0 & 0 & 0 & 0 & 1 & 0 & 0 & 0\\
0 & 0 & 0 & 0 & 1 & 0 & 1 & 0 & 0 & 0 & 0 & 0 & 0 & 1 & 0 & 0\\
0 & 0 & 0 & 0 & 0 & 1 & 0 & 1 & 0 & 0 & 0 & 0 & 0 & 0 & 1 & 0\\
1 & 0 & 0 & 0 & 0 & 0 & 1 & 0 & 0 & 0 & 0 & 0 & 0 & 0 & 0 & 1\\
1 & 0 & 0 & 0 & 0 & 0 & 0 & 0 & 0 & 0 & 0 & 0 & 0 & 0 & 0 & 0\\
0 & 1 & 0 & 0 & 0 & 0 & 0 & 0 & 0 & 0 & 0 & 0 & 0 & 0 & 0 & 0\\
0 & 0 & 1 & 0 & 0 & 0 & 0 & 0 & 0 & 0 & 0 & 0 & 0 & 0 & 0 & 0\\
0 & 0 & 0 & 1 & 0 & 0 & 0 & 0 & 0 & 0 & 0 & 0 & 0 & 0 & 0 & 0\\
0 & 0 & 0 & 0 & 1 & 0 & 0 & 0 & 0 & 0 & 0 & 0 & 0 & 0 & 0 & 0\\
0 & 0 & 0 & 0 & 0 & 1 & 0 & 0 & 0 & 0 & 0 & 0 & 0 & 0 & 0 & 0\\
0 & 0 & 0 & 0 & 0 & 0 & 1 & 0 & 0 & 0 & 0 & 0 & 0 & 0 & 0 & 0\\
0 & 0 & 0 & 0 & 0 & 0 & 0 & 1 & 0 & 0 & 0 & 0 & 0 & 0 & 0 & 0
\end{array}\right).
\end{equation}
The second part of $A_{\{8,3\}}$ describes how the unit cells are connected. For this we define \textit{intercell matrices} $T_{i}$ for $i=1,...,8$, which encode the $\{8,3\}$ edges connecting a unit cell to its neighboring unit cell in the $\gamma_{i}$ direction, where $\gamma_{i}\in\{\gamma_{1},\gamma_{2},\gamma_{3},\gamma_{4},\gamma_{5}$$=$$\gamma_{1}^{-1},\gamma_{6}$$=$$\gamma_{2}^{-1},\gamma_{7}$$=$$\gamma_{3}^{-1},\gamma_{8}$$=$$\gamma_{4}^{-1}\}$ is a translation generator of $\Gamma$ (including inverses). $T_{i}$ are sparse matrices with mostly zeros, so we list only the nonzero elements below:  
\begin{align}
\left(T_{1}\right)_{9,12}&=\left(T_{1}\right)_{16,13}=1,\nn\\
\left(T_{2}\right)_{9,14}&=\left(T_{2}\right)_{10,13}=1,\nn\\
\left(T_{3}\right)_{10,15}&=\left(T_{3}\right)_{11,14}=1,\nn\\
\left(T_{4}\right)_{11,16}&=\left(T_{4}\right)_{12,15}=1,\nn\\
\left(T_{5}\right)_{12,9}&=\left(T_{5}\right)_{13,16}=1,\nn\\
\left(T_{6}\right)_{14,9}&=\left(T_{6}\right)_{13,10}=1,\nn\\
\left(T_{7}\right)_{15,10}&=\left(T_{7}\right)_{14,11}=1,\nn\\
\left(T_{8}\right)_{16,11}&=\left(T_{8}\right)_{15,12}=1.
\end{align}
Note that we obtain $V$ and $T_{i}$ by inspecting the $\{8,3\}$ unit cell, formed by an octagon with one additional edge attached to each vertex (see Fig.~5 of Ref.~\onlinecite{Chen2023symmetry} for a diagram). Then for each pair of $\{8,8\}$ neighboring sites ($n,m)$ such that $\left(A_{\{8,8\}}\right)_{n,m}=1$, we recall which generator\footnote{For sufficiently large clusters, there is a unique generator translating from site $n$ to a neighbor $m$. This is not the case for small clusters. As an extreme example, the adjacency matrix of a single-site $\{8,8\}$ cluster is $A_{\{8,8\}}=8$, and all 8 generators translate the site to itself.} $\gamma_{j}$ translates from $n$ to $m$ (this information is recorded during the construction of $\{8,8\}$ clusters \cite{Tummuru2023}) and add to $A_{\{8,3\}}$ the tensor product $U\otimes T_{j}$, where $U$ is a $N\times N$ matrix with all zero entries except for $U_{nm}=1$.

\begin{table*}
\begin{tabular}{|ccccccccccccc|}
\hline 
Cluster & $N$ & $\langle A^{2}\rangle$ & $\langle A^{4}\rangle$ & $\langle A^{6}\rangle$ & $\langle A^{8}\rangle$ & $\langle A^{10}\rangle$ & $\langle A^{12}\rangle$ & $\langle A^{14}\rangle$ & $\langle A^{16}\rangle$ & $\langle A^{18}\rangle$ & $\langle A^{20}\rangle$ & $\langle A^{22}\rangle$\tabularnewline
\hline 
\multicolumn{13}{|c|}{\ coherent sequence \#1 \ }\\
\hline 
$C_{1,\{8,3\}}^{(1)}$ & 96 & \textbf{3} & \textbf{15} & 88.5 & 575 & 3983 & 28915.5 & 217836 & 1692003 & 13485007.5 & 109838535 & 911212118\tabularnewline
\hline 
$C_{2,\{8,3\}}^{(1)}$ & {768} & \textbf{3} & \textbf{15} & \textbf{87} & \textbf{549} & \textbf{3663} & 25410 & 181334.5 & 1322245 & 9807798 & 73787300 & 561933578.5\tabularnewline
\hline 
$C_{3,\{8,3\}}^{(1)}$ & {7680} & \textbf{3} & \textbf{15} & \textbf{87} & \textbf{549} & \textbf{3663} & 25408.5 & 181285.5 & 1321215 & 9790246.5 & 73524675 & 558345587.5\tabularnewline
\hline 
$C_{4,\{8,3\}}^{(1)}$ & 92160 & \textbf{3} & \textbf{15} & \textbf{87} & \textbf{549} & \textbf{3663} & \textbf{25407} & \textbf{181233} & \textbf{1320117} & \textbf{9772359} & \textbf{73273755} & 555158299\tabularnewline
\hline 
\multicolumn{13}{|c|}{\ coherent sequence \#2 \ }\\
\hline 
$C_{1,\{8,3\}}^{(2)}$ & 160 & \textbf{3} & \textbf{15} & 88.5 & 575 & 3983 & 28879.5 & 216758 & 1672003 & 13187982 & 105959850 & 864592704.5\tabularnewline
\hline 
$C_{2,\{8,3\}}^{(2)}$ & {1600} & \textbf{3} & \textbf{15} & \textbf{87} & \textbf{549} & \textbf{3663} & 25416 & 181537.5 & 1326239 & 9868020 & 74560395 & 570840526\tabularnewline
\hline 
$C_{3,\{8,3\}}^{(2)}$ & 19200 & \textbf{3} & \textbf{15} & \textbf{87} & \textbf{549} & \textbf{3663} & \textbf{25407} & \textbf{181233} & \textbf{1320117} & \textbf{9772359} & 73273785 & 555159770.25\tabularnewline
\hline 
\multicolumn{13}{|c|}{\ coherent sequence \#3 \ }\\
\hline 
$C_{1,\{8,3\}}^{(3)}$ & 112 & \textbf{3} & \textbf{15} & \textbf{87} & \textbf{549} & 3678 & 25845 & 189150 & 1434729 & 11231925 & 90385815 & 744638205\tabularnewline
\hline 
$C_{2,\{8,3\}}^{(3)}$ & {896} & \textbf{3} & \textbf{15} & \textbf{87} & \textbf{549} & \textbf{3663} & 25413 & 181425.5 & 1323921 & 9831988.5 & 74088070 & 565309819.5\tabularnewline
\hline 
$C_{3,\{8,3\}}^{(3)}$ & {8960} & \textbf{3} & \textbf{15} & \textbf{87} & \textbf{549} & \textbf{3663} & \textbf{25407} & \textbf{181233} & 1320121 & 9772539 & 73278390 & 555248284.5\tabularnewline
\hline 
\multicolumn{13}{|c|}{\ coherent sequence \#4 \ }\\
\hline 
$C_{1,\{8,3\}}^{(4)}$ & 96 & \textbf{3} & \textbf{15} & 88.5 & 575 & 3983 & 28915.5 & 217836 & 1692003 & 13485007.5 & 109838535 & 911212118\tabularnewline
\hline 
$C_{2,\{8,3\}}^{(4)}$ & {384} & \textbf{3} & \textbf{15} & \textbf{87} & \textbf{549} & \textbf{3663} & 25446 & 182363.5 & 1340429 & 10064671.5 & 76975010 & 598326726.5\tabularnewline
\hline 
$C_{3,\{8,3\}}^{(4)}$ & 4608 & \textbf{3} & \textbf{15} & \textbf{87} & \textbf{549} & \textbf{3663} & \textbf{25407} & \textbf{181233} & 1320121 & 9772551.75 & 73278977.5 & 555263855\tabularnewline
\hline 
\multicolumn{13}{|c|}{\ exact \ }\\
\hline 
 & $\infty$ & \textbf{3} & \textbf{15} & \textbf{87} & \textbf{549} & \textbf{3663} & \textbf{25407} & \textbf{181233} & \textbf{1320117} & \textbf{9772359} & \textbf{73273755} & \textbf{555158277}\tabularnewline
\hline 
\end{tabular}
\caption{DOS moments of the \{8,3\} PBC clusters constructed from the corresponding \{8,8\} PBC clusters in Table~\ref{table:88_cluster_dos}. The exact DOS moments (printed in boldface)  are obtained from Ref.~\onlinecite{Mosseri2023}.  Odd moments vanish because \{8,3\} is bipartite.} \label{table:83_cluster_dos}
\end{table*}

\section{S4. Probability distribution of $r$} \label{app:Pr}

\begin{figure}
\includegraphics[width=\linewidth]{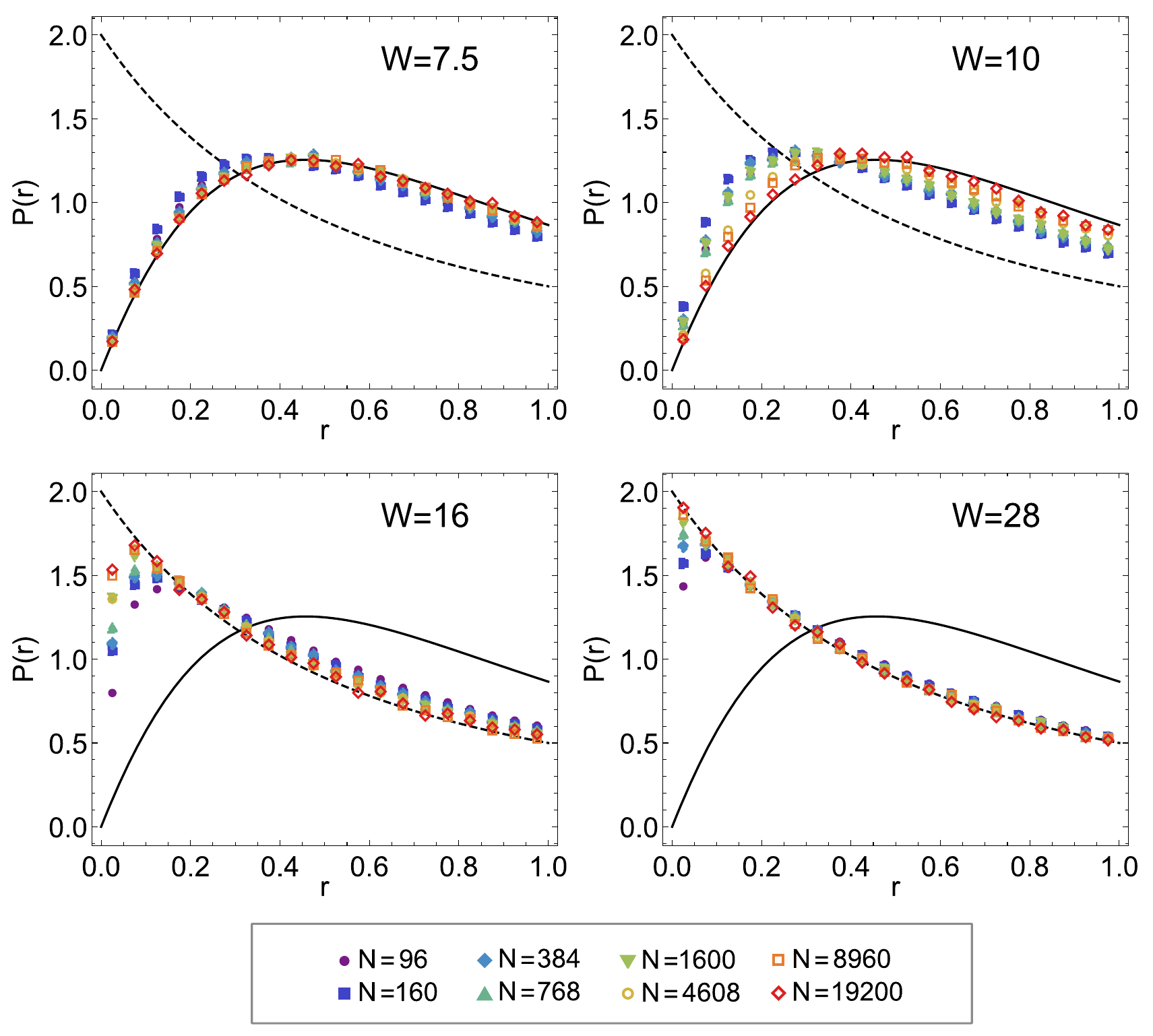}
\caption{The probability distribution $P(r)$ of the gap ratio $r$ computed for the Anderson model implemented on various $\{8,3\}$ PBC clusters. At disorder strength $W=7.5$ and $10$, $P(r)$ converges towards the Wigner surmise of the Gaussian orthogonal ensemble (GOE) (Eq.~\eqref{eq:P_GOE}, solid line) as system size $N$ increases. At disorder strength $W=16$ and $28$, $P(r)$ converges toward the Poisson distribution (Eq.~\eqref{eq:P_Poisson}, dashed line). This indicates that the critical disorder $W_{c}$ occurs somewhere between $W=10$ and $16$, in agreement with $W_{c}\approx13$ obtained through the finite-size scaling analysis. }
\label{fig:r_hist} 
\end{figure}

To study the localization transition in the hyperbolic Anderson model, we compute the ratios of consecutive gaps $r$ (as defined in Eq.~\eqref{eq:r} in the main text) near the center of the energy spectrum over many disorder realizations. The probability distribution $P(r)$ of $r$ in the delocalized phase obeys a simple form derived from the Wigner surmise of the Gaussian orthogonal ensemble (GOE) \cite{Atas2013}: 
\begin{equation} 
P_{\text{GOE}}(r)=\frac{27}{4}\frac{r+r^{2}}{(1+r+r^{2})^{5/2}}. \label{eq:P_GOE}
\end{equation}
In the localized phase, $P(r)$ follows the Poisson distribution: 
\begin{equation} 
P_{\text{Poisson}}(r)=\frac{2}{(1+r)^{2}}. \label{eq:P_Poisson}
\end{equation}
In Fig.~\ref{fig:r_hist}, we plot the probability distribution of $r$ for the Anderson model on various $\{8,3\}$ PBC clusters. At $W=7.5$, all $P(r)$ curves align with $P_{\text{GOE}}(r)$ (solid line), with better agreement observed in the larger systems. At $W=10$, the smaller systems have drifted away from $P_{\text{GOE}}(r)$ while the larger systems still maintain the GOE characteristic. As $W$ increases to $16$, $P(r)$ converges towards $P_{\text{Poisson}}(r)$ (dashed line) as the system size $N$ increases. At $W=28$, all systems follow the Poisson distribution. The above observation indicates that the critical disorder $W_{c}$ occurs somewhere between $W=10$ and $16$, in agreement with $W_{c}\approx15$ obtained through the crossing-drift analysis in S5 and finite-size scaling analysis in S6 and S7.

\section{S5. Finite-size drift of $\langle r\rangle$ crossing} \label{app:crossing_drift}

\begin{figure}
\includegraphics[width=\linewidth]{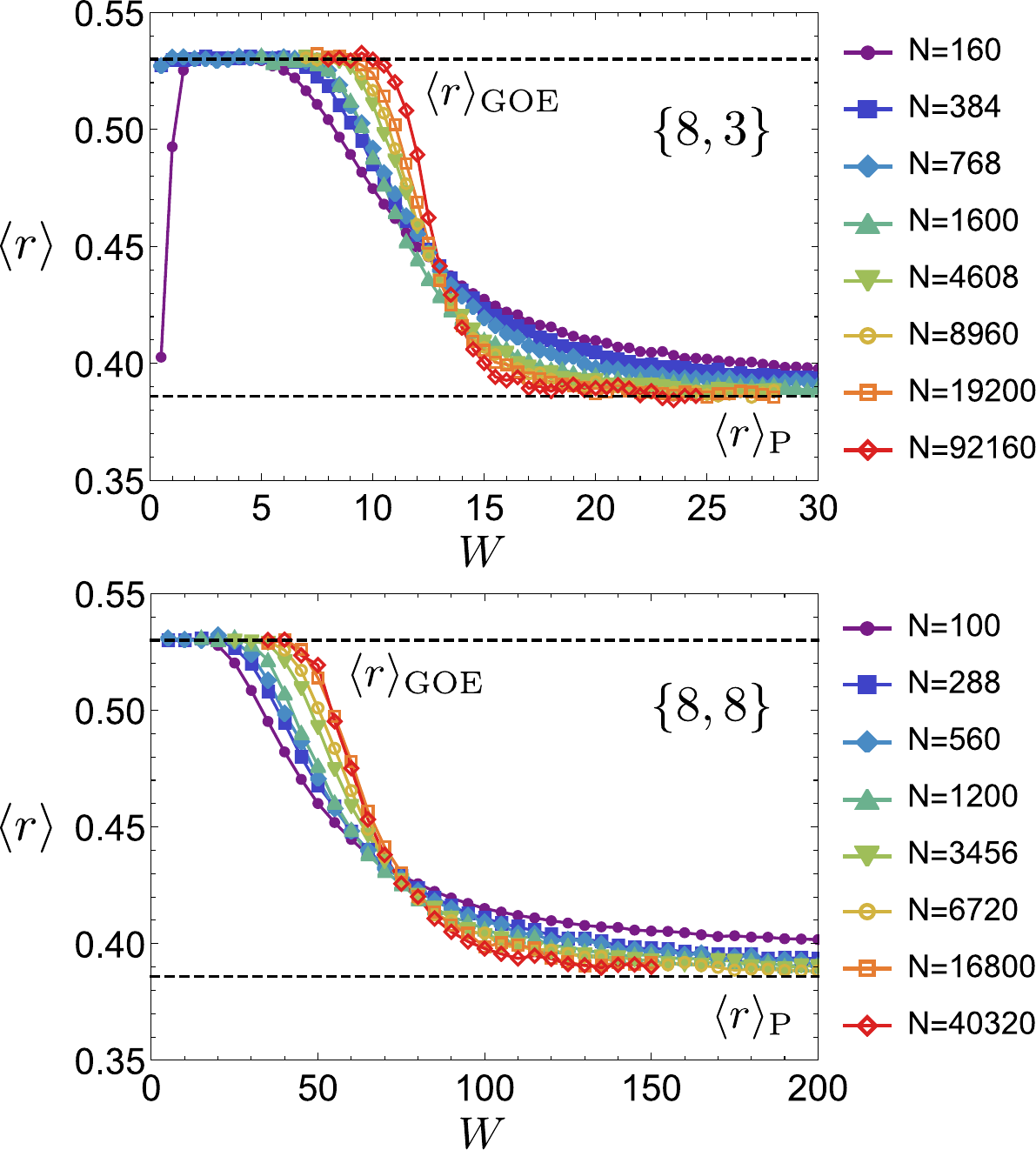}
\caption{The disorder-averaged gap ratio  $\langle r\rangle$ of the hyperbolic Anderson models decreases from the GOE value to the Poisson value as disorder strength $W$ increases. The transition region is narrower for larger systems. The intersection of the $\langle r\rangle$-curves, typically located at the critical point, suffers from a strong finite-size effect such that it drifts toward larger $W$ as system size increases. Fig.~\ref{fig:r_drift} of the main text shows the same plot zoomed-in on the transition region.}
\label{fig:full_r} 
\end{figure}

\begin{figure}
\includegraphics[width=\linewidth]{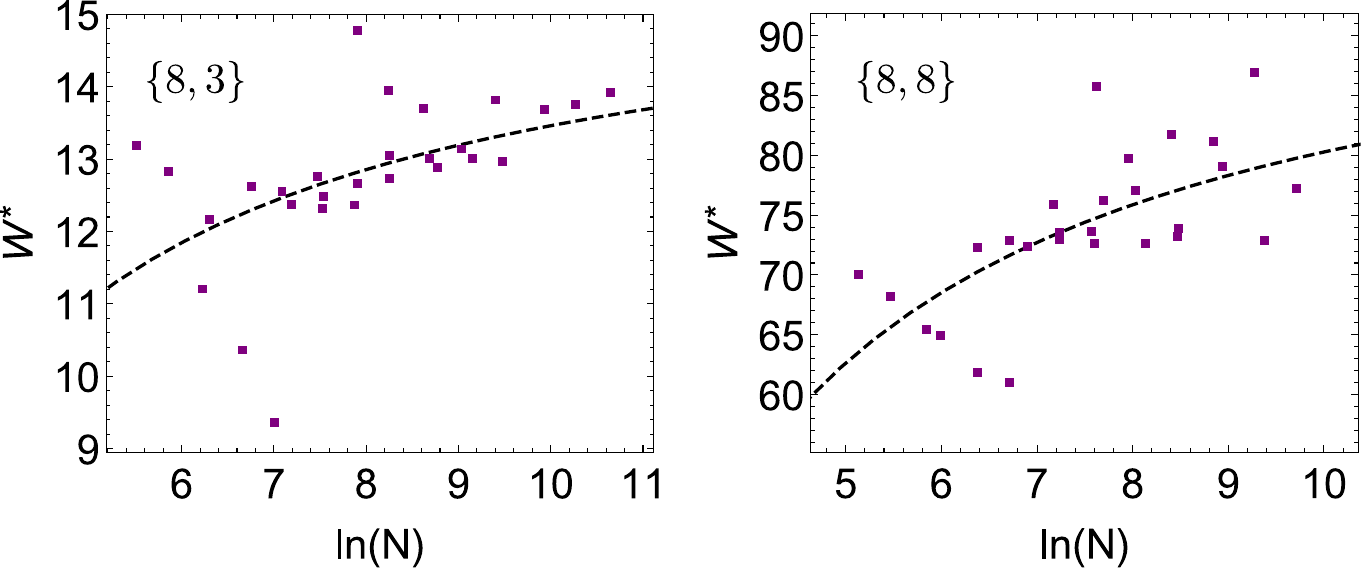}
\caption{The disorder strength $W^*$ at the pairwise $\langle r\rangle$-intersection drifts toward strong disorder as system size $N$ increases. The dashed curves show the extrapolation of $W^*$ to the infinite-$N$ limit using the ansatz in Eq.~\eqref{eq:drift_ansatz}. The best-fitted extrapolation gives $W_c \sim 15$ for $\{8,3\}$ and $W_c\sim100$ for $\{8,8\}$.}
\label{fig:crossing_drift} 
\end{figure}

The disorder-averaged gap ratio of the hyperbolic Anderson models, shown in Fig.~\ref{fig:full_r}, clearly reveals a crossover from the ergodic regime, indicated by the convergence toward $\langle r\rangle_{\text{GOE}}$, to the non-ergodic regime that converges to $\langle r\rangle_{\text{P}}$. This crossover is narrower for larger systems. Note that the $\{8,3\}$ lattices have atypical level statistics in the clean limit due to arithmetic quantum chaos~\cite{AURICH199191,Bolte1993,BOGOMOLNY1997219,PhysRevE.106.034114} that makes the distribution look Poissonian. This effect is well-understood in terms of arithmetic Fuchsian groups~\cite{Katok}, but does not play a role for our numerical study with finite $W>0$.

The intersection of the $\langle r\rangle$-curves, typically occurring at $W_c$, appears to suffer from a strong finite-size effect. Upon zooming in on the transition region (shown in Fig.~\ref{fig:r_drift} of the main text), we observe that the pairwise intersections of the $\langle r\rangle$-curves exhibit an overall drift toward stronger disorder as the system size $N$ increases. Furthermore, the intersections drift toward the Poisson distribution $\langle r\rangle_{\text{P}}$, suggesting that the critical region of the hyperbolic Anderson model is non-ergodic. We note that that such crossing-drift in the $\langle r\rangle$-curves has also been observed in the Anderson models on random regular graphs and Bethe lattices~\cite{Tikhonov2016,Biroli2018arxiv,Tikhonov2021}.

Here we consider the intersection between every pair of $\langle r\rangle$-curves in Fig.~\ref{fig:r_drift}. Specifically, we interpolate our data and compute the intersections of the interpolated curves. In Fig.~\ref{fig:crossing_drift}, we plot the disorder strength at the intersection, denoted by $W^*$, as a function of the averaged linear system size, $\ln(N)\equiv (\ln(N_1) + \ln(N_2))/2 $, of the two systems of sizes $N_1$ and $N_2$. We extrapolate $W^*$ to the thermodynamic limit using the ansatz 
\begin{equation}
W^*=W_c - \frac{\beta}{\ln(N)}.
\label{eq:drift_ansatz}
\end{equation}
The best fits, shown as dashed curves in Fig.~\ref{fig:crossing_drift}, occur at $W_c \sim 15$ for $\{8,3\}$ and $W_c \sim 100$ for $\{8,8\}$.

We note that using another ansatz, e.g., one with a different power of $\ln(N)$, one can also model this finite-size drift. In the case of random regular graphs, various models including Eq.~\eqref{eq:drift_ansatz} have been considered \cite{Sierant2023}. We leave it to future studies to determine the most suitable ansatz for the  finite-size drift exhibited by the hyperbolic Anderson models.

\section{S6. Finite-size scaling analysis -- Method} \label{app:finite_scaling_method}
In this section, we outline our finite-size scaling procedure, adapted from Ref.~\onlinecite{Mata2022}, for collapsing observables  $\eta(W,N)=(\langle r\rangle-\langle r\rangle_{\text{P}})/(\langle r\rangle_{\text{GOE}}-\langle r\rangle_{\text{P}})$ and $I(W,N)=\langle\text{IPR}\rangle$, where the data for $\langle r\rangle$ and $\langle 
\text{IPR} \rangle$ are shown in Supplemental Fig.~\ref{fig:full_r} and Fig.~\ref{fig:IPR_analysis}(a) of the main text respectively. Since the same procedure is applied to $\eta_{\{8,3\}}$, $I_{\{8,3\}}$, $\eta_{\{8,8\}}$, and $I_{\{8,8\}}$, we refer to them collectively as a generic observable $O(W,N)$. The first step is reorganizing the two-parameter data set $O(W,N)$ into single-parameter data sets, $O_{W}(N)$, each corresponding to a different disorder strength $W$. Then we assume a critical disorder strength $W_{c}$ and divide all data sets by the critical data set: $\tilde{O}_{W}(N):=O_{W}(N)/O_{W_{c}}(N)$.

Let us first analyze the delocalized side of the transition, $W<W_{c}$, assuming the volumetric scaling law in Eq.~\eqref{eq:vol_law} of the main text. We select $M$ data sets with $W\lesssim W_{c}$, labeled in ascending order $\{\tilde{O}_{W_{i}}(N)\}_{i=1}^{M}$ with $W_{1}$ being the smallest and $W_{M}$ being the closest (but not equal) to $W_{c}$. Starting with the first data set $\tilde{O}_{W_{1}}(N)$, we rescale the $x$-axis of the second data set by $\tilde{O}_{W_{2}}(N/\Lambda(W_{2}))$, where the value of $\Lambda(W_{2})$ is chosen so that the second data set collapses best onto the first. The goodness of the collapse is measured by a $\chi^{2}$ test (see below for definition) between the two data sets in the log-log scale. The optimal choice of $\Lambda(W_{2})$ and the corresponding  minimal $\chi^{2}$ value are recorded. Next, we rescale the $x$-axis of the third data set as $\tilde{O}_{W_{3}}(N/\Lambda(W_{3}))$ so that it collapses best onto the second rescaled data set $\tilde{O}_{W_{2}}(N/\Lambda(W_{2}))$. The minimal $\chi^2$ between $\tilde{O}_{W_{2}}(N/\Lambda(W_{2}))$ and $\tilde{O}_{W_{3}}(N/\Lambda(W_{3}))$ in the log-log scale and the corresponding $\Lambda(W_{3})$ are recorded. We repeat the pairwise data collapse for all $M$ data sets to obtain the scaling volumes $\Lambda(W_i)$ for $i=2,...,M$ and the total $\chi_{\text{deloc}}^{2}$ value, which is the sum of $(M-1)$ minimal pairwise $\chi^{2}$ values. If the linear scaling law in Eq.~\eqref{eq:lin_law} of the main text is assumed instead, replace $N$ by $\log(N)$ and the scaling volume $\Lambda$ by the scaling length $\xi$ in the above description.

For the localized side $W>W_{c}$, we  select $M$ data sets with $W\gtrsim W_{c}$, labeled in ascending order $\{\tilde{O}_{W_{i}}(N)\}_{i=M+1}^{2M}$ with $W_{2M}$ being the largest and  $W_{M+1}$ being the closest (but not equal) to $W_{c}$. The data sets are collapsed pairwise according to the above procedure, starting with the data set $\tilde{O}_{W_{2M}}(N)$ furthest away from $W_c$. This gives $\Lambda(W_i)$ or $\xi(W_i)$ (depending on the assumed scaling law for the localized side) for $i=M+1, ...,2M-1$ and the total $\chi_{\text{loc}}^{2}$ value that is the sum of $(M-1)$ pairwise $\chi^{2}$ tests. The sum $\chi_{\text{deloc}}^{2}+\chi_{\text{loc}}^{2}$ (shown in Fig.~\ref{fig:IPR_analysis}(b) of the main text) is the smallest for the best data collapses on both sides of the transition, indicating the correctly assumed scaling laws and critical disorder $W_{c}$.

\emph{$\chi^{2}$ test.---}Given two data sets $\{(x_{i},y_{i})\}_{i=1}^{S}$ and $\{(x'_{i},y'_{i})\}_{i=1}^{S'}$ with  similar ranges of $x$ values, we measure how well they collapse together by the following $\chi^{2}$ test. First we interpolate the first data set with a piecewise linear function $g(x)$ (i.e., linear segments between consecutive points). Then for every data point $(x'_{i},y'_{i})$ located within the $x$-range of the first data set, we compute $\lambda_i=(g(x'_{i})-y'_{i})^{2}/|g(x'_{i})|$. If some data point $(x'_{j},y'_{j})$  lies outside the $x$-range of the first data set, it does not contribute to the $\chi^{2}$ test.  We take the average of all $\lambda_i$  to be the $\chi^{2}$ value between these two data sets. Note that in Ref.~\onlinecite{Mata2022}, $\lambda_i$ is defined without the denominator, which deviates from the standard statistical  $\chi^2$ formula. In our analysis, we find that omitting the denominator can sometimes result in a $\chi^2$ test without a local minimum.

\begin{figure}[t!]
\includegraphics[width=\linewidth]{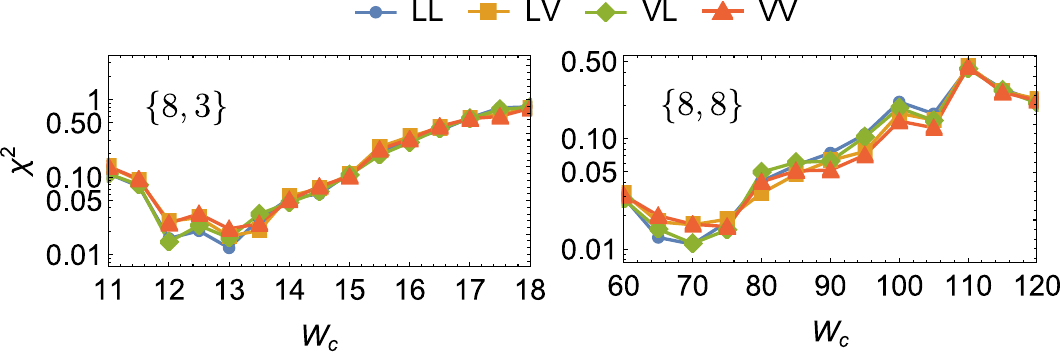}
\caption{The total $\chi^2$ of the $\langle r \rangle$ data collapse appears independent of the assumed scaling laws. Furthermore, the minimum occurs near the naive transition point where the $\langle r \rangle$ curves intersect, as seen in Fig.~\ref{fig:full_r}. This may be due to the noise in the $\langle r \rangle$ data, which renders the $\chi^2$ analysis of $\langle r \rangle$ unsuitable for determining the location of $W_c$.}
\label{fig:r_chi2} 
\end{figure}

\section{S7. Finite-size scaling analysis -- Results} \label{app:finite_scaling_results}

As shown in Fig.~\ref{fig:IPR_analysis}(b) of the main text, the total $\chi^2$ of the scaling collapses of $I(W,N)=\langle\text{IPR}\rangle$ is minimal at $W\sim15$ and $W\sim100$ for $\{8,3\}$ and $\{8,8\}$ respectively, consistent with the crossing-drift analysis. The scaling law which results in the best data collapse is linear on both sides of the phase transition.

On the other hand, we find that the $\chi^2$ obtained from collapsing the level statistics observable $\eta(W,N)=(\langle r\rangle-\langle r\rangle_{\text{P}})/(\langle r\rangle_{\text{GOE}}-\langle r\rangle_{\text{P}})$ is independent of the choice of scaling laws, as shown in Fig.~\ref{fig:r_chi2}. This may be caused by the noise in $\langle r \rangle$, which forms the dominant contribution to $\chi^2$. Therefore we conclude that the transition point cannot be inferred directly from the $\chi^2$ of $\langle r \rangle$ data collapse.

\begin{figure*}[t!]
\includegraphics[width=\linewidth]{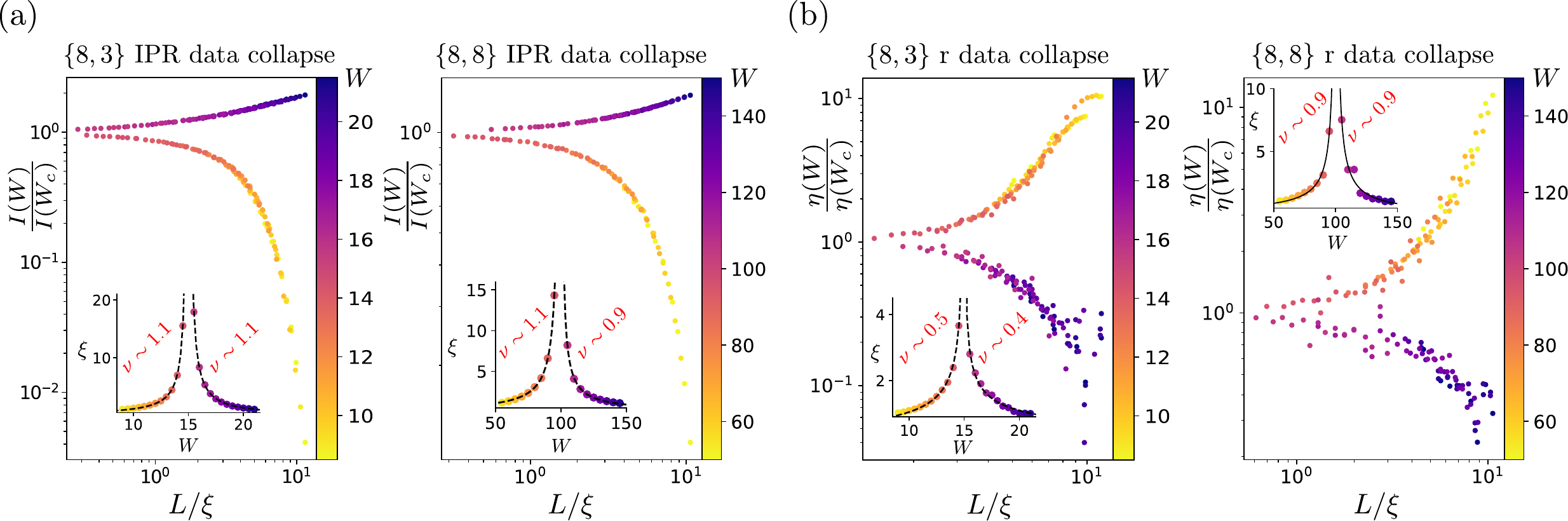}
\caption{The scaling collapses of observables $\eta(W,N)=(\langle r\rangle-\langle r\rangle_{\text{P}})/(\langle r\rangle_{\text{GOE}}-\langle r\rangle_{\text{P}})$ and $I(W,N)=\langle\text{IPR}\rangle$ are performed at $W_c=15\ (100)$ for the $\{8,3\}\ (\{8,8\})$ Anderson models, following the finite-size scaling method detailed in Sec. S6. The values of $W_c$ are estimated from the crossing-drift analysis in Sec. S5 and the $\chi^2$ analysis of $\langle \text{IPR} \rangle$ in Sec. S7. We assume linear scaling laws on both sides of the transition, which is shown to produce the best collapses according to the $\chi^2$ analysis of $\langle \text{IPR} \rangle$. The scaling length $\xi$ follows
$\xi(W)\propto|W-W_c|^{-\nu}$
close to the transition. By fitting the $\xi$ values obtained during the data collapse procedure, we obtain the critical scaling exponents $\nu$ as indicated in the insets.}
\label{fig:all_collapse} 
\end{figure*}

Fig.~\ref{fig:all_collapse} shows the scaling collapses of $I$ and $\eta$ at $W_c=15\ (100)$ for $\{8,3\}\ (\{8,8\}$). We use the linear scaling law on both sides of the transition. The scaling length $\xi$ diverges as
\begin{equation}
    \xi(W)\propto|W-W_c|^{-\nu}
\end{equation}
near the transition. Best-fits of the $\xi$ values, which have been individually determined from the pairwise data collapses, give the critical scaling exponents: $(\nu^{\{8,3\}}_{\eta,\text{deloc}},\nu^{\{8,3\}}_{\eta,\text{loc}})\approx(0.5,0.4)$,  $(\nu^{\{8,8\}}_{\eta,\text{deloc}},\nu^{\{8,8\}}_{\eta,\text{loc}})\approx(0.9,0.9)$, $(\nu^{\{8,3\}}_{I,\text{deloc}},\nu^{\{8,3\}}_{I,\text{loc}})\approx(1.1,1.1)$,  $(\nu^{\{8,8\}}_{I,\text{deloc}},\nu^{\{8,8\}}_{I,\text{loc}})\approx(1.1,0.9)$.

\section{S8. Fractal dimensions} \label{app:fractal_dim}

In disordered Euclidean lattices, the wave functions at the critical point of the localization transition exhibit multifractal structures, characterized by a continuous set of critical exponents governing the scaling behavior of the $q^{\text{th}}$ moment of the wave function, 
\begin{equation} 
\text{IPR}_{q}=\stackrel[i=1]{N}{\sum}|\psi(z_{i})|^{2q}, 
\end{equation} 
where $\psi(z_{i})$ is a wave function at the critical point and $i$ goes over all $N$ sites. This is also known as the generalized IPR and $q$ can be any real number (usually positive). At criticality, the disorder-averaged $\langle\text{IPR}_{q}\rangle$ scales as
\begin{equation} 
\langle\text{IPR}_{q}\rangle\sim L^{-\tau_{q}}, \label{eq:tau2}
\end{equation} 
where $L$ is the linear dimension of the system. The fractal dimensions, defined as $D_{q}\equiv\tau_{q}/(q-1)$, depend nontrivially on the value of $q$, which is a sign of multifractality.

\begin{figure}[ht]
\includegraphics[width=\linewidth]{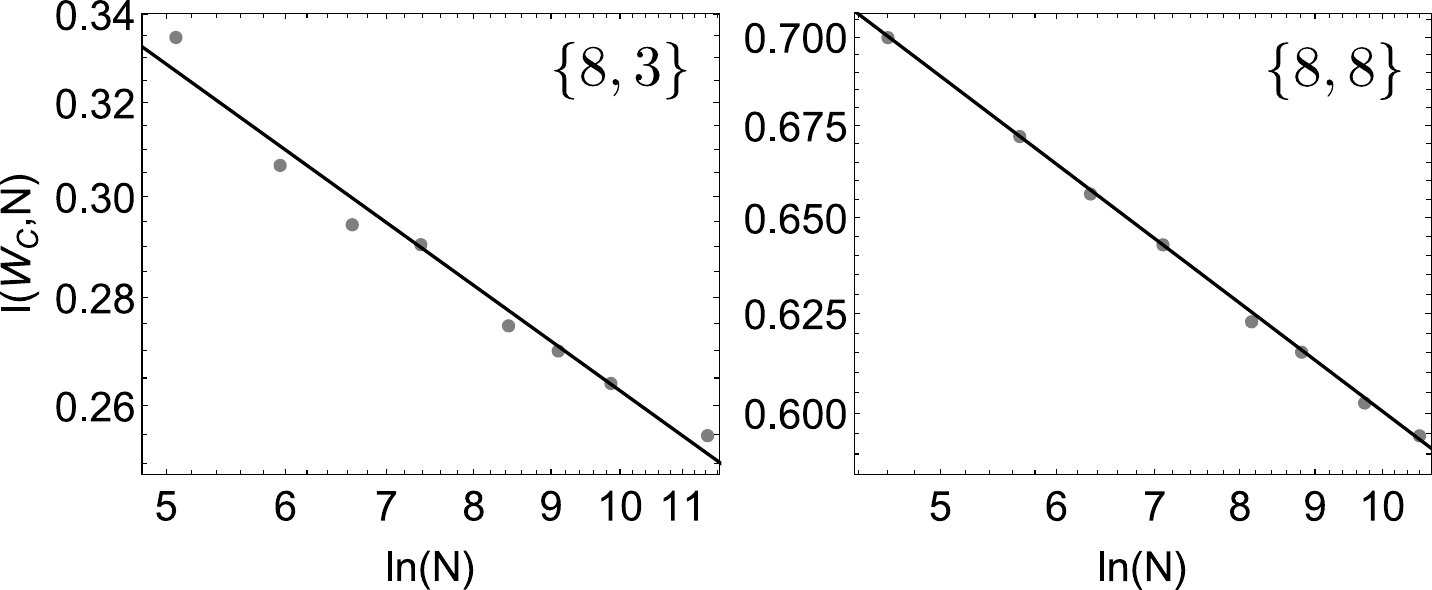}
\caption{The inverse participation ratio $I$ at the critical point $W_c\sim15$ (resp. $W_c\sim100$) of the $\{8,3\}$ (resp. $\{8,8\}$) Anderson model obeys $I \propto \ln(N)^{-\tau_2}$, where the best-fitting line (black) corresponds to $\tau_2^{\{8,3\}}\approx 0.3$ (resp. $\tau_2^{\{8,8\}}\approx 0.2$).}
\label{fig:fractal_dimension} 
\end{figure}

For the case $q=2$ considered in this work, we verify that the Anderson model on both the $\{8,3\}$ and $\{8,8\}$ lattices follow Eq.~\eqref{eq:tau2} for $L=\log(N)$, the diameter of the hyperbolic lattice. The best linear fits of our $\langle\text{IPR}_{2}\rangle$ data at criticality demonstrate that $D_2^{\{8,3\}}=\tau_2^{\{8,3\}}\approx 0.3$ and $D_2^{\{8,8\}}=\tau_2^{\{8,8\}}\approx 0.2$ (see Fig.~\ref{fig:fractal_dimension}).

\vfill


\end{document}